\documentclass[10pt]{article}

\usepackage[superscript,sort]{cite}

\usepackage{ifthen,ifpdf}
\ifpdf
	\usepackage[pdftex]{hyperref}
	\hypersetup{colorlinks=true,linkcolor=black,citecolor=black,urlcolor=blue,backref=page,bookmarks=true,breaklinks=true,plainpages=false }
	\usepackage[pdftex]{graphicx}
	\usepackage[usenames,pdftex]{color}
\else
	\usepackage[colorlinks=true,breaklinks=true,plainpages=false]{hyperref}
	\usepackage{graphicx}
	\usepackage[usenames]{color}
\fi

\usepackage{amsthm,amscd,amsxtra,amsfonts,amsmath,amssymb,multirow,dsfont}
\usepackage{wrapfig}
\usepackage[footnotesize]{caption}
\usepackage[tiny,compact]{titlesec}
\usepackage[textwidth=0.8in,textsize=footnotesize]{todonotes}
\usepackage{algorithm,algorithmic,extarrows}

\begin{document}

\title{A review of geometric, topological and graph theory apparatuses for the modeling and analysis of biomolecular data
}

\author{
Kelin Xia$^1$ \footnote{E-mail: xiakelin@ntu.edu.sg } and
Guo-Wei Wei$^{2,3}$ \footnote{E-mail: wei@math.msu.edu}\\
$^1$Division of Mathematical Sciences, School of Physical and Mathematical Sciences, \\
Nanyang Technological University, Singapore 637371\\
$^2$Department of Mathematics \\
Michigan State University, MI 48824, USA\\
$^3$Department of Biochemistry \& Molecular Biology\\
Michigan State University, MI 48824, USA\\
}

\date{\today}
\maketitle

\begin{abstract}
Geometric, topological and graph theory modeling and analysis of biomolecules are of essential importance in the conceptualization of molecular structure, function, dynamics, and transport. On the one hand, geometric modeling  provides molecular surface and structural representation, and offers the basis for molecular visualization, which is crucial for the understanding of molecular structure and interactions. On the other hand, it bridges the gap between molecular structural data and theoretical/mathematical models. Topological analysis and modeling give rise to atomic critical points and connectivity, and shed light on the intrinsic topological invariants such as independent components (atoms), rings (pockets) and cavities. Graph theory analyzes biomolecular interactions and reveals biomolecular structure-function relationship.
In this paper, we review certain geometric,  topological and graph theory apparatuses for biomolecular data modeling and analysis. These apparatuses are categorized into discrete and continuous ones. For discrete approaches,  graph theory, Gaussian network model, anisotropic network model, normal mode analysis, quasi-harmonic analysis, flexibility and rigidity index, molecular nonlinear dynamics, spectral graph theory, and persistent homology are discussed. For continuous mathematical tools, we present discrete to continuum mapping, high dimensional persistent homology, biomolecular geometric modeling, differential geometry theory of surfaces, curvature evaluation, variational derivation of minimal molecular surfaces,  atoms in molecule theory and quantum chemical topology. Four new approaches, including analytical minimal molecular surface, Hessian matrix eigenvalue  map,  curvature map and virtual particle model,  are introduced for the first time to bridge the gaps in biomolecular modeling and analysis. Emphasis is given to the connections of existing biophysical models/methods to mathematical theories, such as graph theory, Morse theory, Poicar\'e-Hopf theorem, differential geometry,  differential topology, algebraic topology and geometric topology. Potential new directions and standing open problems are briefly discussed.

\end{abstract}


\newpage

{\setcounter{tocdepth}{5} \tableofcontents}

\newpage

\section{Introduction}

Life science is regarded as the last forefront in natural science and the 21$^{\rm st}$ century will be the century of biological sciences. Molecular biology is the foundation of biological sciences and molecular mechanism, which is governed by all the valid mechanics, including quantum mechanics when it is relevant, and is the  ultimate truth of life science.  One trend of biological sciences in the 21$^{\rm st}$ century  is that many traditional disciplines, such as epidemiology, neuroscience, zoology, physiology and population biology,  are transforming from macroscopic and phenomenological to molecular-based sciences.  Another trend is that  biological sciences in the 21$^{\rm st}$ century are transforming from qualitative and descriptive to quantitative and predictive, as many other disciplines in natural science have done in the past.  Such a transformation  creates unprecedented  opportunities for mathematically driven advances in life science \cite{Wei:2016}.

Biomolecules, such as proteins and the nucleic acids,  including  DNA and RNA, are essential for all known forms of life, such as  animals, fungi, protists, archaea,  bacteria and plants. Indeed,  proteins perform a vast variety of biological functions, including membrane channel transport, signal transduction, organism structure supporting,  enzymatic catalysis for transcription and the cell cycle,  and  immune agents. In contrast, nucleic acids function in association with proteins and are essential players in encoding, transmitting and expressing genetic information, which is stored  through nucleic acid sequences, i.e.,  DNA or RNA molecules and  transmitted via transcription and translation processes.  The understanding of biomolecular structure, function, dynamics and transport is a fundamental issue in molecular biology and biophysics.  A traditional dogma is that  sequence determines structure, while structure determines function \cite{Anfinsen:1973}. This, however, has been undermined by the fact that many intrinsically disordered proteins can also be functional  \cite{Onuchic:1997,White:1999,Schroder:2005,Chiti:2006}.  Disordered proteins are associated with sporadic neurodegenerative diseases, including Alzheimer's disease, Parkinson's disease and mad   cow disease \cite{Chiti:2006,Uversky:2008}.  Randomness in disordered proteins is a consequence of protein flexibility, which is an intrinsic protein function. In general,  the understanding of protein structure-function relationship is also crucial for shedding light on protein specification, protein-protein interactions,   protein-drug binding     that are essential to drug design and discovery, and improving human health and wellbeing.

Much of the present understanding of biomolecular structures and functions and their relationship come from experimental data that are collected from a number of means, such as macromolecular X-ray crystallography, nuclear magnetic resonance (NMR),  cryo-electron microscopy (cryo-EM),  electron paramagnetic resonance (EPR), multiangle
light scattering, confocal laser-scanning microscopy, scanning capacitance microscopy, small angle scattering,  ultra fast laser spectroscopy, etc. The major players  for single macromolecules are X-ray crystallography and NMR. For example, advanced X-ray crystallography technology is able to offer decisive structural information at Armstrong and sub-Armstrong resolutions, while an important advantage  of NMR experiments is that they are able to provide biomolecular structural information under physiological conditions. The continuously effort in the past few decades has made  X-ray crystallography and NMR technologically relatively well developed, except for their use in special circumstances, such as the study of membrane proteins. However, theses approaches are not directly suitable for proteasomes, subcellular structures, organelles, cells and tissues, whose study has become increasingly popular in structural biology.  Currently, a unique experimental tool for imaging subcellular structures, organelles,  multiprotein complexes and even cells and tissues  is cryo-EM \cite{Volkmann:2010}.

The rapid advances of experimental technology in the past few decades have led to the accumulation of vast amount of three-dimensional (3D) biomolecular structural data. The  \href{http://www.rcsb.org/pdb/home/home.do}{Protein Data Bank (PDB)} has collected more than one hundred twenty thousands of 3D biomolecular structures.  Biomolecular geometric information holds the key to our understanding of biomolecular structure, function, and dynamics. It has wide spread applications in virtual screening, computer-aid drug design, binding pocket descriptor, quantitative structure activity relationship, protein design, RNA design, molecular machine design, etc. In general, the availability of  biomolecular structural data  has paved the way for the transition from the traditional qualitative description to  quantitative  analysis and prediction in biological sciences. An essential ingredient of quantitative biology is geometric, topological and graph theory modeling, analysis and computation. Aided by increasingly powerful high performance computers, geometric, topological and graph theory modeling, analysis and computation  have become indispensable apparatuses not only  for  the visualization of biological data, but also filling the gap between biological data  and mathematical  models of biological systems  \cite{ZYu:2008, ZYu:2008b,XFeng:2012a,XFeng:2013b,KLXia:2014a,MXChen:2012, Rocchia:2002,PMach:2011,XShi:2011,JLi:2013, Decherchi:2013,LiLin:2014}.

One of the simplest molecular geometric  models, or molecular structural models,   is the space-filling Corey-Pauling-Koltun (CPK) theory, which represents an atom by a solid sphere with a van der Waals (VDW) radius \cite{Koltun:1965}. The outer boundary of CPK model gives rise to the van der Waals (vdW) surface, which is composed of piece-wise  unburied sphere surfaces. Solvent accessible surface (SAS) and solvent-excluded surface (SES)  have also been introduced to create smooth molecular surfaces by rolling a probe molecule over the vdW surface  \cite{Richards:1977,Connolly85}. SESs   have widely been applied to  protein folding \cite{Spolar}, protein surface topography \cite{Kuhn}, protein-protein interactions \cite{Crowley},  DNA binding and bending \cite{Dragan}, macromolecular docking \cite{Jackson}, enzyme catalysis \cite{LiCata}, drug classification \cite{Bergstrom}, and solvation energies \cite{Raschke}. The SES model also plays a crucial role  in  implicit solvent models \cite{Baker:2005,DuanChen:2011a}, molecular dynamics simulations \cite{Geng:2011} and ion channel transports \cite{DuanChen:2011a,QZheng:2011a,QZheng:2011b}. Computationally, efficient algorithms for computing SES are developed or introduced, such as alpha-shapes \cite{WYChen:2010}  and marching tetrahedra \cite{SLChan:1998}. A popular software for the Lagrangian representation of SESs, called MSMS, has been developed \cite{Sanner:1996}. Recently, a software package, called   Eulerian solvent excluded surface (ESES), for the Eulerian representation of SESs \cite{ESES:2015}, has also been developed. However SAS and SES are still not differentiable and have geometric singularities, i.e., cusps and tips. To construct smooth surface representation of macromolecules,  Gaussian  surface (GS) has been proposed to represent each atom by a C$^{\infty}$ Gaussian function, while accounting their overlapping properties \cite{Grant:1995}. Differential geometry theory of surfaces provides a natural approach to describe biomolecular surfaces and boundaries. Utilizing  the Euler-Lagrange variation, a differential geometry based surface model, the minimal molecular surface (MMS), has been introduced for biomolecular geometric modeling \cite{Bates:2006,Bates:2008}. Differential geometry based variational approach has been widely applied to biophysical modeling of solvation \cite{Wei:2009,ZhanChen:2010a,ZhanChen:2010b,BaoWang:2015a}, ion channel \cite{Wei:2012,DuanChen:2012a,DuanChen:2012b} and multiscale analysis \cite{Wei:2009}, in conjugation with other physical models, such as electrostatics, elasticity and molecular mechanics \cite{Wei:2013}.  Geometry modeling and annotation of biomolecular surfaces together with  physical features, such as electrostatics and lipophilicity, provide some of the best predictions of biomolecular solvation free energies \cite{BaoWang:2016FFTS,BaoWang:2016HPK}, protein-drug binding affinities \cite{BaoWang:2016FFTB}, protein mutation energy changes \cite{ZXCang:2016a} and protein-protein interaction hot spots \cite{Darnell:2008,Demerdash:2009}.

Theoretical modeling of the structure-function  relationship of biomolecules is usually based on fundamental laws of physics, i.e., quantum mechanics (QM), molecular mechanism (MM), continuum mechanics, statistical mechanics, thermodynamics, etc. QM methods are indispensable for chemical reactions and protein degradations  \cite{Warshel:1976, Cui:2002, YZhang:2009a}.  Molecular dynamics (MD) \cite{McCammon:1977} is a powerful tool for the understanding of the biomolecular conformational landscapes and elucidating  collective motion and fluctuation. Currently, MD is  a main workhorse in molecular biophysics for biomolecular modeling and simulation.  However, all-electron or all-atom representations and long-time integrations lead to such an excessively large number of degrees of freedom that their application to real-time large-scale dynamics of  large proteins or multiprotein complexes becomes prohibitively expensive. For instance, current computer simulations of protein folding take many months to come up with a very poor copy of what Nature administers  perfectly within a tiny fraction of a second. Therefore, in the past few decades, many graph theory based biomolecular  models, including normal mode analysis (NMA)  \cite{Go:1983,Tasumi:1982,Brooks:1983,Levitt:1985},  elastic network model (ENM) \cite{Bahar:1997,Bahar:1998,Atilgan:2001,Hinsen:1998,Tama:2001,LiGH:2002} become very popular for understanding  protein flexibility and long time dynamics. In these models,  the  diagonalization of the interaction matrix or Hamiltonian of a protein is a required procedure to obtain protein eigenmodes and associated eigenvalues. The low order eigenmodes can be interpreted as the slow motions of the protein around the equilibrium state and the Moore-Penrose pseudo-inverse matrix can be used to  predict the protein thermal factors, or B-factors. However, NMA interaction potentials are quite complicated.    Tirion simplified its complexity by  retaining only the harmonic potential for elasticity, which is the dominant term in the MD Hamiltonian \cite{Tirion:1996}. Network theory \cite{Flory:1976} has had a considerable impact in protein flexibility analysis. The combination of elasticity and coarse-grained network gives rise to  elastic network model (ENM) \cite{Hinsen:1998}. In this spirit,  Gaussian network model (GNM)   \cite{Bahar:1997,Bahar:1998,QCui:2010}  and anisotropic network model (ANM) \cite{Atilgan:2001} have been proposed.  Yang et al. \cite{LWYang:2008} have shown that the GNM is about one order more efficient than most other flexibility approaches. The above graph theory based methods have been improved in a number of aspects, including crystal periodicity and cofactor corrections  \cite{Kundu:2002,Kondrashov:2007,Hinsen:2008,GSong:2007}, and  density - cluster  rotational - translational blocking \cite{Demerdash:2012}. These approaches have many applications in biophysics, including  stability analysis \cite{Livesay:2004},  molecular docking \cite{Gerek:2010},  and viral capsid analysis \cite{Rader:2005,Tama:2005}. In particular, based on spectral graph theory that the behavior of the second eigenmode can be used for clustering, these methods have been utilized to unveil the molecular mechanism  of the protein domain motions of hemoglobin \cite{CXu:2003}, F1 ATPase \cite{WZheng:2003,QCui:2004}, chaperonin GroEL \cite{Keskin:2002,WZheng:2007} and the ribosome \cite{Tama:2003,YWang:2004}. The reader is referred to  reviews for more details          \cite{JMa:2005,LWYang:2008,Skjaven:2009,QCui:2010}.

Note that ENM type of methods is still too expensive for analyzing subcellular organelles and multiportein complexes, such as HIV and Zika virus, and molecular motors, due to  their matrix decomposition procedure which is of the order of ${\cal O}(N^3)$ in computational complexity, where $N$ is the number of network nodes, or protein atoms. An interesting and important mathematical issue is how to reduce the computational complexity of ENM, GNM and ANM for handling excessively large biomolecules. Flexibility-rigidity index (FRI) has been developed as a more accurate and efficient method for biomolecular graph analysis \cite{KLXia:2013d,Opron:2014}. In particular, aided with a cell lists algorithm \cite{Allen:1987}, the  fast FRI (fFRI) is about ten percent  more accurate than GNM on a test set of 364 proteins and is orders of magnitude faster than GNM on a set of 44 proteins, due to its ${\cal O}(N)$ computational complexity. It has been demonstrated that fFRI is able to predict the B-factors of an entire HIV virus capsid with 313,236 residues in less than 30 seconds on a single-core  processor, which would require   GNM more than 120 years  to accomplish if its computer memory were not a problem \cite{Opron:2014}.

Topological analysis of molecules has become very popular since the introduction of the theory of atoms in molecules (AIM) for molecular electron density data by Bader and coworkers \cite{Bader:1985,Bader:1990}. AIM was proposed  to quantitatively define the atomic bonds and interatomic surfaces (IASs) by employing a topology based partition of electron density. It has two main strands: the scalar field topology of molecular electron density maps and the scalar field topology of the local Laplacian of electron density \cite{Popelier:2000,Popelier:2005}. The former characterizes chemical bonds and atoms, and the latter provides a new procedure  to analyze electron pair localization. Electron localization function (ELF) \cite{Silvi:1994} was proposed for the study of electron pairing. ELF  utilizes the gradient vector field topology to partition the electron density map into topological basins. In fact, a general theory called quantum chemical topology (QCT) \cite{Popelier:2005} has been developed for the topological analysis of electron density functions. Apart form  the above mentioned AIM and ELF, QCT also includes the electrostatic potential \cite{Leboeuf:1999}, electron localizability indicator (ELI) \cite{Kohout:2004}, localized orbital locator (LOL) \cite{Schmider:2000}, the virial field \cite{Keith:1996}, the magnetically induced current distribution \cite{Keith:1993}, the total energy (catchment regions) \cite{Mezey:1981} and the intracule density \cite{Cioslowski:1999}. QCT has proved to be very effective in analyzing interactions between atoms in molecular systems, particularly the covalent interactions and chemical structure of  small molecular systems. Many software packages have been developed for  QCT analysis \cite{Biegler:2002,Henkelman:2006}.

The essential idea behind QCT is the scalar field topology analysis \cite{Beketayev:2011,Gunther:2014}, which includes several major components such as  critical points (CPs) and their classification, zero-flux interface (interatomic interface), gradient vector field,  etc. In fact, when vector field topology is applied to the gradient of a scalar function, it coincides with scalar field topology. Mathematically, this  topological analysis is also known as the Morse theory, which describes the topological structure of a closed manifold by means of a nondegenerate gradient vector field. Morse theory is a powerful tool for studying the topology of molecular structural data through critical points of a Morse function. A well-defined Morse function needs to be differentiable and its CPs are isolated and non-degenerated. It can be noticed that all the above-mentioned scalar fields in QCT are Morse functions and more can be proposed as long as they satisfy the Morse function constraints. In the Morse theory, critical points are classified into minima, maxima, and saddle points based on their indices. In AIM, the three types of CPs are associated with chemical meanings. A maximal CP is called a nucleic critical point (NCP). A minimal CP is related with cage critical point (CCP). Finally, saddle points can be further classified  into two types, i.e., bond critical points (BCPs) and ring critical points (RCPs).   Current research issues in QCT include how to reduce the computational complexity and extend this approach for biomolecules  \cite{Gillet:2012,Gunther:2014}. Additionally, its connection to scalar field topology and vector field topology needs to be further clarified so that related mathematical theories, including Poincar\'e - Hopf theorem \cite{Poincare:1890}, Conley index theory \cite{Conley:1978}, Floer homology, etc., and algorithms developed in computer science can be better applied to molecular sciences.

In additional to differential topology, algebraic topology, specifically, persistent homology, has drawn much attention in recent years.  Persistent homology has been developed as a new multiscale representation of topological features.   The 0th dimensional version was originally introduced for computer vision applications under the name ``size function" \cite{Fro90, Frosini:1999} and the idea was also studied by Robins \cite{Robins:1999}.  Persistent homology theory was formulated, together with an algorithm given, by Edelsbrunner et al. \cite{Edelsbrunner:2002}, and a more general theory was developed by Zomorodian and Carlsson \cite{Zomorodian:2005}. There has since been significant theoretical development \cite{BH11,CEH07,CEH09,CEHM09,CCG09,CGOS11,Carlsson:2009theory,CSM09,SMV11,zigzag}, as well as various computational algorithms \cite{OS13,DFW14,Mischaikow:2013,javaPlex,Perseus, Dipha}. Often, persistent homology can be visualized through barcodes \cite{CZOG05,Ghrist:2008}, in which various horizontal line segments or bars are the homology generators that survive over filtration scales. Persistence diagrams are another equivalent representation \cite{edelsbrunner:2010}.  { Computational homology and persistent homology  have been applied to a variety of domains, including image analysis \cite{Carlsson:2008,Pachauri:2011,Singh:2008,Bendich:2010,Frosini:2013}, chaotic dynamics verification \cite{Mischaikow:1999,kaczynski:mischaikow:mrozek:04}, sensor network \cite{Silva:2005}, complex network \cite{LeeH:2012,Horak:2009}, data analysis \cite{Carlsson:2009,Niyogi:2011,BeiWang:2011,Rieck:2012,XuLiu:2012}, shape recognition \cite{DiFabio:2011,AEHW06} and computational biology \cite{Kasson:2007,Gameiro:2014,Dabaghian:2012,Perea:2015a,Perea:2015b}.} Compared with traditional computational topology \cite{Krishnamoorthy:2007,YaoY:2009,ChangHW:2013}  and/or computational homology, persistent homology  inherently has an additional dimension, the filtration parameter, which can be utilized to embed some crucial geometric or quantitative information into  topological invariants. The importance of retaining  geometric information in topological analysis has been recognized  \cite{Biasotti:2008}, and  topology has been advocated as a new approach for tackling  big datasets \cite{BVP15, BHPP14,Fujishiro:2000,Carlsson:2009,Ghrist:2008}. Most recently, persistent homology has been developed as a powerful tool for analyzing biomolecular topological fingerprints  \cite{KLXia:2014c,KLXia:2015d,KLXia:2015e}, quantitative fullerene stability analysis \cite{KLXia:2015a}, topological transition in protein folding \cite{KLXia:2015c}, cryo-EM structure determination \cite{KLXia:2015b},  and in conjugation with machine learning for protein classification  \cite{ZXCang:2015}  and protein-ligand/drug binding affinity prediction \cite{ZXCang:2016b}.   Differential  geometry based topological persistence \cite{BaoWang:2016a} and  multidimensional persistence \cite{KLXia:2015c} have also been developed for biomolecules analysis and modeling. It is worthy to mention that persistent topology along is able to outperform all the eminent methods in computational biophysics for the blind binding affinity prediction of  protein-ligand complexes from massive data sets \cite{ZXCang:2016b}.

The objective of this paper is threefold. First, the main objective is to provide a  review of some widely used geometric, topological and graph theory 	apparatuses for the modeling and analysis of biomolecular data. We keep our description concise, elementary and accessible to upper level undergraduate students in mathematics and most researchers in computational biophysics. We point out some open problems and potential topics in our discussions. Our goal is to provide a reference for mathematicians who are interested in mathematical molecular bioscience and biophysics (MMBB), an emergent field in mathematics \cite{Wei:2016}, and for biophysicists and theoreticians who are interested in mathematical foundations of many theoretical approaches in molecular biology and biophysics.  Obviously, our topic selection is limited by our knowledge,  experience and understanding, and for the same reason, we might have missed many important results and references on the selected topics as well.  Additionally, inspired by the success of QCT and persistent homology, the density filtration for Hessian matrix eigenvalue maps and molecular curvature maps has been introduced. Both maps are constructed from molecular rigidity density obtained via a discrete to continuum mapping (DCM) technique, which transfers atomic information in a molecule to atomic density distribution, a continuous scalar function.  In this approach, a series of isosurfaces are generated and systematically studied for eigenvalue and curvature maps. Geometric and topological (Geo-Topo) fingerprints are identified to characterize unique patterns within eigenvalue and curvature maps, specifically,  the maps of three eigenvalues  derived from local Hessian matrix at each location of the rigidity density and the maps of  Gaussian, mean, maximal and minimal curvatures computed  everywhere of the rigidity density. Topological properties of eigenvalue and curvature maps are classified by their critical points. The evolution of isosurfaces during the filtration process is found to be well characterized by CPs. Different behaviors are found in different types of maps. Persistent homology analysis is also employed for eigenvalue map analysis to reveal intrinsic topological invariants of three Hessian matrix eigenvalues. Finally, a new minimal molecular surface, called analytical minimal molecular surface (AMMS) via  the zero-value isosurface of the  mean curvature map, has been introduced. It is found that this new surface definition can capture the topological property, such as the  inner bond information. It also offers an efficient geometric modeling of biomolecules.

The rest of this paper is organized as following: Section \ref{sec:discrete} is devoted to a review of some discrete mathematical apparatuses, namely, graph theory and persistent homology,  for the analysis and modeling of  biomolecular data. More specifically,  we illustrate the applications of graph theory, Gaussian network model, anisotropic network model, normal mode analysis, flexibility and rigidity index,  spectral graph theory,  and persistent homology to biomolecular data analysis, such as protein B-factor prediction, domain separation, anisotropic motion, topological fingerprints, etc.   The review of some continuous geometric and topological apparatuses for scalar field topology and geometry  are given in Section \ref{sec:continuous}. We examine the basic concepts in differential geometry, biomolecular surfaces, curvature analysis, and theory of atoms in molecules. Discrete to continuum mapping and two algorithms for curvature evaluation are discussed. Further, we introduce two new approaches, i.e., Hessian matrix eigenvalue maps and curvature maps, for geometric-topological fingerprint analysis of biomolecular data. The relation between scalar field geometry and topological CPs are discussed in   detail. A new analytical minimal molecular surface is also introduced. Virtual particle model is proposed to analyze the anisotropic motions of continuous scalar fields, such as cryo-EM maps.
Finally, persistent homology analysis for eigenvalue scalar field is discussed. This paper ends with some concluding remarks.

\section{Discrete apparatuses for biomolecules}\label{sec:discrete}

One of the major challenges in the biological sciences is the prediction of protein functions from  protein structures. One function prediction is about protein flexibility, which strongly correlates with biomolecular enzymatic activities, such as allosteric transition, ligand binding and catalysis, as well as protein stiffness and rigidity for structural supporting. For instance, in enzymatic processes, protein flexibility enhances protein-protein interactions, which in turn reduce the activation energy barrier. Additionally, protein flexibility and motion amplify the probability of barrier crossing in enzymatic chemical reactions. Therefore, the investigation of protein flexibility at a variety of energy spectra and time scales is vital to the understanding and prediction of other protein functions. Currently, the most important technique for protein flexibility analysis is X-ray crystallography. Among more than one hundred twenty thousand structures in the protein data bank (PDB), more than eighty percent structures are collected by X-ray crystallography. The Debye-Waller factor, or B-factor, can be directly computed from  X-ray diffraction or other diffraction data. In the PDB, biomolecular structures are recorded in terms of (discrete) atomic types, atomic positions, occupation numbers, and B-factors.
Although atomic B-factors are directly associated with atomic flexibility, they can be influenced by variations in atomic diffraction cross sections and chemical stability during the diffraction data collection. Therefore, only the B-factors for specific types of atoms, say C$_\alpha$ atoms, can be directly interpreted as their relative flexibility without corrections. Another important method for accessing protein flexibility is nuclear magnetic resonance (NMR) which often provides structural flexibility information under physiological conditions. NMR spectroscopy allows the characterization of protein flexibility in diverse spatial dimensions and a large range of time scales. About six percent of structures in the PDB are determined by electron microscopy (EM) which does not directly offer the flexibility information at present. Therefore, it is important to have mathematical or biophysical methods to predict their flexibility.

\subsection{Graph theory related methodologies} \label{sec:graph}

With the development of  experimental tools, vast amount of data for biomolecular structures and interaction networks are available and provide us with unprecedented opportunities in mathematical modeling. The graph theory and network models have been widely used in the study of biomolecular structures and interactions and found many applications in drug design, protein function analysis, gene identification, RNA structure representation, etc. \cite{DasGupta2016,Gan:2004rag,Fera:2004rag} Generally speaking, biomolecular graph and network models can be divided into two major categories, namely, abstract graph/network models, which include  biomolecular interaction-network models, and geometric graph/network models, where the distance geometry plays an important role.

The geometric graph models or biomolecular structure graph models construct unique graphs based on biomolecular 3D structural data. The graph theory is then employed to analyze biomolecular properties in four major aspects: flexibility and rigidity analysis, protein mode analysis, protein domain decomposition and biomolecular nonlinear dynamics.  Many other network based approaches,  including GNM  \cite{Bahar:1997,Bahar:1998}  and ANM \cite{Atilgan:2001}, have been developed for protein flexibility analysis.
More recently,  FRI has been proposed as a matrix-decomposition-free method for flexibility analysis,
\cite{KLXia:2013d,Opron:2014}.
The fundamental assumptions of the FRI method are as follows. Protein functions, such as flexibility, rigidity, and energy, are fully determined by the structure of the protein and its environment, and the protein structure is in turn  determined by   the relevant interactions. Therefore, whenever a native protein structure is available, there is no need to analyze protein flexibility and rigidity by tracing back to the protein interaction Hamiltonian.
Consequently, the FRI bypasses the  ${\cal O}(N^3)$ matrix diagonalization. In fact,   FRI does not even require the 3D geometric information of the protein structure. It assesses graphic connectivity of the protein distance geometry and analyzes the geometric compactness of the protein structure. It can be regarded as a kernel generalization of the local density model \cite{Halle:2002,DWLi:2009,CPLin:2008}. 

Another very important application of biomolecular structure graph model is the protein mode analysis. As stated above, the low order eigenmodes provide information of the protein dynamics at equilibrium state. Normal mode analysis (NMA) plays important roles in mode analysis. However, its potential function involves too many interactions and it is very inefficient for large biomolecular systems. Anisotropic network model (ANM) has dramatically reduced the complexity of the potential function by representing the biological macromolecule as an elastic mass-and-spring network. In the network each node is a C$_\alpha$ atom of the associated residue and  springs represent the interactions between the nodes. The overall potential is the sum of harmonic potentials between interacting nodes. To describe the internal motions of the spring connecting  two atoms, there is only one degree of freedom. Qualitatively, this corresponds to the compression and expansion of the spring in a direction given by the locations of the two atoms. In other words, ANM is an extension of the GNM to three coordinates per atom, thus accounting for directionality.

The biomolecular structure graph models can also be used in protein domain decomposition. The biomolecular structural domains are stable and compact units of the structure that can fold independent of the rest of the protein and perform a specific function.  A domain usually contains a hydrophobic core and a protein is usually formed by the combination of two or several domains. There are many methods that decompose a protein structure into domains  \cite{alexandrov:2003,Guo:2003,holm:1996,SKundu:2004,murzin:1995,orengo:1997,veretnik:2007}. Some of them are done manually through structure visualization. One of them is the structural classification of proteins (SCOP) database, where  data are largely manually classified into protein structural domains based on similarities of their structures and amino acid sequences. With the surge of protein structure data, efficient and robust computational algorithms are developed. They have demonstrated a high level of consistency and robustness in the process of partitioning a structure into domains. Graph theory is also used in RNA structure analysis, particularly in RNA motif representation and RNA classification\cite{Gan:2004rag,Fera:2004rag,Kim2004:candidates}. Spectral graph theory is widely used for clustering. The essential idea is to study and explore graphs through the eigenvalues and eigenvectors of matrices naturally associated with these graphs.

Molecular nonlinear dynamics (MND) models can be naturally derived from biomolecular graph models \cite{KLXia:2014b}. Essentially, each node in the graph is an atom and  represented by a nonlinear oscillator. These nonlinear oscillators are connected through the graph connectivity. In this manner, one can study protein structure and function through the nonlinear dynamics theory widely used in chaos, synchronization, stability, pattern formation, etc.

\begin{figure}
\begin{center}
\begin{tabular}{c}
\includegraphics[width=0.3\textwidth]{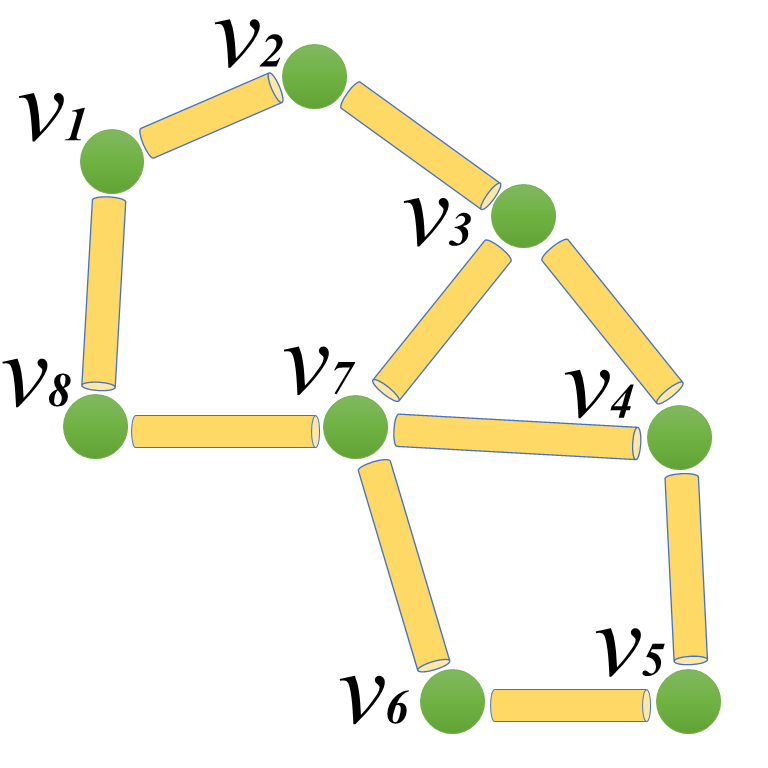}
\end{tabular}
\end{center}
\caption{Illustration of a graph. The associate adjacent, weight and Laplacian matrix can be found in Eq. (\ref{eq:graph_matrix}).
}
\label{fig:GRAPH}
\end{figure}

\subsubsection{Elementary graph theory} \label{sec:Graph}

Graph theory deals with a set of discrete  vertices (or atoms) and their  connectivity (or bonds). 
Normally, an undirected graph $G$ can be denoted as a pair $G(V,E)$, where $V= \{v_i;i=1,2,...,N \}$ denotes its set of  $N$ vertices (or protein atoms), $N=|V|$. Here $E=\{e_i=(v_{i_1},v_{i_2});1\leq i_1 \leq N, 1\leq i_2 \leq N \}$ denotes its set of edges, which can be understood as certain covalent or noncovalent bonds among atoms in a   molecule. Each edge in $E$ is an unordered pair of vertices, with the edge connecting distinct vertices $v_{i_1}$ and $v_{i_2}$ written as $e_i=(v_{i_1},v_{i_2})$. Then the adjacency matrix $A$ of $G$ is given by \cite{ChungOverview, Mohar:1991laplacian, Mohar:1997some,von:2007tutorial}
\begin{eqnarray}\label{eq:couple_matrix28}
A_{ij}=\begin{cases} \begin{array}{ll}
	        1 & (v_i,v_j) \in E\\
            0 & (v_i,v_j) \not \in E.\\
	      \end{array}
\end{cases}
\end{eqnarray}
The degree of a vertex $v_i$ is defined as $d_i=\sum_{i \neq j}^N A_{ij}$, which is the total number of edges that are connected to node $v_i$. The degree matrix $D$ can be defined as
\begin{eqnarray}\label{eq:couple_matrix27}
D_{ij}=\begin{cases} \begin{array}{ll}
	        \sum_{i \neq j}^N A_{ij} & i=j\\
            0 & i \neq j.
	      \end{array}
\end{cases}
\end{eqnarray}
With these two matrices, one can defined Laplacian matrix as $L=D-A$. The Laplacian matrix is also known as admittance matrix, Kirchhoff matrix or discrete Laplacian. It is widely used to represent a graph. More specifically, it can be expressed as,
\begin{eqnarray}\label{eq:couple_matrix26}
L_{ij}=\begin{cases} \begin{array}{ll}
            -1 & i \neq j~{\rm and} ~ (v_i,v_j) \in E\\
						-\sum_{i \neq j}^N L_{ij} &i=j\\
            0 &{ \rm otherwise.}
	      \end{array}
\end{cases}
\end{eqnarray}
For example, the  adjacency, degree and Kirchhoff matrices for the graph in Fig. \ref{fig:GRAPH} are, respectively
{\small
\begin{eqnarray}\label{eq:graph_matrix}
A=\left( \begin{array}{llllllll}
	        0  &1  &0 &0  &0  &0  &0  &1  \\
	        1  &0  &1 &0  &0  &0  &0  &0  \\
	        0  &1  &0 &1  &0  &0  &1  &0  \\
	        0  &0  &1 &0  &1  &0  &1  &0  \\
	        0  &0  &0 &1  &0  &1  &0  &0  \\
	        0  &0  &0 &0  &1  &0  &1  &0  \\
	        0  &0  &1 &1  &0  &1  &0  &1  \\
	        1  &0  &0 &0  &0  &0  &1  &0
	      \end{array}
\right),
D=\left( \begin{array}{llllllll}
	        2  &0  &0 &0  &0  &0  &0  &0  \\
	        0  &2  &0 &0  &0  &0  &0  &0  \\
	        0  &0  &3 &0  &0  &0  &0  &0  \\
	        0  &0  &0 &3  &0  &0  &0  &0  \\
	        0  &0  &0 &0  &2  &0  &0  &0  \\
	        0  &0  &0 &0  &0  &2  &0  &0  \\
	        0  &0  &0 &0  &0  &0  &4  &0  \\
	        0  &0  &0 &0  &0  &0  &0  &2
	      \end{array}
\right),
L=\left( \begin{array}{llllllll}
	        2  &$ -1$  &0 &0  &0  &0  &0  &$-1$   \\
	        $ -1$  &2  &$-1$ &0  &0  &0  &0  &0  \\
	        0  &$ -1$  &3 &$-1$  &0  &0  &$-1$   &0  \\
	        0  &0  &$-1$  &3  &$-1$   &0  &$-1$   &0  \\
	        0  &0  &0 &$-1$   &2  &-1  &0  &0  \\
	        0  &0  &0 &0  &$-1$  &2  &$-1$   &0  \\
	        0  &0  &$-1$  &$-1$   &0  &$-1$   &4  &$-1$   \\
	        $-1$   &0  &0 &0  &0  &0  &$-1$   &2
	      \end{array}
\right)
\end{eqnarray}
}
The Laplacian matrix has several basic properties. It is a symmetric and semi-positive definite. The rank of the Laplacian matrix is $N-N_0$ with $N_0$ the number of connected components. Its second smallest eigenvalue is known as the algebraic connectivity (or Fiedler value)\cite{ChungOverview,von:2007tutorial}.

More generally, one can assign weights to edges to construct a weighted graph $G(V,E,W)$. Here $G(V,E)$ is the associated unweighted graph, and $W=\{w_{ij}; 1\leq i \leq N, 1\leq j \leq N, w_{ij}\geq 0\}$ is the weighted adjacent matrix. The weight is also known as pairwise distance or pairwise affinity. The new degree of vertex $v_i$ is $d_i=\sum_{j=1}^N w_{ij}$. The weighted degree matrix $D$ and weighted Laplacian matrix $L$ can be defined accordingly.


Normally, a graph structure is not given. Instead, one may have the information of nodes and general weight functions. In this case there are several general ways to construct a graph\cite{von:2007tutorial}:
\begin{enumerate}
\item
 $\epsilon$-neighborhood graph: connect all points whose pairwise distances are smaller than $\epsilon$;
\item
 $k$-nearest neighbor graph: connect vertex $v_i$ with vertex $v_j$, if $v_j$  is among the $k$-nearest neighbors of $v_i$; and
\item
 fully connected graph: connect all points with positive similarity with each other.
\end{enumerate}

In biomolecular structure graph models, coordinates for atoms in molecules are available. Therefore, distances and distance-based functions can be used to construct structure graphs. The simplest way is to use a cutoff distance $r_c$ and build up edges between atoms or residues within the cutoff distance only. This approach has been used in GNM, which is an important tool for the study of protein flexibility and rigidity.

To unify the notation, in the following discussion, one can consider an $N$-particle representation of a biomolecule. Here a particle can be an ordinary atom in a full atomic representation or a C$_\alpha$ atom in a coarse-grained representation. One can denote $\{ {\bf r}_{i}| {\bf r}_{i}\in \mathbb{R}^{3}, i=1,2,\cdots, N\}$ the coordinates of these particles and $r_{ij}=\|{\bf r}_i-{\bf r}_j\|$ the Euclidean space distance between $i$th  and $j$th particles. More specifically, the coordinate is a position vector ${\bf r}_i=(x_i, y_i, z_i)$.

\subsubsection{Gaussian network model (GNM)} \label{sec:GNM}
\begin{figure}
\begin{center}
\begin{tabular}{c}
\includegraphics[width=0.6\textwidth]{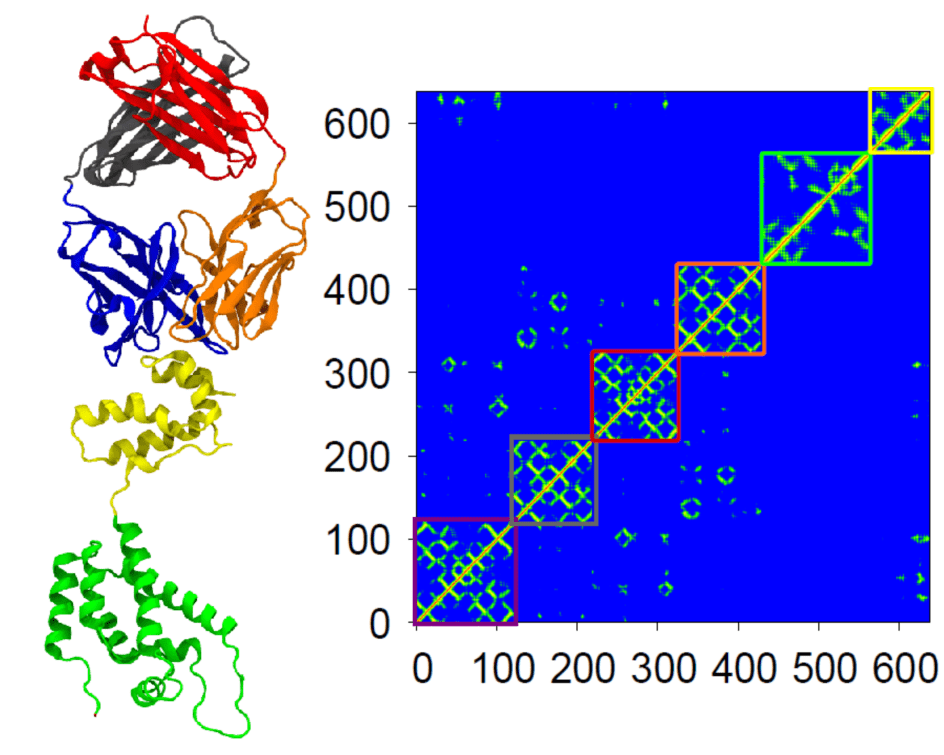}
\end{tabular}
\end{center}
\caption{Illustration of a weighted graph  Laplacian matrix for HIV capsid protein 1E6J. Left: six subdomains. Right: correlation map for residue C$_{\alpha}$ atoms indicating domain separations.}
\label{fig:correlation_matrix}
\end{figure}
Gaussian network model (GNM) \cite{Bahar:1997,Bahar:1998,QCui:2010,LiGH:2002,Yang:2006}  can be viewed as a special graph model using the Kirchhoff matrix. It was proposed for biomolecular flexibility and long-time scale dynamics analysis, particularly, the prediction of the Debye-Waller factor or B-factor. Experimentally, B-factor is an indication of the relative thermal fluctuations of different parts of  a structure. Atoms with small B-factors belong to a part of the structure that is very rigid. Atoms with large B-factors generally belong to part of the structure that is very flexible. The B-factor information can be found in the structural data downloaded from the Protein Data Bank (PDB).

As stated above, the graph or network in GNM is constructed by using a cutoff distance $r_c$. If the distance between two atoms are less than the cutoff distance, an edge is formed between them. Otherwise, no edge is built. The corresponding discrete Laplacian matrix describes the relative connectivity within a protein structure, and thus, it is also called a connectivity matrix.
\begin{eqnarray}\label{eq:couple_matrix25}
L_{ij}=\begin{cases} \begin{array}{ll}
            -1 & i \neq j~{\rm and} ~ r_{ij} \leq r_c\\
						-\sum_{i \neq j}^N L_{ij} &i=j\\
            0 &{ \rm otherwise}
	      \end{array}.
\end{cases}
\end{eqnarray}

In a nutshell, the GNM prediction of the $i$th B-factor of the biomolecule can be expressed as \cite{Bahar:1997,Bahar:1998,JKPark:2013}
\begin{eqnarray}\label{eqn:GNM}
B_i^{\rm GNM}=a \left(L^{-1} \right)_{ii}, \forall i=1,2,\cdots, N,
\end{eqnarray}
where $a$ is a fitting parameter that can be related to  the thermal energy and $\left(L^{-1} \right)_{ii}$ is the $i$th diagonal element of the Moore-Penrose pseudo-inverse of graph Laplacian matrix $L$. More specifically, $\left(L^{-1} \right)_{ii}=\sum_{k=2}^N  \lambda_k^{-1}\left[{\bf q}_k {\bf q}_k^T \right]_{ii}$, where $T$ denotes the transpose and $\lambda_k$ and ${\bf q}_k$ are the $k$th eigenvalue and eigenvector of $\Gamma$, respectively. The summation omits the first eignmode whose eigenvalue is zero.

\subsubsection{Anisotropic network model (ANM)}\label{sec:ANM}

In Gaussian network model \cite{Bahar:1997,Bahar:1998,QCui:2010,LiGH:2002,Yang:2006}, only the distance information is used with no consideration about the anisotropic  properties in different directions. It should be noticed that in GNM, the Kirchhoff matrix is of the dimension  $N*N$ with $N$ being the total number of atoms. In order to introduce the anisotropic information, one has to discriminate the distance between atoms in three different directions. To this end, at each label of $ij$, a local $3*3$ Hessian matrix is constructed \cite{Atilgan:2001}
\begin{eqnarray}\label{eq:multi-kirchoff1}
 H_{ij} = -\frac{1}{r_{ij}^2}\left[ \begin{array}{ccc}
	        (x_j-x_i)(x_j-x_i) &(x_j-x_i)(y_j-y_i) &(x_j-x_i)(z_j-z_i)\\
            (y_j-y_i)(x_j-x_i) &(y_j-y_i)(y_j-y_i) &(y_j-y_i)(z_j-z_i)\\
            (z_j-z_i)(x_j-x_i) &(z_j-z_i)(y_j-y_i) &(z_j-z_i)(z_j-z_i)
	      \end{array}\right]  ~ \forall ~ i \neq j ~{\rm and} ~ r_{ij}\leq r_c.
 \end{eqnarray}
As the same in the GNM, the diagonal part is the negative summation of the off diagonal elements:
\begin{eqnarray}\label{eq:multi-kirchoff1_diagonal}
 H_{ii} = -\sum_{i\neq j} H_{ij}.
 \end{eqnarray}
This approach, called anisotropic network model (ANM), can be used to generate the anisotropic motion of biomolecules. It is noticed that the dimension of the Hessian matrix is no longer $N*N$, instead it is $3N * 3N$. The dimension of an eigenvector is $3N$. Therefore, for each atom, one now has a vector associated with it, which gives an direction in the ${\mathbb R}^3$. The norm of this vector gives a relative amplitude. This eigenvector is also called eigenmode. It describes the relative motion of the protein near its equilibrium state.

\paragraph{Generalized GNM and generalized ANM}

In both Gaussian network model and anisotropic network model, a cutoff distance is used to construct their connectivity matrices, i.e., Laplacian matrix and Hessian matrix, respectively. However, physically, the correlation between any two particles normally decays with respect to distance. To account for this effect, a correlation function $\Phi(r_{ij}; \eta_{ij}) $ is introduced. In general, it is a  real-valued monotonically decreasing radial basis function satisfying \cite{KLXia:2013d,Opron:2014},
\begin{eqnarray}\label{eq:couple_matrix1-1}
\Phi( r_{ij};\eta_{ii})&=&1 \\ \label{eq:couple_matrix1-12}
  \Phi( r_{ij};\eta_{ij})&=&0 \quad {\rm as }\quad  r_{ij} \rightarrow\infty.
\end{eqnarray}
In this function, the parameter $\eta_{ij}$ is a characteristic distance between particles $v_i$ and $v_j$. It can also be simplified to atomic parameter $\eta_{j}$, which depends only on the atomic type. In coarse-grained models, only $C_{\alpha}$ atom is considered. Therefore, one can further simplify it to a constant $\eta$. This parameter can also be viewed as a resolution parameter. 

\begin{figure}
\begin{center}
\begin{tabular}{c}
\includegraphics[width=0.4\textwidth]{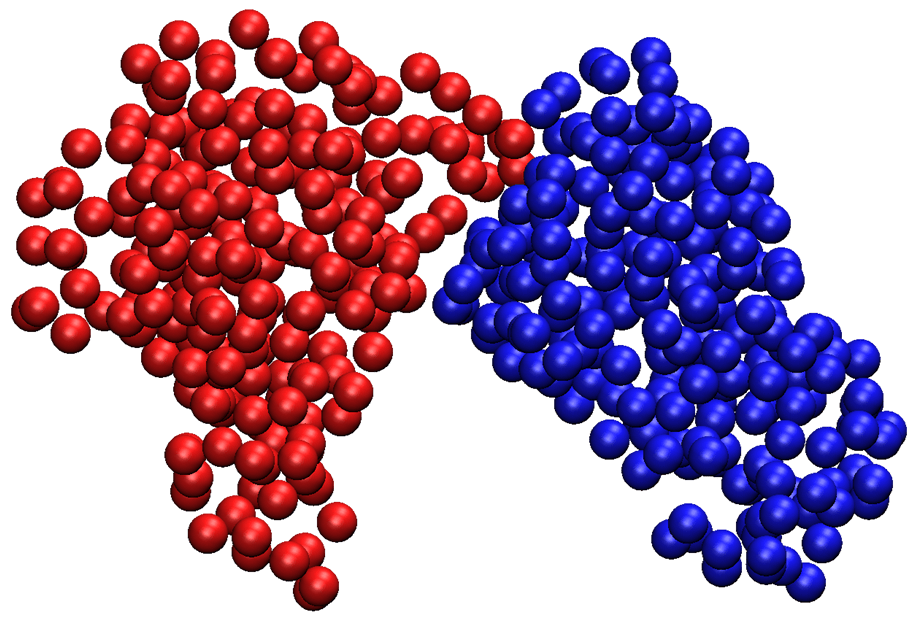}
\end{tabular}
\end{center}
\caption{ Protein domain separation of protein 3PGK C$_\alpha$ atoms using a new spectral clustering method, gGNM.  The separation is carried out with the eigenvector corresponding to the second lowest eigenvalue.}
\label{fig:cp1}
\end{figure}

Delta sequences of the positive type discussed in an earlier work \cite{GWei:2000} are all good choices.  For example, one can use generalized exponential functions
\begin{eqnarray}\label{eq:couple_matrix1}
\Phi(r_{ij};\eta_{ij}) =    e^{-\left(r_{ij}/\eta_{ij}\right)^\kappa},    \quad \kappa >0
\end{eqnarray}
or   generalized Lorentz functions
\begin{eqnarray}\label{eq:couple_matrix24}
 \Phi(r_{ij};\eta_{ij}) = \frac{1}{1+ \left( r_{ij}/\eta_{ij}\right)^{\upsilon}},  \quad  \upsilon >0.
 \end{eqnarray}
Using these correlation functions, one can obtain a weighted graph representation or  weighted graph Laplacian as   \cite{KLXia:2015f},
\begin{eqnarray}\label{eq:couple_matrix23}
L_{ij}=\begin{cases} \begin{array}{ll}
            -\Phi(r_{ij};\eta_{ij})  & i \neq j\\
						 - \sum_{i \neq j}^N L_{ij} &i=j\\
	      \end{array}.
\end{cases}
\end{eqnarray}
It is found that the weighted graph can deliver a better prediction of B-factors. This weighted graph approach is called the generalized GNM (gGNM). Figure \ref{fig:cp1} shows the protein domain separation obtained with gGNM.

\begin{figure}
\begin{center}
\begin{tabular}{c}
\includegraphics[width=0.3\textwidth]{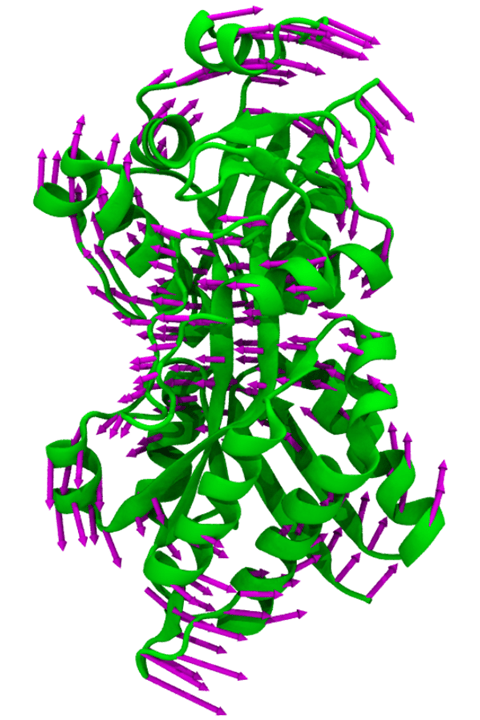}
\end{tabular}
\end{center}
\caption{Illustration of eigenmode for protein 2ABH. The eigenmode can be used to describe the biomolecular dynamics near equilibrium state. The eigenmode is generated by  the generalized anisotropic normal model method.}
\label{fig:2ABH}
\end{figure}

The local Hessian matrix in Eq. (\ref{eq:multi-kirchoff1})  can also be generated to consider the distance effect to obtain a generalized form \cite{KLXia:2015f},
\begin{eqnarray}\label{eq:multi-kirchoff12}
 H_{ij} = -\frac{\Phi( r_{ij};\eta_{ij})}{r_{ij}^2}\left[ \begin{array}{ccc}
	        (x_j-x_i)(x_j-x_i) &(x_j-x_i)(y_j-y_i) &(x_j-x_i)(z_j-z_i)\\
            (y_j-y_i)(x_j-x_i) &(y_j-y_i)(y_j-y_i) &(y_j-y_i)(z_j-z_i)\\
            (z_j-z_i)(x_j-x_i) &(z_j-z_i)(y_j-y_i) &(z_j-z_i)(z_j-z_i)
	      \end{array}\right]  ~ \forall ~ i \neq j.
 \end{eqnarray}
Again the diagonal part is the negative summation of the off-diagonal elements the same as Eq. (\ref{eq:multi-kirchoff1_diagonal}). Note that Hinsen \cite{Hinsen:1998} has proposed a special case: $\Phi( r_{ij};\eta_{ij} )= e^{-\left(\frac{r_{ij}}{\eta }\right)^2}$, where $\eta$ is a constant.
It was shown that gGNM and generalized anisotropic network model (gANM) outperform the original GNM and ANM respectively in B-factor predictions  \cite{KLXia:2015f}. Figure \ref{fig:2ABH} illustrates an eigenmode for protein 2ABH obtained with gANM.
There will be a continuous interest in design new and optimal graph theory approaches for biomolecular analysis.

\subsubsection{Normal mode analysis (NMA) and quasi-harmonic analysis}

Normal mode analysis (NMA) is one of the major tools for the study of biomolecular motions \cite{Go:1983,Levitt:1983normal,Brooks:1983harmonic,Lopez:2014normal,Hayward:2008normal}. It is found that protein  normal modes with the largest fluctuation or the lowest frequency are functionally relevant. Mathematically, NMA has its root in harmonic analysis. It assumes that conformational energy surface at an energy minimum can be approximated by some harmonic functions.

In normal mode analysis, one usually needs the atomic coordinates and a force field describing the interactions between constituent atoms.
Typically, there are three major steps in applying NMA \cite{Hayward:2008normal}. Firstly, one needs to carry out molecular dynamics simulations to minimize the conformational potential energy to obtain an equilibrium state. Secondly, one needs to calculate the second derivatives of the potential energy to construct the Hessian matrix. Finally, one needs to perform the eigenvalue decomposition of the Hessian matrix.

Originally, NMA uses exactly the same force fields as used in molecular dynamics simulations. Due to the computational inefficiency, elastic network models (ENMs) was proposed. Generally speaking, ENM is just the NMA with a simplified force field and associated coarse-grained representation. It has two major advantages. Firstly, there is no need for energy minimization as the distances of all of the elastic connections are taken to be at their minimal energy lengths. Secondly, due to the coarse-grained representation, the eigenvalue decomposition is much efficient. Even through ENM is a much simplified model, it is found that ENM is able to reproduce the NMA results with a respectable degree of similarity.

\paragraph{Standard NMA}

For a mechanical system consisting of $N$ atoms ${\bf r}=({r}_1,{ r}_2,\cdots, { r}_{3N})$, its Hamiltonian $\mathcal{H}({\bf r})$ is given by the sum of kinetic energy $\mathcal{K}({\bf r})$ and potential energy $\mathcal{U}({\bf r})$:
\begin{eqnarray}
\mathcal{H}({\bf r})=\mathcal{K}({\bf r})+\mathcal{U}({\bf r}).
\end{eqnarray}

If the structure has an equilibrium conformation ${\bf r}^0=({r}_1^0,{ r}_2^0,...,{ r}_{3N}^0)$, one can have the Taylor expansion of the potential energy
\begin{eqnarray}
\mathcal{U}({\bf r}) \approx \mathcal{U}({\bf r}^0)+ \sum_{i}^{3N}\frac{\partial  \mathcal{U}}{\partial r_i} \Big| _{\bf r=r^0}(r_i-r_i^0)+ \frac{1}{2}\sum_{i}^{3N}\sum_{j}^{3N}\frac{\partial^2  \mathcal{U}}{\partial r_i \partial r_j} \Big|_{\bf r=r^0}(r_i-r_i^0)(r_j-r_j^0)+\cdots.
\end{eqnarray}
Since the biomolecular system achieves a minimum of the energy at the equilibrium conformation ${\bf r}^0$, the related derivative functions $\frac{\partial  \mathcal{U}}{\partial r_i}\Big| _{\bf r=r^0}$ vanishes. If one uses the mass-weighted coordinates $X_i=m_i^{\frac{1}{2}} (r_i-r_i^0)$, the potential function becomes
\begin{eqnarray}
&&  \mathcal{U}({\bf r}) \approx \frac{1}{2}\sum_{i}^{3N}\sum_{j}^{3N}\frac{\partial^2 \mathcal{U}}{\partial r_i \partial r_j} \Big|_{\bf r=r^0}(r_i-r_i^0)(r_j-r_j^0) \\ \nonumber
&& \qquad \approx \frac{1}{2}\sum_{i}^{3N}\sum_{j}^{3N}\frac{\partial^2  \mathcal{U}}{\partial r_i \partial r_j} \Big|_{\bf r=r^0}(r_i-r_i^0)(r_j-r_j^0).
\end{eqnarray}
The related kinetic energy  is
\begin{eqnarray}
\mathcal{K}({\bf r})=\frac{1}{2}\sum_{i}^{3N}m_i\left(\frac{dr_i}{dt}\right)^2=\frac{1}{2}\sum_{i}^{3N}\dot{X}_i^2.
\end{eqnarray}
The Hamiltonian is
\begin{eqnarray}
\mathcal{H}({\bf r}) \approx \frac{1}{2}\sum_{i}^{3N}\sum_{j}^{3N} \dot{X}_i^2+\frac{1}{2}\sum_{i}^{3N}\sum_{j}^{3N}\frac{\partial^2\mathcal{ U}}{\partial X_i \partial X_j} \Big|_{ X=X^0}(X_i-X_i^0)(X_j-X_j^0),
\end{eqnarray}
where $\dot X$ indicates the derivative of $  X$ with respect to  time.

It can be seen that the oscillatory motions in this system are coupled and thus the movement of one atom depends on that of others. However, one can decompose the motion into independent harmonic oscillators with an appropriate normal mode coordinates. This is done by the eigenvalue decomposition of the Hessian matrix $H=Q\Lambda Q^T$. Here $H$ is obtained from the second order derivative of the potential function. The matrix $Q=\{{ {\bf q}_1,{\bf q}_2,...,{\bf q}_{3N}} \}$ contains the eigenvectors and $Q^T Q=I$. The diagonal matrix $\Lambda$ contains the corresponding eigenvalues. The mass-weighted Cartesian and normal mode coordinates are linearly related by $X=QY$. Finally, the Hamiltonian is expressed in the form,
\begin{eqnarray}
\mathcal{H} \approx \frac{1}{2}\sum_{i}^{3N} \dot{Y}_i^2+\frac{1}{2}\sum_{i}^{3N}\lambda_i {Y}_i^2.
\end{eqnarray}

\paragraph{Essential dynamics and quasi-harmonic analysis}
Due to the complexity of biomolecular systems, it is notoriously difficult to carry out molecular dynamics simulations over the relevant biological time scales. However, it has been found that the vast majority of protein dynamics can be described by a surprisingly low number of collective degrees of freedom. In this manner, a principal components analysis (PCA) is often employed to analyze the simulation results\cite{Garcia:1992large,Kitao:1991effects}. Mathematically, similar to NMA, PCA also employs the eigenvalue decomposition as it assume that that the major collective modes of fluctuation dominate the functional dynamics. In contrast to NMA, PCA of a molecular dynamics simulation trajectory does not rest on the assumption of a harmonic potential. Modes in PCA are usually sorted according to variance rather than frequency. As the collective motion is highly related to biomolecular functions, the dynamics in the low-dimensional subspace spanned by these modes was termed ``essential dynamics'' \cite{Amadei:1993essential}. A major advantage of PCA is that individual modes can be visualized and studied separately.

PCA on the mass weighted MD trajectory is also called quasi-harmonic analysis \cite{Brooks:1995harmonic}, which typically consists of three steps \cite{Hayward:2008normal}. Firstly, one can superimpose all biomolecular configurations from the simulation trajectory to remove the internal rotation and translation. Secondly, one can perform an average over the regularized trajectory to construct a covariance matrix. Thirdly, an eigenvalue decomposition is employed on the covariance matrix. The original trajectory can then be analyzed in terms of principal components.

For a $3N$-dimensional vector trajectory $\bf {r}(t)$, the correlation between atomic motions can be expressed in the covariance matrix $C$:
\begin{eqnarray}
C={\rm cov}({\bf r})=<\left({\bf r}(t)-<{\bf r}(t)>\right)\cdot\left({\bf r}(t)-<{\bf r}(t)>\right)>
\end{eqnarray}
where ${\rm cov}$ is the statistical covariance and $< >$ denote the average over time. The correlation matrix is symmetric  and can be diagonalized by an orthogonal transformation,
\begin{eqnarray}
C=Q'\Lambda (Q')^T
\end{eqnarray}
with $Q'=\{{\bf q}'_1,{\bf q}'_2,...,{\bf q}'_{3N} \}$ being  eigenvectors. The original configurations can be projected into principal components ${\bf q}_i$, i.e., ${q}'_i(t)=({\bf r}(t)-<{\bf r}(t)>)\cdot {\bf q}_i$. For visualization, one can transform principal components into the Cartesian coordinates: $ {\bf r}'(t)={q}'_i(t){\bf q}_i+<{\bf r}(t)>$.


\subsubsection{Flexibility rigidity index (FRI)} \label{sec:FRI}

Due to the involved matrix diagonalization, the computational complexity of GNM is of the order of ${\cal O}(N^3)$, which is intractable for large biomolecules, such as viruses and subcellular organelles. Therefore, it is both important and desirable to have a method whose computational complexity scales as  ${\cal O}(N^2)$ or better, as   ${\cal O}(N)$. This order reduction is a standard mathematical issue and is mathematically challenging. However, by examining  Eq. (\ref{eqn:GNM}), one notices that what is used in the GNM theoretical prediction is the diagonal elements of the inverse of the graph Laplacian matrix.  Mathematically, a good approximation is given by the inverse of the diagonal elements of the graph Laplacian matrix, providing that  the matrix  is diagonally dominant.  Flexibility rigidity index (FRI)  \cite{KLXia:2013d,Opron:2014} is such a method and has several major characteristics.  FRI provides a more straightforward and computationally-efficient way to predict B-factors.  A major advantage of the FRI method is that it does not resort to  mode decomposition and its computational complexity can be reduced to ${\cal O}(N)$ by means of the cell lists algorithm used in  fast FRI (fFRI) \cite{Opron:2014}.

The fundamental assumptions of the FRI method are as follows. Protein functions, such as flexibility, rigidity, and energy, are fully determined by the structure of the protein and its environment, and the protein structure is in turn  determined by   the relevant interactions. Therefore, whenever the protein structural data is available, there is no need to analyze protein flexibility and rigidity by tracing back to the protein interaction Hamiltonian. Consequently, the FRI bypasses the  ${\cal O}(N^3)$ matrix diagonalization.

In a nutshell, the FRI prediction of the $i$th B-factor of the biomolecule can be given by \cite{KLXia:2013d,Opron:2014}
\begin{eqnarray}\label{eqn:FRI}
B_i^{\rm FRI}=a \frac{1}{\sum_{j,j\neq i}^N w_j\Phi(r_{ij};\eta_{ij})} + b, \forall i=1,2,\cdots, N,
\end{eqnarray}
where $a$ and $b$ are fitting parameters,   $f_i=\frac{1}{\sum_{j,j\neq i}^N w_j\Phi(r_{ij};\eta_{ij})}$ is the $i$th flexibility   index and
\begin{eqnarray}\label{rigidity1}
 \mu_i=\sum_{j,j\neq i}^N w_j\Phi(r_{ij};\eta_{ij})
\end{eqnarray}
is the $i$th rigidity index. Here, $w_j$ is an atomic number depended weight function that can be set to $w_j=1$ and $\eta_{ij}=\eta$ for a C$_{\alpha}$ network. The correlation function $\Phi( r_{ij};\eta)$ can be chosen from any monotonically decreasing function satisfying Eqs. (\ref{eq:couple_matrix1}) and (\ref{eq:couple_matrix2}).
FRI was shown to outperform GNM and ANM in  B-factor predictions based on hundreds of biomolecules \cite{KLXia:2013d,Opron:2014}.

\paragraph{Multiscale FRI}\label{sec:MFRI}
Biomolecules are inherently multiscale in nature due to their multiscale interactions. For example, proteins involve covalent bonds, hydrogen bonds, van der Walls bonds, electrostatic interactions, dipolar and quadrupole interactions, hydrophobic interactions,  domain interactions, and protein-protein interactions.  Therefore, their thermal motions are influenced by the multiscale interactions among their particles.  Multiscale FRI (mFRI) was proposed to capture biomolecular multiscale behavior \cite{Opron:2015a}.
Essential idea is to build multiscale kernels, i.e., kernels parametrized at multiple scales. Multiscale flexibility index can be expressed as
\begin{eqnarray}\label{eq:flexibility3}
 f^{n}_i  =  \frac{1}{\sum_{j=1}^N w^{n}_{j} \Phi^{n}( \|{\bf r}_i - {\bf  r}_j \|;\eta^{n} )},
 \end{eqnarray}
where  $w^{n}_{j}$, $\Phi^{n}( \|{\bf r}_i - {\bf  r}_j \|;\eta^{n}) $ and $\eta^{n}$ are the corresponding quantities associated with the $n$th kernel. Then, one organizes these kernels in a multi-parameters  minimization procedure
\begin{eqnarray}\label{eq:regression2}
{\rm Min}_{a^{n},b} \left\{ \sum_i \left| \sum_{n}a^n f^{n}_i + b-B^e_i\right|^2\right\}
\end{eqnarray}
where $\{B^e_i\}$ are the experimental B-factors. In principle, all parameters can be optimized. For simplicity and computational efficiency, one only needs to determine $\{a^n\}$ and $b$ in the above  minimization process. For each kernel $\Phi^n$,  $w^n_j$  and $\eta^n_j$ will be selected according to the type of particles.

Specifically, for a simple C$_\alpha$ network (graph), one can set  $w^n_j=1$ and choose a single kernel function parametrized at different scales. The predicted B-factors can be expressed as
\begin{eqnarray}\label{eq:flexibility4}
 B^{\rm mFRI}_i  = b+ \sum_{n=1}\frac{a^n}{\sum_{j=1}^N  \Phi( \|{\bf r}_i - {\bf  r}_j \|;\eta^{n} )}.
 \end{eqnarray}
The difference between Eqs. (\ref{eq:flexibility3}) and (\ref{eq:flexibility4}) is that, in Eqs. (\ref{eq:flexibility3}), both the kernel and the scale can be changed for different $n$. In contrast, in Eq.  (\ref{eq:flexibility4}), only the scale is changed. One can use a given kernel, such as
\begin{eqnarray}\label{eq:couple_matrixn}
 \Phi(\|{\bf r} - {\bf r}_j \|;\eta^n) = \frac{1}{1+ \left( \|{\bf r} - {\bf r}_j \|/\eta^n\right)^{3}},
 \end{eqnarray}
to achieve good multiscale predictions. It was demonstrated that mFRI is about 20\% more accurate than GNM in the B-factor predictions \cite{Opron:2015a}. Parameters learned from mFRI were incorporated in GNM and ANM to create multiscale GNM (mGNM) and multiscale ANM (mANM)  \cite{Opron:2015a}.

\paragraph{Consistency between GNM and FRI}\label{sec:MFRI2}
\begin{figure}
\begin{center}
\begin{tabular}{c}
\includegraphics[width=0.7\textwidth]{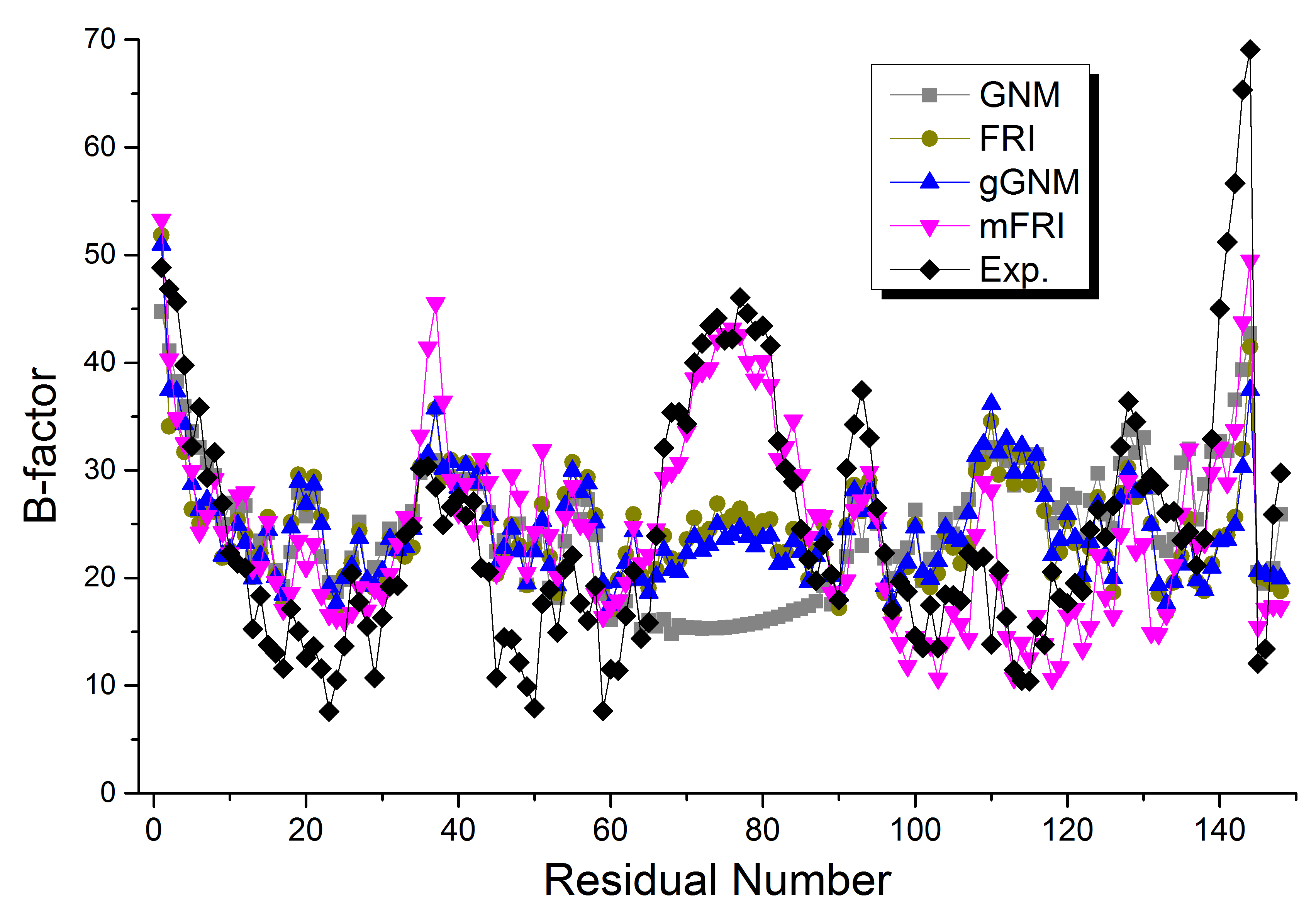}
\end{tabular}
\end{center}
\caption{A comparison of B-factor prediction of protein 1CLL by various models, including flexibility rigidity index (FRI), Gaussian network model (GNM), generalized GNM (gGNM) and multiscale FRI (mFRI). Experimental results (Exp.) are given as a reference.}
\label{fig:bfactor}
\end{figure}
To further explore the relation between GNM and FRI, let us examine the parameter limits of   generalized exponential functions (\ref{eq:couple_matrix1}) and generalized Lorentz functions (\ref{eq:couple_matrix2})
\begin{eqnarray}\label{eq:Asmpt1}
e^{-\left( r_{ij} /\eta\right)^\kappa} \rightarrow \Phi( r_{ij};r_c)  & {\rm as } & \kappa\rightarrow\infty\\ \label{eq:Asmpt2}
\frac{1}{1+ \left( r_{ij} /\eta\right)^{\upsilon}} \rightarrow \Phi( r_{ij};r_c) &  {\rm as } &\upsilon\rightarrow\infty,
\end{eqnarray}
where $r_c=\eta$ and $\Phi( r_{ij};r_c) $ is  the ideal low-pass  filter (ILF) used in the GNM Kirchhoff matrix
\begin{eqnarray}\label{eqn:IdealLF}
\Phi( r_{ij};r_c)  = \begin{cases}\begin{array}{ll}
       1, &  r_{ij} \leq r_c \\
       0,  & r_{ij}  > r_c  \\
			\end{array}
       \end{cases}.
\end{eqnarray}
Relations (\ref{eq:Asmpt1}) and (\ref{eq:Asmpt2}) unequivocally connect FRI correlation functions to the GNM Kirchhoff matrix.

It has been observed that GNM-ILF and FRI-ILF provide essentially identical predictions when the cutoff distance is equal to or larger than 20\AA  \cite{KLXia:2015f}.  This phenomenon indicates that when the cutoff is sufficiently large, the diagonal elements of the gGNM inverse matrix and the direct inverse of the diagonal elements of the FRI correlation matrix become linearly strongly dependent. To understand this dependence at a large cutoff distance,  an extreme case is considered when the cutoff distance is equal to or even larger than the protein size so that all the particles within the network are fully connected. In this situation, one can analytically calculate $i$th diagonal element of the GNM  inverse matrix
\begin{eqnarray}\label{eqn:proprotion0}
\left(L (\Phi( r_{ij};r_c\rightarrow\infty)) \right)_{ii} = \frac{N-1}{N^2},
\end{eqnarray}
and the FRI inverse of the $i$th diagonal element
\begin{eqnarray}  \label{eqn:proprotion01}
 \frac{1}{\sum_{j,j\neq i}^N  \Phi(r_{ij};r_c\rightarrow\infty)}=  \frac{1}{N-1}.
\end{eqnarray}
These two expressions have the same asymptotic behavior as $N\rightarrow\infty$, which explains numerical results. However, mathematically, it is still an open problem
to estimate the error bound between gGNM and FRI methods, i.e., the difference between the direct inverse of a diagonal element of a given weighted graph Laplacian matrix and the diagonal element of its matrix inverse.

A comparison of the performances of GNM, FRI, gGNM and mFRI is illustrated in Fig.  \ref{fig:bfactor}. The hinge around 75th residue is well captured by mFRI. Indeed, mFRI has the best accuracy in B-factor prediction  based on the test of 364 proteins \cite{Opron:2015a}.
The possible  application to FRI in other systems, such as social networks, genetic networks, cellular networks and tissue networks, is still an open problem.

\paragraph{Anisotropic FRI}
The anisotropic FRI (aFRI) has been proposed for mode analysis \cite{Opron:2014}. In this model, depending on one's interest, the size of the Hessian matrix can vary from $3\times 3$ for a completely local aFRI to $3N\times 3N$ for a completely global aFRI. To construct such a Hessian matrix, one can partition all $N$ atoms in a molecule into a total of $M$ clusters $\{c_1,c_2,\dots,c_M\}$. Each cluster $c_k$ with $k=1,\dots,M$ has $N_k$ atoms so that $N=\sum_{k=1}^{M} N_k$. For convenience, one can denote
\begin{eqnarray}\label{eq:Anisorigidity1}
	\Phi^{ij}_{uv}  = \frac{\partial}{\partial u_i} \frac{\partial}{\partial v_j} \Phi( \|{\bf r}_i - {\bf  r}_j \|; \eta_{ij} ), \quad  u,v= x, y, z; i,j =1,2,\cdots,N.
\end{eqnarray}
Note that  for each given $ij$, one can define $\Phi^{ij}=\left( \Phi^{ij}_{uv} \right)$ as a local anisotropic matrix
\begin{equation}\label{eq:afri_local_Hessian}
	\Phi^{ij}=\left(
	\begin{array}{ccc}
	\Phi^{ij}_{xx} & \Phi^{ij}_{xy}& \Phi^{ij}_{xz}\\
	\Phi^{ij}_{yx} & \Phi^{ij}_{yy}& \Phi^{ij}_{yz}\\
	\Phi^{ij}_{zx} & \Phi^{ij}_{zy}& \Phi^{ij}_{zz}
	\end{array}
	\right).
\end{equation}

In the anisotropic flexibility and rigidity (aFRI) approach, a flexibility Hessian matrix ${\bf F}^{1}(c_k)$ for cluster $c_k$ is defined by
\begin{eqnarray}\label{eq:Anisoflexibility}
	{\bf F}^{1}_{ij}(c_k)     =&  - \frac{1}{w_{j}} {\rm adj}(\Phi^{ij}),                &\quad   i,j \in c_k; i\neq j;  u,v= x, y, z \\ \label{eq:Anisoflexibilityy3}
	{\bf F}^{1}_{ii}(c_k)=&   \sum_{j=1}^N \frac{1}{w_{j}} {\rm adj}(\Phi^{ij}),  &\quad   i \in c_k;  u,v= x, y, z \\ \label{eq:Anisoflexibility4}
	{\bf F}^{1}_{ij}(c_k)=&  0,                                     &\quad   i,j \notin  c_k; u,v= x, y, z,
\end{eqnarray}
where ${\rm adj}(\Phi^{ij})$ denotes the adjoint of matrix $\Phi^{ij}$ such that $\Phi^{ij} {\rm adj}(\Phi^{ij})=| \Phi^{ij}|I$, here $I$ is identity matrix.

Another representation for the flexibility Hessian matrix ${\bf F}^{2}(c_k)$ can be defined as follows
\begin{eqnarray}\label{eq:Anisoflexibility5}
	{\bf F}^{2}_{ij}(c_k)     =&  - \frac{1}{w_{j}} |\Phi^{ij}|(J_{3} - \Phi^{ij}),                &\quad   i,j \in c_k; i\neq j;  u,v= x, y, z \\ 
	{\bf F}^{2}_{ii}(c_k)=&   \sum_{j=1}^N \frac{1}{w_{j}} |\Phi^{ij}|(J_3 - \Phi^{ij}),  &\quad   i \in c_k;  u,v= x, y, z \\ 
	{\bf F}^{2}_{ij}(c_k)=&  0,                                     &\quad   i,j \notin  c_k; u,v= x, y, z,
	\end{eqnarray}
where $J_3$ is a $3\times3$ matrix with every element being  one.
	
One can achieve $3N_k$ eigenvectors for $N_k$ atoms in cluster $c_k$ by diagonalizing ${\bf F}^{\alpha}(c_k)$, $\alpha=1,2$. Note that, the diagonal part ${\bf F}^{\alpha}_{ii}(c_k)$, $\alpha=1,2$, has inherent information of all atoms in the system. As a result, the B-factors can be predicted by the following form:
\begin{eqnarray}\label{eq:Anisoflexibility2}
	f_i^{\rm AF_\alpha} &=&{\rm Tr} \left({\bf F}^{\alpha}(c_k)\right)^{ii},   \\
	&=&  \left({\bf F}^{\alpha}(c_k)\right)^{ii}_{xx}+ \left({\bf F}^{\alpha}(c_k)\right)^{ii}_{yy}+ \left({\bf F}^{\alpha}(c_k)\right)^{ii}_{zz}, \quad \alpha=1,2.
\end{eqnarray}
It was found that aFRI is much more accurate than ANM in protein B-factor prediction \cite{Opron:2014}.
The anisotropic cluster analysis was found to play a significant role in study the local motion of RNA polymerase II translocation \cite{Opron:2016a}.

\subsubsection{Spectral graph theory } \label{sec:spectral}

Spectral graph theory \cite{ChungOverview, Mohar:1991laplacian, Mohar:1997some,von:2007tutorial} concerns the study and exploration of graphs through the eigenvalues and eigenvectors of matrices naturally associated with those graphs \cite{shi:2000,meila:2001,ng:2002,azran:2006,zelnik:2005}. Therefore, widely used GNM and ANM methods make use of spectral graph theory.

For a given graph $G(V,E)$ with $N$ nodes, one is interested in its matrix representation.  Matrices $A$ and $D$ correspond to weighted adjacent matrix and weighted degree matrix, respectively. With this notation, one has the unnormalized graph Laplacian
$$L=D-A.$$ It is often called admittance matrix, Kirchhoff matrix or discrete Laplacian.
The graph matrix has several interesting properties. Firstly, it is symmetric and positive semi-definite. Secondly, it has $N$ non-negative, real-valued eigenvalues. Thirdly, the smallest eigenvalue is 0 and the corresponding eigenvector is constant vector $\mathds{1}$. Fourthly, for every vector ${\bf c} \in \mathbb{R}^N$, one has ${\bf c}^T L {\bf c}=\frac{1}{2}\sum_{i,j}w_{ij}(c_i-c_j)^2$, which can be derived from
\begin{eqnarray}\label{eq:couple_matrix22}
&&  {\bf c}^T L {\bf c}= {\bf c}^TD{\bf c} -{\bf c}^T A{\bf c}= \sum_{i}c_i^2 d_i- \sum_{i,j}c_i c_i w_{ij}\\\nonumber
&& =\frac{1}{2}\left( \sum_{i}c_i^2 d_i-2\sum_{i,j}c_i c_j w_{ij}+ \sum_{j}c_j^2 d_j \right)=\frac{1}{2}\sum_{i,j}w_{ij}(c_i-c_j)^2,
\end{eqnarray}
where $d_i=\sum_{j=1}^N w_{ij}$ is the degree of the $i$th vertex.

In spectral graph theory, two other kinds of Laplacian matrices  \cite{ChungOverview} are also widely used.
They are the normalized Laplacian matrix
$$L_{\rm sym}=I-D^{-\frac{1}{2}}AD^{-\frac{1}{2}},$$
and random-walk normalized Laplacian matrix \cite{meila:2001,Lovasz:1993,Aldous:2002}
$$L_{\rm rw}=I-D^{-1}A,$$
where $I$ is an identity matrix.
Three different Laplacian matrices are tightly related to different ways of graph decompositions.

\paragraph{Graph decomposition and graph cut}

 A protein may have different domains. Identifying protein domains and analyzing their relative motions are important for studying protein functions. A protein complex involves different proteins. The study of  protein complex can often be formulated as a graph decomposition problem as well.

In many situations, for a given graph $G(V,E)$, one wants to partition it into subgraphs, so that nodes within a subgroup are of similar properties and nodes in different subgroups are of different properties. Mathematically, if the weight is a measurement of similarity, i.e., large weight means a great similarity, an optimized partition means that edges within the same subgroup should have large weights and edges across subgroups should have small weights. State differently, one wants to find a way to cut the graph so that it will minimize the weights of edges connecting vertices in different subgroups.

Let us first consider a simple situation, divide $G$ into two subgroup $G_1$ and $\bar{G_1}$. The notation $\bar{G_1}$ denotes the complementary of ${G_1}$.  One can define a partition ${\bf c}=\{c_i; i=1,2,...,N\}$ as,
\begin{eqnarray}\label{eq:couple_matrix2}
c_i=\begin{cases} \begin{array}{ll}
	        1 & {\rm if} ~i \in G_1,\\
         -1 & {\rm if} ~i \in \bar{G_1}.\\
	      \end{array}
\end{cases}
\end{eqnarray}
Therefore, if two nodes $v_i$ and $v_j$ are in the same subgroup, one will have $(c_i-c_j)^2=0$, otherwise $(c_i-c_j)^2=4$. In this way, a ${\rm Cut}(G_1,\bar{G_1})$, which is the total weights of edges connecting two subgroups, can be defined as following,
\begin{eqnarray}\label{eq:couple_matrix222}
&&  {\rm Cut}(G_1,\bar{G_1})={\bf c}^T L {\bf c}= {\bf c}^TD{\bf c} -{\bf c}^T A{\bf c}= \sum_{i}c_i^2 d_i- \sum_{i,j}c_i c_i w_{ij}\\\nonumber
&& =\frac{1}{2}\left( \sum_{i}c_i^2 d_i-2\sum_{i,j}c_i c_j w_{ij}+ \sum_{j}c_j^2 d_j \right)=\frac{1}{4}\sum_{i,j}w_{ij}(c_i-c_j)^2.
\end{eqnarray}
It can be also noticed that if one defines $W(G_1,\bar{G_1})=\sum_{i\in G_1,j \in \bar{G_1}} w_{ij}$, then one has $ {\rm Cut}(G_1,\bar{G_1})=W(G_1,\bar{G_1})$.
%

To obtain an optimized partition means to minimize the value of $ {\rm Cut}(G_1,\bar{G_1})$. However, the way of cut stated above does not consider the size of the subgraphs. In this way, the cut or graph composition can be very uneven in terms of the number of nodes in subgroup. For example, one extreme situation is that one of the subgraph may only have a few nodes (i.e., one or two nodes), while the other subgroup may have all the rest of nodes. To avoid this problem, three commonly defined cuts, namely, ratio cut, normalized cut and min-max cut, are proposed in the literature \cite{Hagen:1992new,shi:2000,Ding:2001min}. To facilitate the description, one can define  $|G|$ as the total number of nodes in graph $G$ and ${\rm vol}(G)$ is the summation of all weights in $G$, i.e., ${\rm vol}(G)=\sum_{ij}w_{ij}$.
\begin{itemize}
\item
Ratio cut  is defined as \cite{Hagen:1992new}
\begin{eqnarray}
{\rm Rcut}(G_1,\bar{G_1})=\frac{W(G_1,\bar{G_1})}{|G_1|} + \frac{W(G_1,\bar{G_1})}{|\bar{G_1}|}.
\end{eqnarray}

\item
Normalized cut is defined as \cite{shi:2000}
\begin{eqnarray}
{\rm Ncut}(G_1,\bar{G_1})=\frac{W(G_1,\bar{G_1})}{W(G_1,G_1)+W(G_1,\bar{G_1})} + \frac{W(G_1,\bar{G_1})}{W(\bar{G_1},\bar{G_1})+W(\bar{G_1},G_1)}.
\end{eqnarray}

\item
Min-Max cut is given by \cite{Ding:2001min}
\begin{eqnarray}
{\rm Mcut}(G_1,\bar{G_1})=\frac{W(G_1,\bar{G_1})}{W(G_1,G_1)} + \frac{W(G_1,\bar{G_1})}{W(\bar{G_1},\bar{G_1})}.
\end{eqnarray}
\end{itemize}
These decompositions have found many applications in image segmentation \cite{shi:2000}. However, the impact of these cuts to protein domain partition is yet to be examined. An important issue is how to cut a given biomolecule to elucidate its biological function and predict its chemical and biological behavior.

\paragraph{Ratio cut and Laplaician matrix}
To solve the optimization problem
\begin{eqnarray}
\min \limits_{G_1 \subset G} {\rm Rcut}(G_1,\bar{G_1}),
\end{eqnarray}
one can define the vector $\bf {c}$ as
\begin{eqnarray}\label{eq:indicator}
c_i=\begin{cases} \begin{array}{ll}
	         \sqrt{\frac{\bar{|G_1|}}{|G_1|}}  & {\rm if} ~i \in G_1,\\
             -\sqrt{\frac{|G_1|}{\bar{|G_1|}}} & {\rm if} ~i \in \bar{G_1}.\\
	      \end{array}
\end{cases}
\end{eqnarray}
One can have
\begin{eqnarray}
{\bf c}^T L {\bf c}&=&\frac{1}{2}\sum_{i,j}w_{ij}(c_i-c_j)^2  \\\nonumber
&=& \frac{1}{2}\sum_{i\in G_1,j \in \bar{G_1}} \left( \sqrt{\frac{\bar{|G_1|}}{|G_1|}} + \sqrt{\frac{|G_1|}{\bar{|G_1|}}} \right)^2 + \frac{1}{2}\sum_{i\in \bar{G_1},j \in G_1} \left( -\sqrt{\frac{\bar{|G_1|}}{|G_1|}} - \sqrt{\frac{|G_1|}{\bar{|G_1|}}}  \right)^2 \\\nonumber
&=& W(G_1,\bar{G_1}) \left( \frac{\bar{|G_1|}}{|G_1|} + \frac{|G_1|}{\bar{|G_1|}} +2  \right) \\\nonumber
&=& W(G_1,\bar{G_1}) \left( \frac{\bar{|G_1|}+|G_1|}{|G_1|} + \frac{|G_1|+\bar{|G_1|}}{\bar{|G_1|}} \right) \\\nonumber
&=& |G| {\rm Rcut}(G_1,\bar{G_1})
\end{eqnarray}
One also has
\begin{eqnarray}
 \sum_{i} c_i= \sum_{i\in G_1} \sqrt{\frac{\bar{|G_1|}}{|G_1|}} - \sum_{i \in \bar{G_1}}\sqrt{\frac{|G_1|}{\bar{|G_1|}}}=0
\end{eqnarray}
Therefore, the vector ${\bf c}$ is orthogonal to the vector with  common components (leading eigenvector of the Laplacian matrix). It is noted that
\begin{eqnarray}
\|f\|^2= \sum_{i} c_i^2= \sum_{i\in G_1} \frac{\bar{|G_1|}}{|G_1|} - \sum_{i \in \bar{G_1}}\frac{|G_1|}{\bar{|G_1|}}=N.
\end{eqnarray}
In this way, the minimization    is equivalent to
\begin{eqnarray}\label{Eq:min}
\{ \min {\bf c}^T L {\bf c}~|~ {\bf c}~ {\rm satisfies ~ Eq. ~(\ref{eq:indicator})};~ {\bf c} \bot \mathds{1};~ \|f\|^2=N \}.
\end{eqnarray}
As the entries of the solution vector are only allowed to take two particular values,  Eq. (\ref{Eq:min}) is a discrete optimization problem. One can discharge the discreteness condition and allow the vector to take any arbitrary values. This results in a relaxed problem
\begin{eqnarray}
\{ \min \limits_{{\bf c}\in \mathbb{R}^N}{\bf c}^T L {\bf c}~|~ {\bf c} \bot \mathds{1};~ \|f\|^2=N \}.
\end{eqnarray}
From the Rayleigh-Ritz theorem, it can be seen that the solution of this problem is the second smallest eigenvector ${\bf q}_2$ of the Laplacian matrix.  In order to obtain a partition of the graph, one can choose,
\begin{eqnarray} \label{eq:cluster_eigenv2}
\begin{cases} \begin{array}{ll}
	          i \in G_1 (c_i= 1)    &{\rm if} ~ ({\bf q}_2)_i\geq 0\\
              i \in \bar{G_1} ( c_i=-1) &{\rm if} ~({\bf q}_2)_i<0
	      \end{array}.
\end{cases}
\end{eqnarray}
Mathematically, the second smallest eigenvalue $\lambda_2$ is known as the algebraic connectivity (or Fiedler value) of a graph. The corresponding eigenvector of the second eigenvalue offers a near optimized partition. 
It becomes an interesting issue to design certain weighted Laplacian matrix so that a protein  domain partition is optimal with respect to protein functions. Transferring the discrete Laplacian matrix  to a continuous Laplacian operator and casting  the domain separation problem into an optimization one are promising approaches.  Certainly, these issues are also biologically significant in the exploration of  protein structure-function relationship.

\paragraph{Normalized cut and normalized Laplacian matrix}

One can define the vector ${\bf c}$ as
\begin{eqnarray}\label{eq:normalized_indicator}
c_i=\begin{cases} \begin{array}{ll}
	         \sqrt{\frac{{\rm vol}(\bar{G_1})}{{\rm vol}(G_1)}}  & {\rm if} ~i \in G_1\\
             -\sqrt{\frac{{\rm vol}(G_1)}{{\rm vol}(\bar{G_1})}}  & {\rm if} ~i \in \bar{G_1}
	      \end{array}.
\end{cases}
\end{eqnarray}
One can have
\begin{eqnarray}
{\bf c}^T L {\bf c}&=&\frac{1}{2}\sum_{i,j}w_{ij}(c_i-c_j)^2  \\\nonumber
&=& \frac{1}{2}\sum_{i\in G_1,j \in \bar{G_1}} \left( \sqrt{\frac{{\rm vol}(\bar{G_1})}{{\rm vol}(G_1)}} + \sqrt{\frac{{\rm vol}(G_1)}{{\rm vol}(\bar{G_1})}} \right)^2 + \frac{1}{2}\sum_{i\in \bar{G_1},j \in G_1} \left( - \sqrt{\frac{{\rm vol}(\bar{G_1})}{{\rm vol}(G_1)}} - \sqrt{\frac{{\rm vol}(G_1)}{{\rm vol}(\bar{G_1})}} \right)^2 \\\nonumber
&=& |G| {\rm Ncut}(G_1,\bar{G_1}).
\end{eqnarray}
Additionally, it is easy to see that
\begin{eqnarray}
 (D {\bf c})^T\mathds{1}=\sum_{i} d_i c_i= \sum_{i\in G_1} d_i \sqrt{\frac{{\rm vol}(\bar{G_1})}{{\rm vol}(G_1)}} - \sum_{i \in \bar{G_1}} d_i \sqrt{\frac{{\rm vol}(G_1)}{{\rm vol}(\bar{G_1})}}=0
\end{eqnarray}
This means that vector ${\bf c}$ is orthogonal to constant one vector. Moreover, one can evaluate
\begin{eqnarray}
{\bf c}^T D {\bf c} = \sum_{i} d_i c_i^2= \sum_{i\in G_1} d_i  \frac{{\rm vol}(\bar{G_1})}{{\rm vol}(G_1)} + \sum_{i \in \bar{G_1}} d_i  \frac{{\rm vol}(G_1)}{{\rm vol}(\bar{G_1})}={\rm vol}(G)
\end{eqnarray}
In this way, the minimization process is equivalent to
\begin{eqnarray}
\{\min {\bf c}^T L {\bf c}~|~ {\bf c}~ {\rm satisfies ~ Eq.~ (\ref{eq:normalized_indicator}) };~ D{\bf c} \bot \mathds{1};~ {\bf c}^T D {\bf c}={\rm vol}(G) \}
\end{eqnarray}
This is also a discrete optimization problem, because the entries of the solution vector are only allowed to take two particular values. By discarding the discreteness condition and allowing  the vector to be any arbitrary values,  one results in a relaxed problem
\begin{eqnarray}
\{ \min \limits_{{\bf c} \in \mathbb{R}^N}{\bf c}^T L {\bf c}~|~ D{\bf c} \bot \mathds{1};~ {\bf c}^T D {\bf c}={\rm vol}(G) \}
\end{eqnarray}
Now one can substitute ${\bf c}'=D^{\frac{1}{2}}{\bf c}$. After substitution, the problem is
\begin{eqnarray}
\{ \min \limits_{{\bf c}' \in \mathbb{R}^N}({\bf c'})^T D^{-\frac{1}{2}}L D^{\frac{1}{2}} {\bf c'}~|~ {\bf c}' \bot D^{\frac{1}{2}} \mathds{1};~ \|{\bf c'}\|={\rm vol}(G) \}.
\end{eqnarray}
It can been seen that $D^{-\frac{1}{2}}L D^{\frac{1}{2}}=L_{\rm sym}$ and $D^{\frac{1}{2}} \mathds{1}$ is the first eigenvector of $L_{\rm sym}$. From the Rayleigh-Ritz theorem, it can be seen that the solution of this problem is the second smallest eigenvalue of $L_{\rm sym}$.

\paragraph{Graph Laplacian and continuous Laplace operator}

It has been found that there is a connection between graph Laplacian and the continuous Laplace operator \cite{Belkin:2003,Lafon:2004,Belkin:2005,Hein:2005graphs,Gine:2006empirical}.
Roughly speaking, if one chooses $w_{ij}=\frac{1}{r_{ij}^2}$ with $r_{ij}$ as the distance between node $v_i$ and node $v_j$,   one can have
\begin{eqnarray}
w_{ij}(c_i-c_j)^2=\left(\frac{c_i-c_j}{r_{ij}}\right)^2.
\end{eqnarray}
The term $\frac{c_i-c_j}{r_{ij}}$ can be roughly viewed as a discretization of $\nabla c({\bf r})$. In this way, there is a connection between graph Laplacian and the continuous Laplace operator through this functional formulation,
\begin{eqnarray}
{\bf c}^T L {\bf c}=<{\bf c}, L {\bf c}> \approx \int |\nabla c({\bf r})|^2 d{\bf r}.
\end{eqnarray}

More specifically, one can set  $w_{ij}= \Phi(r_{ij},\eta_{ij})$ as defined in Eqs. (\ref{eq:couple_matrix1-1}) and (\ref{eq:couple_matrix1-12})
\begin{eqnarray}
 L_N { c({\bf r}_i)}=c({\bf r}_i) \sum_j \Phi(r_{ij},\eta_{ij})- \sum_j c({\bf r}_i) \Phi(r_{ij},\eta_{ij}),
\end{eqnarray}
where $c({\bf r}_i)=c_i$ and ${\bf r}_i$ is the coordinate of $i$th node.
This operator can be naturally extended to an integral operator
\begin{eqnarray}
 L_N {c({\bf r})}=c({\bf r}) \sum_j \Phi(|{\bf r}-{\bf r}_j|,\eta_{ij}) - \sum_j c({\bf r})\Phi(|{\bf r}-{\bf r}_j|,\eta_{ij}).
\end{eqnarray}

If data points are sampled from a uniform distribution on a $k$-dimensional manifold $\mathcal{M}$, let set $w_{ij}= \Phi(r_{ij},\eta_{ij})
=e^{-\frac{r_{ij}^2}{4t}}$, with $t=t_N=N^{-\frac{1}{k+2+\alpha}}$, where $\alpha >0$, and assume $c(\bf {r}) \in C^{\infty}(\mathcal{M})$. Belkin \cite{Belkin:2005} found that there is a constant $C$, such that in probability \cite{Belkin:2003,Belkin:2005}
\begin{eqnarray}
\lim \limits_{N \rightarrow \infty}C\frac{(4\pi t_N)^{-\frac{k+2}{2}}}{N} L_N^{t_N}c({\bf r})=\Delta_{\mathcal{M}} c({\bf r}).
\end{eqnarray}

The continuous Laplace operator has been widely utilized in many biophysical models, such as Poisson-Boltzmann theory for electrostatics \cite{Holst:1994}, Laplace-Beltrami equation for molecular surface modeling \cite{Bates:2008,Wei:2009}, Poisson-Nernst-Planck equation for ion channel modeling \cite{Hyon:2010,Wei:2012}, and elasticity equation  for  macromolecular conformational change induced by electrostatic forces \cite{Zhou:2008d}. Obviously, these issues are associated with a graph problem. The modeling of biomolecular structure, function, dynamics and transport by combining graph theory and partial differential equation (PDE) is  an open problem.

\paragraph{Modularity} Modularity is total summation of the weights within the group  minus the expected one in an equivalent network with weight randomly placed.

\begin{figure}
\begin{center}
\begin{tabular}{c}
\includegraphics[width=0.99\textwidth]{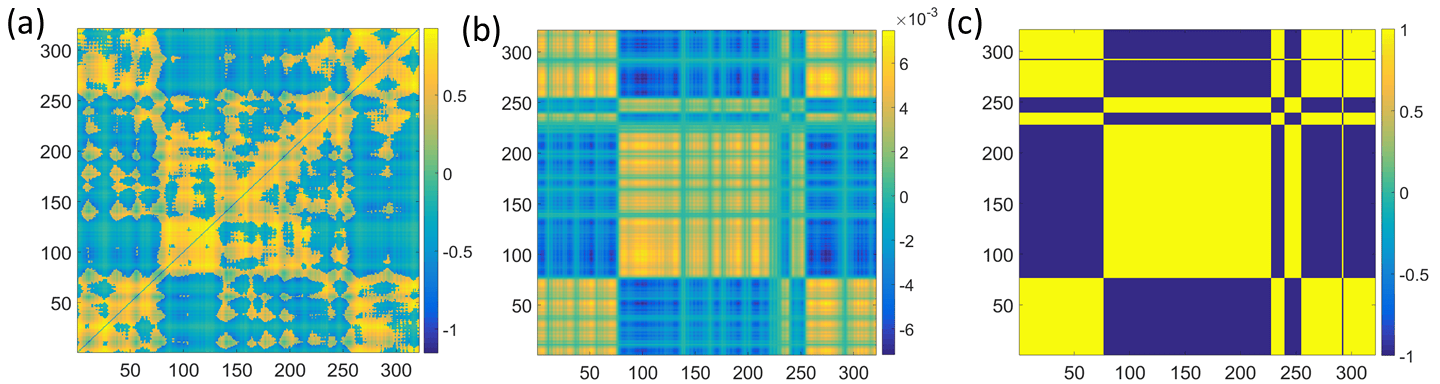}
\end{tabular}
\end{center}
\caption{An illustration of modularity matrix, second-eigenvector formed matrix and domain indication matrix. The protein network for 2ABH C$_{\alpha}$ is constructed by using Gaussian network model with cut off distance 23 \AA.~ (a) The illustration of modularity matrix in Eq. (\ref{eq:modularity_matrix}); (b) The illustration of the matrix formed by the second eigenvector, i.e., ${\bf q}_2 {\bf q}_2^T$; (c) The illustration of the index matrix ${\bf c} {\bf c}^T$. The vector ${\bf c}$ is generated from ${\bf q}_2$ by $\{c_i=1;~if~({\bf q}_2)_i \geq0\}$ and $\{c_i=-1;~if~({\bf q}_2)_i<0\}$. The parameter $\gamma=1$ is used in the modularity model.}
\label{fig:2abh_domain}
\end{figure}

Mathematically, the modularity  is defined as \cite{Newman:2004,Newman:2006,Fortunato:2010,Jain:2010}
\begin{eqnarray}
Q=\frac{1}{2{\rm vol}(G)} \sum_{ij} \left(A_{ij}-\gamma \frac{d_i d_j}{{\rm vol}(G)}\right) (\delta(c_i,c_j)+1)=\frac{1}{2{\rm vol}(G)} \sum_{ij} \left(A_{ij}- \gamma \frac{d_i d_j}{{\rm vol}(G)}\right) \delta(c_i,c_j),
\end{eqnarray}
where $\gamma$ is a resolution parameter, which is designed to change the scale at which a network is clustered \cite{Fortunato:2010}. Here $\delta(c_i,c_j)=1$ if $c_i$ and $c_j$ are in the same subgroup ($c_i=c_j$), otherwise it equals to 0 ($c_i \neq c_j$). The term $A_{ij}$ is the weighted adjacent matrix component.

The modularity matrix is defined as:
\begin{eqnarray}\label{eq:modularity_matrix}
B_{ij}=A_{ij}-\frac{d_i d_j}{{\rm vol}(G)},
\end{eqnarray}
then the above equation can be further simplified as
\begin{eqnarray}
Q=\frac{1}{2{\rm vol}(G)}{\bf c}^T  B {\bf c}.
\end{eqnarray}

Again one assume that graph $G$ can be divided into two parts $G_1$ and $\bar{G_1}$
\begin{eqnarray}
&& Q =\frac{1}{{\rm vol}(G)}\left[ \left({\rm vol}(G)-\sum_{c_i \neq c_j}w_{ij} \right) -\frac{\gamma}{{\rm vol}(G)} \left(\sum_{c_i=c_j} d_i d_j \right)  \right] \\ \nonumber
&& \quad  =1-\frac{1}{{\rm vol}(G)} \left({\rm Cut}(G_1,\bar{G_1}) + \frac{\gamma}{{\rm vol}(G)} {\rm vol}(G_1)^2\right)  \\ \nonumber
&& \quad =1-\gamma-\frac{1}{{\rm vol}(G)} \left({\rm Cut}(G_1,\bar{G_1}) -\frac{\gamma}{{\rm vol}(G)}{\rm vol}(G_1){\rm vol}(\bar{G_1})\right).
\end{eqnarray}
One can define the total variation (TV): $|c|_{\rm TV}=\frac{1}{2} \sum_{i,j}w_{ij}|c_i-c_j|$, weighted $\ell_2$-norm $\|c\|^2_{\ell_2}= \sum_{i}d_{i}|c_i|^2$, and mean  ${\rm mean}(c)=\frac{1}{{\rm vol}(G)} \sum_{i}d_{i}|c_i|$.

\begin{figure}
\begin{center}
\begin{tabular}{c}
\includegraphics[width=0.4\textwidth]{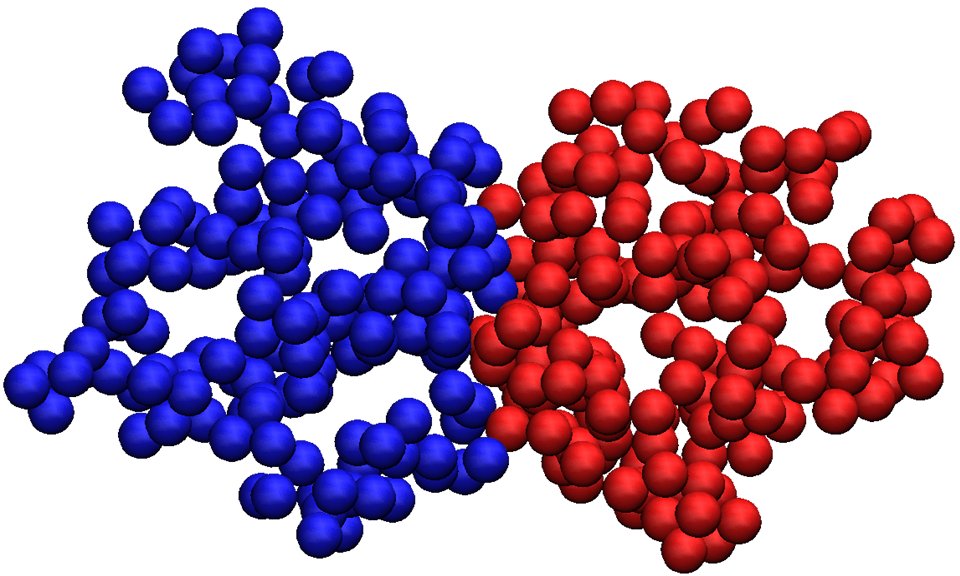}
\end{tabular}
\end{center}
\caption{An illustration of graph decomposition with modularity based eigenvectors. The protein network of 2ABH C$_{\alpha}$ atoms is constructed by using the Gaussian network model with cut off distance 23 \AA. Based on this network,  modularity matrix is constructed with parameter $\gamma=1$. The modularity eigenvector corresponding to the second lowest eigenvalue is used for protein domain decomposition. }
\label{fig:2abh_modularity2}
\end{figure}

One can also define $c$ to be a function $\chi_{G_1}: G \rightarrow \{0,1\}$. This is the indicator function of a subsect $G_1 \subset G$. In this manner, one has
\begin{eqnarray}
&& \quad |c|_{\rm TV}-\gamma\|c-{\rm mean}(c)\|_{\ell_2}^2 \\ \nonumber
&&= |\chi_{G_1}|_{\rm TV} -\gamma \|\chi_{G_1}-{\rm mean}(\chi_{G_1})\| \\ \nonumber
&&={\rm Cut}(G_1,\bar{G_1})-\gamma \left( \sum_i d_i \left| \chi_{G_1}-\frac{{\rm vol}(G_1)}{{\rm vol}(G)} \right|^2 \right)  \\ \nonumber
&&={\rm Cut}(G_1,\bar{G_1})-\gamma \left( {\rm vol}(G_1)\left(1-\frac{{\rm vol}(G_1)}{{\rm vol}(G)}\right)^2+{\rm vol}(\bar{G_1}) \left(\frac{{\rm vol}(G_1)}{{\rm vol}(G)}\right)^2 \right) \\ \nonumber
&&={\rm Cut}(G_1,\bar{G_1})- \frac{\gamma}{{\rm vol}(G)} {\rm vol}(G_1) {\rm vol}(\bar{G_1}).
\end{eqnarray}
Figure \ref{fig:2abh_domain} illustrates the modularity matrix, second-eigenvector formed matrix and domain indication matrix for protein 2ABH. Figure \ref{fig:2abh_modularity2} shows the protein domain decomposition using a modularity matrix.

One can have the theorem: maximizing the modularity functional $Q$ over all the partitions is equivalent to minimize$|c|_{\rm TV}-\gamma\|c-{\rm mean}(c)\|_{\ell_2}^2$  \cite{Hu:2013method} .
Essentially, this is equal to a balance cut problem
\begin{eqnarray}
\min \frac{{\rm Cut}(G_1,\bar{G_1})}{{\rm vol}(G_1){\rm vol}(\bar{G_1})}
\end{eqnarray}

One can let $c=(c^1,c^2)$ by $c: G\rightarrow V^2$ with $V^2=\{(1,0),(0,1)\}$  \cite{Hu:2013method} . In this way, for each $c_i$, it has only a single entry equals 1. The minimizing problem can be solved through the following equation \cite{Merkurjev:2013,Hu:2013method}
\begin{eqnarray}
\frac{\partial c}{\partial t}=-(L c^1,L c^2)-\frac{1}{\epsilon^2}\nabla W_{\rm multi}(c)+\frac{\delta}{\delta c}\left( \gamma \| c-{\rm mean}(c) \| \right).
\end{eqnarray}
Here  $\nabla W_{\rm multi}(c)$ is the composition of function $W_{\rm multi}$ and $c$. Normally, the function $W_{\rm multi}$ is a multi-well potential \cite{Merkurjev:2013}.

The field of spectral graph, modularity and related variation formation for biomolecular systems is completely open. There is much to be done on this interesting field.  For example, one can formulate molecular design, such as drug design, protein design, and the design of protein-DNA and/or protein-RNA complexes, as graph cut problems. In drug design, one would like to optimize the protein-drug binding affinity for a given drug candidate and its target protein. If the selection of graph notes is also a part of optimization, one then would like to optimize protein-drug binding affinity, drug target selectivity, drug pharmacokinetics, drug toxicity, etc. To be more specific, one needs to consider a minimization process,
\begin{eqnarray} \label{eq:cutfreeenergy}
\min \limits_{{\bf r} \in {\mathbb R}^{3N}} {\rm Cut}(G_1({\bf r}),\bar{G_1}({\bf r}))+\gamma \Delta F ({\bf r}).
\end{eqnarray}
Here $N$ is the total number of atoms in the complex, the parameter $\gamma$ is a scale parameter and $ \Delta F ({\bf r})$ is the free energy change. The function $G({\bf r})$ is denoted as the graph representation of protein-protein or protein-ligand complexes. The functions $G_1({\bf r})$ and $\bar{G_1}({\bf r})$ are graph models for the corresponding protein or ligand. They are all position-dependent and the minimization process is to find the best fitting configuration so that one can achieve a cut that  minimizes the free energy change $\Delta F ({\bf r})$. In fact, biomolecular free energy minimization often leads to PDE based models for solvation, ion channel, membrane protein interaction, molecular machine assembly, to name only a few \cite{Wei:2009}. It should be noticed that in Eq. (\ref{eq:cutfreeenergy}), all the three types of graph cuts might be used. This approach can be combined with techniques in other mathematical disciplines, such as those in dynamical systems, stochastic analysis, and differential equation, to address complex design problems, namely, drug design, protein design, RNA design, molecular machine design etc, in biomolecular systems.


\subsubsection{Molecular nonlinear dynamics} \label{sec:MND}

To introduce molecular nonlinear dynamics, one can consider a folding protein that constitutes $N$ particles and has the spatiotemporal complexity of ${\mathbb{R}}^{3N}\times \mathbb{R}^{+} $. Assume that the molecular mechanics of the protein is described by  molecular nonlinear dynamics having a set of $N$ nonlinear oscillators of dimension ${\mathbb{R}}^{nN}\times \mathbb{R}^{+}$, where $n$ is the dimensionality of a single nonlinear oscillator.
Let us consider an $n\times N$-dimensional nonlinear system for $N$ interacting chaotic oscillators  \cite{KLXia:2014b}
\begin{eqnarray}
\label{eq:couple_matrix}
\frac{d{\bf u}}{dt} &=& {\bf F}({\bf u})+ {G}{\bf u}, ~~~
\end{eqnarray}
where ${\bf u}=({\bf u}_1,{\bf u}_2,\cdots, {\bf u}_N )^T$ is an array of state functions for $N$ nonlinear oscillators, ${\bf u}_j=(u_{j1},u_{j2}, \cdots, u_{jn})^T$ is an $n$-dimensional nonlinear function for the $j$th oscillator,  \linebreak $ {\bf F}({\bf u})=(f({\bf u}_1), f({\bf u}_2), \cdots,  f({\bf u}_N))^T$ is an array of nonlinear functions of $N$  oscillators, and ${G}=\varepsilon L \otimes \Gamma$.  Here,  $\varepsilon$ is the overall  interaction strength,  $L$ is an $N\times N$ weighted Laplacian matrix and $\Gamma$ is an $n\times n$ linking matrix. Essentially, for each node in the biomolecular graph, one has an $n$-dimensional nonlinear oscillator. These oscillators are connected by the Laplacian matrix and a fixed $n\times n$ linking matrix.

For example, one can choose a set of  $N$ Lorenz attractors \cite{Lorenz:1963} and a simple $3*3$ link matrix as following:
 $\mathbf{u}_i=(u_{i1},u_{i2},u_{i3})^T$,
\begin{eqnarray} \begin{cases}  \nonumber
   \frac{du_{i1}}{dt}=\alpha(u_{i2}-u_{i1}) \\  \label{oscillator}
   \frac{du_{i2}}{dt}=\gamma u_{i1}-u_{i2}-u_{i1}u_{i3} \\ \nonumber
   \frac{du_{i3}}{dt}=u_{i1}u_{i2}-\beta u_{i3}, i= 1, 2, \cdots, N
   \end{cases},~\Gamma= \left( \begin{array}{ccc}
                      0               & 0             & 0 \\
                      1               & 0             & 0 \\
                      0               & 0             & 0
                      \end{array}
                      \right).
\end{eqnarray}
If the Laplacian matrix shown in Fig. \ref{fig:GRAPH} is used as the connectivity matrix, the nonlinear dynamic system for the first node is
\begin{eqnarray} \begin{cases}  \nonumber
   \frac{du_{11}}{dt}=\alpha(u_{12}-u_{11}) \\  \label{oscillator2}
   \frac{du_{12}}{dt}=\gamma u_{11}-u_{12}-u_{11}u_{13}+\varepsilon (2u_{11}-u_{21}-u_{81})\\ \nonumber
   \frac{du_{13}}{dt}=u_{11}u_{12}-\beta u_{13}
\end{cases}.
\end{eqnarray}

\paragraph{Stability analysis}\label{Sec:Stability}


One can use the FRI kernel weighted Laplacian matrix to define the driving and response relation of nonlinear chaotic oscillators. Due to the synchronization of chaotic oscillators, an $N$-time reduction in the spatiotemporal complexity  can be achieved, leading to an intrinsically low dimensional manifold (ILDM) of dimension ${\mathbb{R}}^{n}\times \mathbb{R}^{+}$.  Formally, the $n$-dimensional ILDM is defined as
\begin{eqnarray}
{\bf u}_1(t)={\bf u}_2(t)=\cdots = {\bf u}_N(t)={\bf s} (t),
\end{eqnarray}
where ${{\bf s} (t)}$ is a synchronous state or reference state.

To understand the stability of the ILDM of protein chaotic dynamics, one can define a transverse state function as ${\bf w}(t)={\bf u}(t)-{\bf S}(t)$, where ${\bf S}(t)$ is a vector of $N$ identical components $({\bf s}(t),{\bf s}(t),\cdots, {\bf s}(t) )^T$. Obviously, the invariant  ILDM is given by  ${\bf w}(t)={\bf u}(t)-{\bf S}(t)={\bf 0}$. Therefore, the stability of the ILDM can be analyzed by $\frac{d {\bf w}(t)}{dt} =\frac {d {\bf u}(t)}{dt} -\frac {d {\bf S}(t)}{dt}$, which can be studied by the following linearized equation \cite{Pecora:1997,GHu:1998}
\begin{eqnarray}\label{eqn:trans}
\frac{d{\bf w}}{dt} = ({\bf DF} ({\bf s}) + {G}){\bf w}, ~~~
\end{eqnarray}
where ${\bf DF}({\bf s})$ is the Jacobian of ${\bf F}$.

To further analyze the stability of Eq.    (\ref{eqn:trans}),
{one can diagonalize connectivity matrix ${L}$ }
\begin{eqnarray}\label{eqn:trans2}
{L}{\bf \varphi}_j(t)  =\lambda_j {\bf \varphi}_j(t), \quad  j=1,2,\cdots, N,
\end{eqnarray}
{where  $\{{\bf \varphi}_j\}_{j=1}^N$ are eigenvectors and ${\lambda_j}_{j=1}^N$ are the associated eigenvalues. These eigenvectors
span a vector space in which  a transverse state vector has the expansion \cite{Pecora:1997,GHu:1998}}
\begin{eqnarray}
{\bf w}(t)=\sum_{j}{\bf q}_j(t)\phi_j(t).
\end{eqnarray}
{Therefore, the stability problem of the ILDM  is equivalent to the following  stability problem}
\begin{eqnarray}\label{eqn:trans3}
\frac{d{\bf q}_j(t)}{dt} &=& ( Df ({\bf s}) + \varepsilon \lambda_j\Gamma){\bf q}_j(t), \quad j=1,2,\cdots,N,
\end{eqnarray}
where $Df ({\bf s})$ is the diagonal component of ${\bf DF} ({\bf s})$.
The stability of Eq.    (\ref{eqn:trans3}) is determined by the largest Lyapunov exponent $L_{\rm max}$, namely, $L_{\rm max} < 0$, which can be decomposed into two contributions
$$L_{\rm max}=L_{\rm f}+L_{\rm c},$$
where $L_{\rm f}$ is the largest  Lyapunov exponent of the original $n$ dimensional  chaotic system $\frac{d{\bf s}}{dt} =  { f}({\bf s})$, which can be easily computed for most chaotic systems. Here, $L_{\rm c}$ depends on   $\lambda_j$ and $\Gamma$. The largest eigenvalue  $\lambda_1$ equals  0,  and its corresponding eigenvector represents the homogeneous motion of the ILDM, and all of other eigenvalues $\lambda_j, j=2,3,\cdots, N$ govern the transverse stability of the ILDM. Let us consider a simple case in which the linking matrix is the unit matrix ($\Gamma={\bf I}$). Then stability of the ILDM is determined by  the second largest eigenvalue $\lambda_{2}$ (algebraic connectivity, or Fiedler value), which enables us to estimate the critical interaction strength $\varepsilon_{c}$   in terms of $\lambda_{2}$ and $L_{f}$  \cite{KLXia:2014b},
\begin{equation} \label{eq:expect}
\varepsilon_{c}=\frac{L_f}{- \lambda_{2}}.
\end{equation}
The dynamical system reaches the ILDM when $\varepsilon > \varepsilon_{c}$ and is unstable when $\varepsilon \leq \varepsilon_{c}$.
The eigenvalues of protein connectivity matrices  are obtained with a standard matrix diagonalization algorithm. Molecular nonlinear dynamics has been recently developed as an efficient means for protein B-factor prediction   \cite{KLXia:2014b}. It can be potentially used for protein domain separation without resorting to the matrix diagonalization.

The availability of more than a hundred thousands of interacting protein networks in the PDB provides living example problems for analyzing dynamical systems. Indeed, the connection between dynamical systems and spectral graph theory in mathematics, and the structure and function of macromolecules gives rise to  exciting opportunities to further study dynamical systems and graph theory, and better analyze biomolecules.

\subsection{Persistent homology } \label{sec:PHA}

\begin{figure}
\begin{center}
\begin{tabular}{c}
\includegraphics[width=0.6\textwidth]{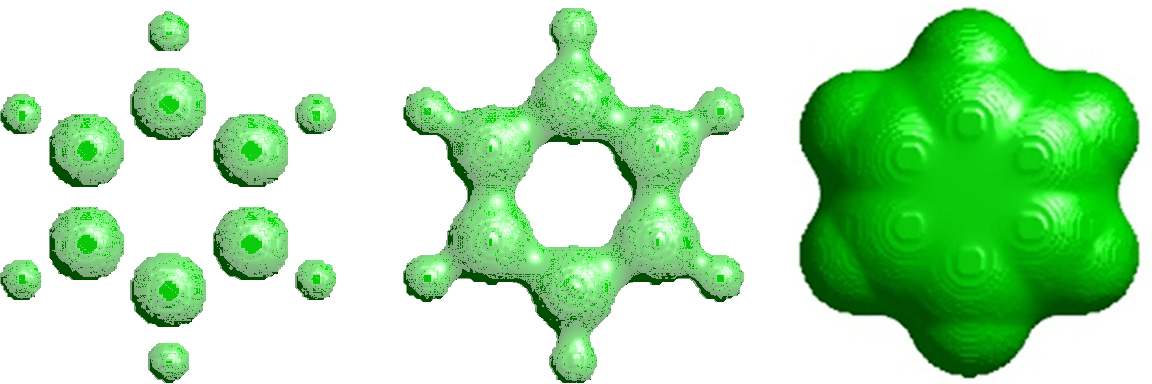}
\end{tabular}
\end{center}
\caption{An illustration of topological change over different dense thresholds for a  benzene molecule. From left to right, the density threshold is decreased.}
\label{fig:density}
\end{figure}

As a branch of algebraic topology, persistent homology is a topological approach that utilizes algebraic algorithms to compute topological invariants in data\cite{Carlsson:2009,Niyogi:2011,BeiWang:2011,Rieck:2012,XuLiu:2012}. It is a working horse in the popular topological data analysis and has found its  success in  qualitative characterization or classification. Specifically,  persistent homology describes geometric features of biomolecular date with  topological invariants that persist over the systematic change of a scale parameter relevant  to topological events. Such a change is called filtration. Figure \ref{fig:density} illustrates the change   of the topology of  benzene molecule over different density thresholds.
The idea is to  capture topological structures continuously over a range of spatial  scales during the filtration process. Unlike computational homology that is based on  truly metric free or coordinate free representations, persistent homology can embed geometric information from protein data into topological invariants so that ``birth"  and ``death" of  geometric features, such as circles, rings, loops, pockets, voids and  cavities can be monitored by topological measurements during the filtration process\cite{Edelsbrunner:2002,Zomorodian:2005}

\subsubsection{Simplicial homology and persistent homology}\label{sec:SimplicialHomology}
An essential ingredient of persistent homology is simplicial homology which is built on simplicial complex. 
Simplicial complex is a finite set that consists of discrete vertices (nodes or atoms in a protein), edges (line segments or bonds in a biomolecule), triangles, and their high dimensional counterparts. Simplicial homology can be defined on simplicial complex to analyze and extract topological invariants. Then a filtration process is used to establish topological persistence from simplicial homology analysis\cite{Edelsbrunner:2002,Zomorodian:2005}.

\paragraph{Simplicial complex}
A key component of simplicial complex $K$ is a $k$-simplex, $\sigma^k$, defined as the convex hall of $k+1$ affine independent nodes in $\mathbb{R}^N$ ($N>k$). Let $v_0,v_1,v_2,\cdots,v_k$ be $k+1$ affine independent points (or atoms in a biomolecule) and express a $k$-simplex $\sigma^k=\{v_0,v_1,v_2,\cdots,v_k\}$ as
\begin{eqnarray}\label{eq:couple_matrix12}
\sigma^k=\left\{\tau_0 v_0+\tau_1 v_1+ \cdots +\tau_k v_k \mid \sum^{k}_{i=0}\tau_i=1;0\leq \tau_i \leq 1,i=0,1, \cdots,k \right\}.
\end{eqnarray}
Moreover, let us define an $i$-dimensional face of $\sigma^k$ as the convex hall formed by the nonempty subset of $i+1$ vertices from $\sigma^k$ ($k>i$). Clearly, a 0-simplex is a vertex, a 1-simplex is an edge, a 2-simplex is a triangle, and a 3-simplex represents a tetrahedron. One can also define the empty set as a (-1)-simplex.

A simplicial complex is constructed to combine these geometric components, including vertices, edges, triangles, and tetrahedrons together under certain rules.  More specifically, a simplicial complex $K$ is a finite set of simplicies that satisfy two conditions. One is that any face of a simplex from  $K$  is also in  $K$ and  the other is that the intersection of any two simplices in  $K$ is either empty or  shared faces. The dimension of a simplicial complex is defined as the maximal dimension of its simplicies. The underlying topological space $|K|$ is a union of all the simplices of $K$, i.e.,  $|K|=\cup_{\sigma^k\in K} \sigma^k$. Further,  the concept of chain is introduced to  associate this topological space with algebra groups.

\paragraph{Homology}
One can denote a linear combination $\sum^{k}_{i}\alpha_i\sigma^k_i$ of $k$-simplex $\sigma^k_i$ as a $k$-chain $[\sigma^k]$. The coefficients $\alpha_i$ can be chosen from different fields, such as rational field $\mathbb{Q}$,  real number field and complex number field, and from integers $\mathbb{Z}$ and prime integers $\mathbb{Z}_p$ with prime number $p$. For simplicity, one can consider the coefficients $\alpha_i$ are  chosen from $\mathbb{Z}_2$, for which  the addition operation between two chains is the modulo 2 addition for the coefficients of their corresponding simplices. The set of all $k$-chains of simplicial complex $K$ and the addition operation form an Abelian group $C_k(K, \mathbb{Z}_2)$. Therefore, the homology of a topological space is represented by a series of Abelian groups.

A boundary operation $\partial_k$ is defined  as $\partial_k: C_k \rightarrow C_{k-1}$. Without  orientation, the boundary of a $k$-simplex $\sigma^k=\{v_0,v_1,v_2,\cdots,v_k\}$ is
\begin{eqnarray}
\partial_k \sigma^k = \sum^{k}_{i=0} \{ v_0, v_1, v_2, \cdots, \hat{v_i}, \cdots, v_k \},
\end{eqnarray}
where the notation $\{v_0, v_1, v_2, \cdots ,\hat{v_i}, \cdots, v_k\}$ means that  the $(k-1)$-simplex is generated by eliminating vertex $v_i$ from the sequence. When a boundary operator is applied twice, any $k$-chain will be mapped to a zero element, i.e.,  $\partial_{k-1}\partial_k= \emptyset$. As a special case, one has $\partial_0= \emptyset$.  The $k$th cycle group $Z_k$ and the $k$th boundary group $B_k$ are the subgroups of $C_k$ and can be defined  by means of the boundary operator,
\begin{eqnarray}
&& Z_k={\rm Ker}~ \partial_k=\{c\in C_k \mid \partial_k c=\emptyset\}, \\
&&{ B_k={\rm Im} ~\partial_{k+1}= \{ c\in C_k \mid \exists d \in C_{k+1}: c=\partial_{k+1} d\}.}
\end{eqnarray}
An element in the $k$th cycle group $Z_k$ or the $k$th boundary group $B_k$ is called the $k$th cycle or the $k$th boundary. One has $B_k\subseteq Z_k \subseteq C_k$ since the boundary of a boundary is always empty $\partial_{k-1}\partial_k= \emptyset$. Geometrically, the $k$th cycle is a $k$ dimensional loop or hole.

With all the above definitions, one can define the homology group. Specifically, the $k$th homology group $H_k$ is defined as the quotient group of  the $k$th cycle group $Z_k$ and $k$th boundary group $B_k$: $H_k=Z_k/B_k$. Two $k$th cycle elements are  called homologous if they are different by a $k$th boundary element. The $k$th Betti number represents the rank of the $k$th homology group,
\begin{eqnarray}
\beta_k = {\rm rank} ~H_k= {\rm rank }~ Z_k - {\rm rank}~ B_k.
\end{eqnarray}
From the fundamental theorem of finitely generated Abelian groups, the $k$th homology group $H_k$ can be given as a direct sum,
\begin{eqnarray}
H_k= {Z}\oplus \cdots \oplus {Z} \oplus {Z}_{p_1}\oplus \cdots \oplus {Z}_{p_n}= {Z}^{\beta_k} \oplus {Z}_{p_1}\oplus \cdots \oplus {Z}_{p_n},
\end{eqnarray}
where $\beta_k$ is the rank of the subgroup and is $k$th Betti number. Here  $ {Z}_{p_i}$ is torsion subgroup with torsion coefficients $\{p_i| i=1,2,...,n\}$, the power of prime number.

Topologically, cycle element in $H_k$ forms a $k$-dimensional loop or ring that is not from the boundary of a higher dimensional chain element. The geometric meanings of  Betti numbers in $\mathbb{R}^3$ are the follows: $\beta_0$ represents the number of isolated components (i.e., protein atoms), $\beta_1$ is the number of one-dimensional loop or ring, and $\beta_2$ describes the number of two-dimensional voids or cavities. Together, the Betti number sequence { $\{\beta_0,\beta_1,\beta_2,\cdots \}$} gives the intrinsic topological property of  biomolecular data.

\paragraph{$\check{\rm C}$ech complex, Rips complex and alpha complex}
A key concept for the construction of simplicial complex from a point set of a given topological space is nerve. Basically,   given an index set $I$ and  open set ${\bf U}=\{U_i\}_{i\in I}$ that is a cover of  a point set $X \in \mathbb{R}^N$, i.e., $X \subseteq \{U_i\}_{i\in I}$, the nerve {\bf N} of {\bf U} should satisfy two basic conditions. One   is that $\emptyset \in {\bf N}$. Additionally,  if $\cap_{j \in J} U_j \neq \emptyset $ for $J \subseteq I $, then  $J \in {\bf N}$. Usually, for a given biomolecular dataset, the simplest way to construct a cover is to assign a ball of certain radius around each atom. If the biomolecular dataset is dense enough and the radius is large enough, then the union of all the  balls has the capability to recover the underlying space for the biomolecule.

The nerve of a cover of the biomolecule constructed from the union of atomic balls is a $\check{\rm C}$ech complex for the biomolecule. More specifically, for a biomolecular dataset $X \in \mathbb{ R}^N$, one defines a cover of closed atomic balls ${\bf B}=\{B (x, r)\mid x \in X \}$ with radius $r$ and centered at $x$. The $\check{\rm C}$ech complex of $X$ with radius  $r$ is denoted as $\mathcal{C}(X,r)$, which is the  nerve of the closed ball set {\bf B},
\begin{eqnarray}
\mathcal{C}(X,r) = \left\{ \sigma \mid \cap_{x \in \sigma} B (x,r) \neq \emptyset \right\}.
\end{eqnarray}
One can relax $\check{\rm C}$ech complex  conditions   to generate a Vietoris-Rips complex, in which, a simplex $\sigma$ is constructed if the largest distance between any two atoms is at most $2r$. One can denote  $\mathcal{R}(X,r)$ the    Vietoris-Rips complex, or Rips complex \cite{Edelsbrunner:1994}. There is a sandwich relation for these abstract complexes,
\begin{eqnarray}\label{eq:SandwichRelation}
\mathcal{C}(X,r)\subset \mathcal{R}(X,r) \subset \mathcal{C}(X,\sqrt{2}r).
\end{eqnarray}
In practical applications, Rips complex is  preferred due to  its computational convenience.

Another important geometric concept in computational geometry is alpha complex.
Let $X$ be a biomolecular dataset in Euclidean space $\mathbb{R}^d$ and define the Voronoi cell of a point $x \in X$ as
\begin{eqnarray}
V_x = \{ u\in R^d \mid |u-x|\leq |u-x'|, \forall x'\in X \}.
\end{eqnarray}
Then the collection of all Voronoi cells for the biomolecule forms a Voronoi diagram. Further, the nerve of the biomolcular Voronoi diagram generates a Delaunay complex.

One can define $R(x,r)$ as the intersection of Voronoi cell $V_x$ with ball $B(x,\epsilon)$, i.e., $R(x,r)= V_x \cap B(x,r)$. The alpha complex $\mathcal{A}(X,r)$ of the dataset $X$ is defined as the nerve of cover $\cup_{x\in X} R(x,r)$,
\begin{eqnarray}
\mathcal{A}(X,r) = \left\{ \sigma \mid \cap_{x \in \sigma} R (x,r) \neq \emptyset \right\}.
\end{eqnarray}
Therefore, an alpha complex is a subset of the Delaunay complex.

\begin{figure}
\begin{center}
\begin{tabular}{c}
\includegraphics[width=0.6\textwidth]{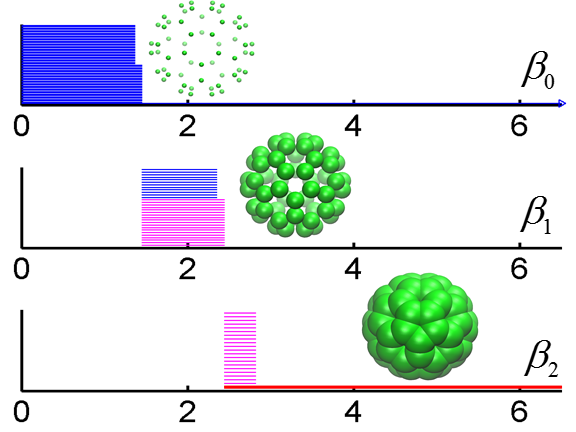}
\end{tabular}
\end{center}
\caption{Barcode representation of persistent homology analysis for fullerene molecule $C_{60}$.}
\label{fig:c60}
\end{figure}

\paragraph{General filtration processes}
In order to construct a simplicial homology from a dataset, a special parameter (like the radius $r$ mentioned above) is commonly used. However, to find a ``suitable" value for this parameter, so that it can reveal the underlying manifold, is not straightforward. An elegant alternative approach is to carry out a filtration process  \cite{Bubenik:2007, edelsbrunner:2010,Dey:2008,Dey:2013,Mischaikow:2013}. A suitable filtration is  vital  to the resulting persistent homology, which will be described in the following section. In practice, two commonly used filtration algorithms are the Euclidean-distance or correlation matrix based and density based ones.  These basic filtration algorithms can be modified  to achieve different goals in data analysis \cite{KLXia:2014c,KLXia:2015c,KLXia:2015d}.

In the  Euclidean-distance based filtration,  one associates each atom with an ever-increasing  radius to form an ever-growing ball for each atom. Various aforementioned complex construction algorithms can be utilized to identify the corresponding complexes. 
This filtration process can be formalized by  the use of a distance matrix $\{d_{ij}\}$. Here the matrix element $d_{ij}$ represents the distance between atom $i$ and atom $j$. For diagonal terms, one assumes $d_{ij}=0$.   With a filtration threshold $\varepsilon$,  a 1-simplex is generated between atoms $i$ and $j$ if $d_{ij}\leq\varepsilon$. Higher dimensional complexes can also be created similarly.

Another important filtration process is the density based filtration process. In this process, the filtration goes along the increase or decrease of the density value. In this way, a series of isosurfaces are generated. Morse complex is used for the characterization of their topological invariants. 


\paragraph{Persistent homology}
Persistent homology is an elegant mathematical theory to describe topological invariants from a series of topological spaces in various scales, that are generated by the filtration process.
Persistent homology concerns a family of homologies, in which the connectivity of the given dataset is systematically reset according to a (scale) parameter. 
For a simplicial complex $K$, the filtration is defined as a nested sub-sequence of subcomplexes,
\begin{eqnarray}
\varnothing = K^0 \subseteq K^1 \subseteq \cdots \subseteq K^m=K.
\end{eqnarray}
The introduction of filtration   leads to the creation of persistent homology. When the filtration parameter is a scale parameter, simplicial complexes generated from a filtration give a multiscale representation of the corresponding topological space, from which related homology groups can be evaluated to reveal  topological features of the given dataset. Furthermore, the concept of persistence is introduced to measure the persistent length of topological features.  The $p$-persistent $k$th homology group $K^i$ is
\begin{eqnarray}
H^{i,p}_k=Z^i_k/(B_k^{i+p}\bigcap Z^i_k).
\end{eqnarray}
Through the study of the persistent pattern of these topological features, the so called persistent homology is capable of capturing the intrinsic properties of the underlying protein topological space solely from the protein atomic coordinates.

To visualize the persistent homology results, many elegant representation methods have been proposed, including persistent diagram\cite{Edelsbrunner:2008persistent}, persistent barcode\cite{Ghrist:2008}, persistent landscape \cite{Bubenik:2015statistical}, etc. In this paper,   a barcode representation is used. The persistent barcode of an Euclidean-distance based filtration process of a fullerene molecular $c_{60}$ is shown in Fig. \ref{fig:c60}.

The combination of optimization and persistent homology was discussed in a recent work for biomolecular data analysis \cite{BaoWang:2016a}. The essentially idea to create  an object functional for extracting certain geometric features in data. Then the use of variational principle to result in a differential equation, which is subsequently utilized to filtrate the biomolecular data. In this work, the minimization of the surface energy of biomolecules was the objective, which leads to the Laplace-Beltrami flow for filtration.  In this manner, one can have connected persistent homology to other important mathematical subjects, such as partial differential equations, optimizations, and differential geometry  \cite{BaoWang:2016a}.

Object-oriented persistent homology is expected to play an important role in massive   data analysis. In particular, this approach can be combined with a deep learning strategy to automatically extract desirable information in a semi-supervised or  unsupervised learning framework.

\subsubsection{Multiscale   persistent homology}\label{sec:mPHA}

\begin{figure}
\begin{center}
\begin{tabular}{c}
\includegraphics[width=0.9\textwidth]{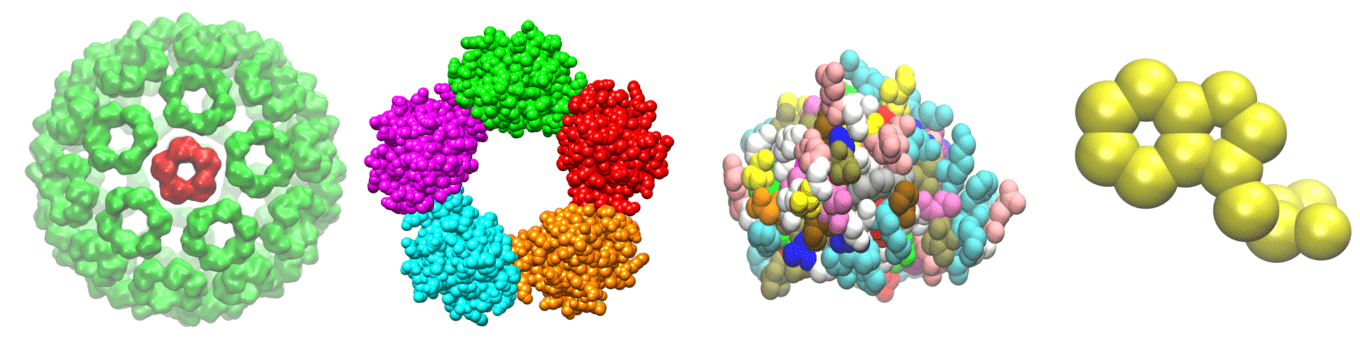}
\end{tabular}
\end{center}
\caption{Illustration of multiscale properties in an icosahedral viral particle capsid, which consists many hexagons and pentagons as shown on the left chart.  A pentagon shown on the second left consists of five proteins. For a protein shown on second right, there are many residues indicated by different colors. Finally,  for residue shown on the right, it has many atoms, including    hexagonal and pentagonal rings. }
\label{fig:Multiscale_illustration}
\end{figure}

The emergence of complexity in self-organizing biological systems frequently requires more comprehensive topological descriptions. Therefore, multiscale persistent homology, multiresolution persistent homology and multidimensional persistence become valuable for biological systems as well as many other complex systems.

It is noted that there is no need for a simplicial complex to be built exclusively on a set of atoms. It can be constructed for a biomolecule from its coarse-grained representation, namely, a set of amino acid residues. For protein complexes, such as a viral particle as shown in Fig. \ref{fig:Multiscale_illustration}, a
 basic vortex in a simplicial complex can be either a protein, a residue (i.e., C$_{\alpha}$ representation) or an atom. Each of these selections gives rise to a different persistent homology at a given scale.   To   demonstrate the utility of persistent homology for cryo-EM data analysis, a microtubule intermediate structure, protein complex 1129 from  electron microscopy data (EMD) is considered \cite{KLXia:2015b}. In this study, each point in the simplicial complex is a protein (tubulin).

An interesting problem in persistent homology studies is how to select appropriate scales and appropriate types of molecular information in filtration to better analyzing the structure, function and dynamics of subcellular organelles, molecular motors and multiprotein complexes.

\subsubsection{Topology based quantitative modeling}

Traditional topological analysis analyzes data in terms of topological invariants, such as Euler characteristic, winding number, Betti numbers and so on, thus leads to so much reduction that the resulting information is hardly useful for complex real world problems. Geometric tools often become computationally intractable  for macromolecules and their interactions due to the involved high degrees of freedom. As one can see,  persistent homology embeds geometric information in topological invariants and bridges the gap between geometry and topology. However, persistent homology has  been mainly employed for qualitative analysis, namely, characterization and classification. Only recently,  persistent homology   has been devised for quantitative analysis,  mathematical modeling, and physical prediction \cite{KLXia:2014c,KLXia:2015c,KLXia:2015d,BaoWang:2016a}.
It has been shown that the length of intrinsic Betti 2 bar provides an excellent model for fullerene thermal energies \cite{BaoWang:2016a}. It has also been shown that accumulated Betti 0 and Betti 1 bar lengths offer highly accurate predictions of unfolding protein bond and total energies, respectively \cite{KLXia:2015c}. It is expected that persistent homology will continue to play a significant role in quantitative modeling and prediction.

Additionally, earlier work regard short-lived barcodes or topological invariants as {\it noise}. It has also been pointed out that, for biomolecular datsets, these short-lived topological invariants or non-persistent topological features, are part of topological fingerprints and have  meaningful biophysical interpretations \cite{KLXia:2014c}.

In general, topology based quantitative modeling and analysis will be a new trend in molecular bioscience and biophysics. Particularly, topological features are ideally suitable for machine learning based quantitative predictions of biomolecular functions.

\section{Continuous apparatuses for biomolecules}\label{sec:continuous}

As stated above, the topological study of scalar fields, particularly electron density data, has advanced the understanding of molecules tremendously. The theory of atoms in molecules (AIM) has provided an elegant and feasible partition of electron density field, so that one can mathematically rigorously define the atoms in molecule  \cite{Bader:1985,Bader:1990}. AIM can be generalized into a more general theory called quantum chemical topology (QCT), which employs the topological analysis in AIM for the study of other physically meaningful scalar fields  \cite{Popelier:2005}. Additionally, geometric modeling and analysis, particularly the surface curvature analysis, has contributed a lot to the molecular visualization and structure characterization \cite{Whitley:2012}. Thanks to the geometric analysis, the establishment and further deeper understanding of structure-function relationship have been achieved. In this section,  a brief discuss of geometric and topological analysis of scalar fields is presented. 

Continuum representation of biomolecules plays an important role in their modeling, analysis and simulation. Volumetric biomolecular data are typically obtained from cryo-EM maps, quantum mechanical simulations and mathematical models that transform discrete datasets originally generated by X-ray crystallography or other means into continuous ones. Therefore, continuous mathematical approaches for analysis and modeling of biomolecules in the volumetric data form are as important as their discrete counterparts.

\subsection{Geometric representation}\label{Sec:GeometricRep}
	

Geometric modeling is a crucial  ingredient of biophysics. Due to   increasingly powerful high performance computers, geometric modeling  has become an essential 	apparatus    for  biomolecular surface representation,  visualization, surface and volumetric meshing, area and volume estimation, curvature analysis and  filling the gap between macromolecular structural information and their theoretical models \cite{ZYu:2008, XFeng:2012a,XFeng:2013b,KLXia:2014a,JLi:2013,Quine:2006intensity}.  The visualization of macromolecules  sheds light on biomolecular structure, function and interaction, including ligand-receptor binding sites, protein specification, drug binding, macromolecular assembly, protein-nucleic acid interactions, protein-protein binding hot spots, and enzymatic mechanism \cite{GRASP2,Rocchia:2002,NKWH07, Decherchi:2013}.

\paragraph{Non-smooth biomolecular surface representations}

A number of  molecular surface models has been proposed. Among them, the van der Waals surface (vdWS)  is defined as the union of the atomic surfaces under a given atomic radius for each type of atoms. Solvent accessible surface (SAS) is defined as the trajectory of the center of a probe sphere moving around the van der Waals surface \cite{Lee:1971}. Because  vdWs and SASs are non-smooth at intersection areas where two or more atoms join together,  solvent excluded surface (SES)  was introduced to generate relatively more smooth surfaces   \cite{Richards:1977,Connolly85}. The SES can be obtained by tracing the inward moving surface of a probe sphere rolling around the vdW surface. Connolly divided SES into two major parts, the contact areas formed by the subsets of the vdWs surface and the re-entrant surfaces, which contain toroidal patches and concave spherical triangles.

\paragraph{Smooth biomolecular surface representations}
The SES of proteins admits geometric singularities, such as tips, sharp edges and self-intersecting surfaces \cite{Sanner:1996,Yu:2007a}
The construction of smooth biomolecular surfaces has been of considerable interest  \cite{Blinn:1982,Duncan:1993, QZheng:2012,ZYu:2008,MXChen:2011}.  The rigidity index in Eq. (\ref{rigidity1}) has been extended into a continuous rigidity density \cite{KLXia:2013d,Opron:2014}
\begin{align}\label{eq:rigidity3}
	\mu^1(\mathbf{r})=\sum_{\substack{j=1}}^{N} w_{j} \Phi\left(\|\mathbf{r}-\mathbf{r}_j\|;\eta_{j}\right).
\end{align}
Rigidity density (\ref{eq:rigidity3}) serves as an excellent representation of molecular surfaces \cite{KLXia:2015e}.
 Gaussian  surface  was proposed with the Gaussian kernel  \cite{Zap,Grant:2007, LLi:2013, LinWang:2015}.
Recently,  Gaussian  surface has been extended to a new class of surface densities equipped with a wide variety of FRI correlation kernels  ($\Phi\left(\|\mathbf{r} -\mathbf{r}_j\|;\eta_{ j}\right)$)  \cite{DDNguyen:2016b}
\begin{align}\label{rigidity2}
\mu^2(\mathbf{r}) =1-\prod_{\substack{j=1}}\left[1-w_{ j}\Phi\left(\|\mathbf{r} -\mathbf{r}_j\|;\eta_{ j}\right)\right].
\end{align}
Two rigidity densities $\mu^\alpha(\mathbf{r}), ~\alpha=1,2$ may behave very differently. Therefore, one can normalize these densities by their maximal values
\begin{align}\label{normalization}
	\bar{\mu}^{\alpha}(\mathbf{r})=\frac{\mu^{\alpha}(\mathbf{r})}{\max\limits_{\mathbf{r}\in \mathbb{R}^3} \mu^\alpha(\mathbf{r})}, \quad \alpha =1,2.
\end{align}
As a result, the behaviors of two rigidity surfaces can be compared.

\paragraph{Discrete to continuum mapping}

Many geometric and topological  apparatuses are invented for continuous volumetric data. A typically examples include differential geometry and  differential topology that  deal with differentiable functions on differentiable manifolds. Cryo-EM data and electron quantum densities can be directly treated by   mathematical tools devised from  differentiable manifolds. However, a large variety of discrete  macromolecular data originate from X-ray crystallography, NMR etc are not directly differentiable. Therefore, it desirable to transform discrete biomolecular datasets into continuous ones.

The rigidity  densities defined in Eq.  (\ref{eq:rigidity3}) is differentiable for  $\eta_{ j}>0$. Therefore,  rigidity  densities also serve as a discrete to continuum mapping. The resolution parameter can be exploited for generating multidimensional persistence as illustrated in Section \ref{sec:resultionPH}.  As  a result, many mathematical techniques developed for continuous datasets can be employed to analyze discrete biomolecular datasets, such as X-ray crystallography data.

Additionally, the normalized rigidity density $\bar{\mu}^{1}(\mathbf{r})$ given in Eq. (\ref{normalization})  can be used as solute domain indicators for  implicit solvent models \cite{Holst:1994,Baker:2001,Geng:2007a,Geng:2011}, such as those used in the differential geometry based  Poisson-Boltzmann theory \cite{Wei:2009,ZhanChen:2010a,ZhanChen:2010b}.

Mathematically, the aforementioned discrete to continuum mapping is an interpolation using kernels. Many other techniques, such as splines, polynomials, wavelets,  and Pad\'{e} approximation can be used as well. Currently, there is little  numerical analysis of the mapping in biomolecular context and further mathematical study is needed to improve the stability and efficiency of the mapping for large data sets.

\subsection{Multiresolution and multidimensional persistent homology}\label{sec:resultionPH}

\begin{figure}
\begin{center}
\begin{tabular}{c}
\includegraphics[width=0.8\textwidth]{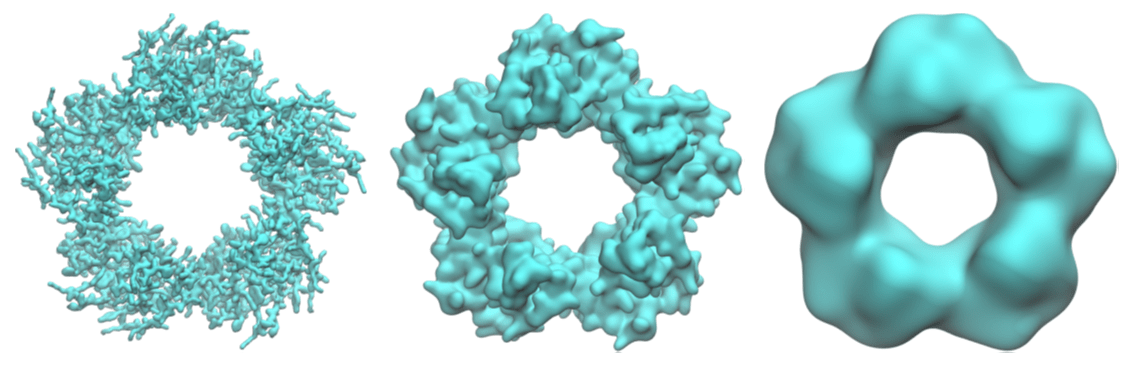}
\end{tabular}
\end{center}
\caption{Multiresolution representations of protein 1DYL. Different values of resolution parameter $\eta$ are used to generate rigidity density functions in different scales.}
\label{fig:1DYL}
\end{figure}
Although  persistent homology was originally built over simplicial complex of discrete data sets,  homology and persistent homology in the cubical complex setting have been developed   \cite{Kaczynski:2004,Strombom:2007}. Therefore, one can apply these techniques to volumetric datasets directly, particularly, with an available software package,  \href{http://www.sas.upenn.edu/~vnanda/perseus/index.html}{Perseus} \cite{Mischaikow:2013}. The reader is referred to the literature  for more comprehensive  discussion  and treatment  \cite{Kaczynski:2004,Strombom:2007}.

Recently, persistent homology analysis of macromolecular volumetric datasets have been demonstrated \cite{KLXia:2014c,KLXia:2015a,KLXia:2015b}. In this paper, the multiresolution persistent homology \cite{KLXia:2015d,KLXia:2015e} and multidimensional persistent homology \cite{KLXia:2015c}  developed for volumetric macromolecular datasets are discussed.

\paragraph{Multiresolution persistent homology}

As stated earlier, the basic idea of persistent homology is to exploit the topological changes of a given dataset at different scales of representations \cite{KLXia:2014c,KLXia:2015a,KLXia:2015b}.   For the geometric representation given by  Eq.  (\ref{eq:rigidity3}),  rigidity density $\mu^1(\mathbf{r})$ depends on the resolution parameter $\eta$. The resolution parameter can be turned to emphasize  the molecular features of   scale $\eta$, see Fig. \ref{fig:1DYL}. Resolution based continuous coarse-grained representations can be constructed for excessively large  datasets in the spirit of wavelet multiresolution analysis. This approach is particularly valuable for representing viruses, protein complexes and subcellular organelles.

 Since the geometric representation is controlled by the resolution parameter $\eta$  in Eq.  (\ref{eq:rigidity3}), one can develop  multiresolution persistent homology (MPH) for macromolecular analysis. The essential idea is to match the scale of interest with appropriate resolution in the topological analysis. In contrast to the original persistent homology that is based on a uniform resolution of the point cloud data over the filtration domain, the MPH provides a mathematical microscopy of the topology at a given scale through a corresponding resolution. MPH can be utilized to reveal the topology of a given geometric scale and employed as a topological focus of lens. It becomes powerful when it is applied in conjugation with the data that has a multiscale nature, such as a multiprotein complex as shown in Fig. \ref{fig:Multiscale_illustration}. In this case,  MPH can be used to extract the topological fingerprints either at  atomic scale, residue scale, alpha helix and beta sheet scale, domain scale or at the protein scale.

Another very interesting multiresolution model is Mapper \cite{singh:2007,Carlsson:2009}. This method is proposed for qualitative analysis, simplification and visualization of high dimensional data sets. It manages to reduce the complexity by using fewer points which can capture topological and geometric information at a specified resolution. Interestingly, Mapper also uses kernel functions. However, it should be noticed that it does not utilize the resolution parameter in the filtration process for persistent homology. 

It is expected that related subjects, such co-homology, Floer homology, Sheaf and K-theory,  will find interesting applications in biomolecular systems.

\paragraph{Multidimensional persistent homology}

\begin{figure}
\begin{center}
\begin{tabular}{c}
\includegraphics[width=0.8\textwidth]{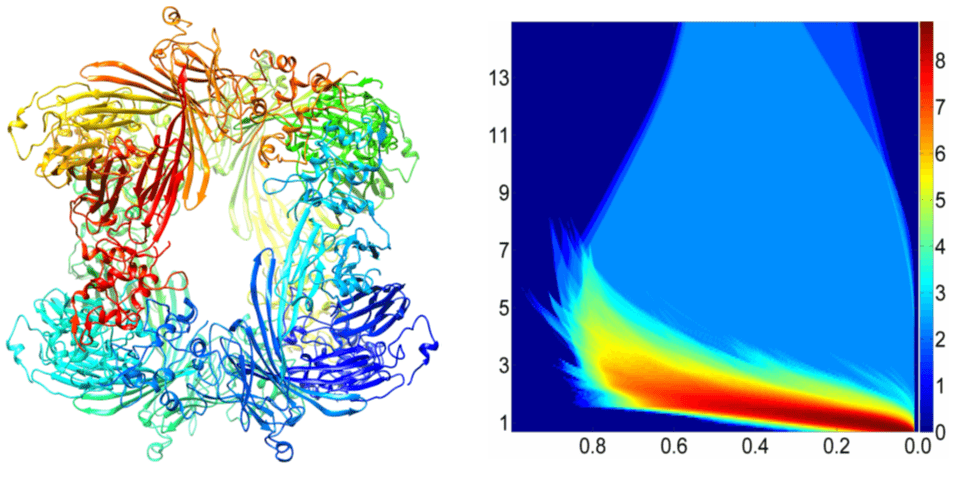}
\end{tabular}
\end{center}
\caption{An illustration of  multidimensional persistent homology analysis of protein 2YGD. Left chart: protein 2YGD; Right chart: two-dimensional persistence of 2YGD. The horizontal axis denotes density and the vertical axis represents the logarithmic values of persistent Betti numbers.}
\label{fig:2ygd}
\end{figure}

There have been considerable interest in developing  multidimensional persistent homology or multidimensional persistence. Here, two classes of multidimensional persistence  algorithms for biomolecular data are discussed. One class of multidimensional persistence is generated by repeated applications of 1D persistent homology to high-dimensional data, such as those from protein folding, molecular dynamics, geometric partial differential equations, etc. The resulting high-dimensional persistent homology is a pseudo-multidimensional persistence and has been applied to protein folding analysis  to identify topological transitions \cite{KLXia:2015c}.

Another class of multidimensional persistence is created by the geometric representation given in  Eq.  (\ref{eq:rigidity3}). As   $\eta$ is an independent variable that can modify the geometry and topology of the underlying data set, one can carry out filtration with respect to both the original density and the resolution $\eta$ to obtain genuine 2D persistent homology \cite{KLXia:2015c}. Indeed,  each  $\eta$ value leads to a family of new simplicial complexes. Similarly, for an $N$-dimensional  persistent homology, $N$-independent variables should be introduced for the filtration. Higher dimensional persistence has been demonstrated for macromolecular data  \cite{KLXia:2015c}. Resolution induced persistent homology and multidimensional persistence have been applied to many  biomolecular  systems, including protein flexibility analysis, protein folding characterization, topological denoising, noise removal from cryo-EM data, and analysis of fullerene molecules. An example of  multidimensional persistent homology is depicted in Fig. \ref{fig:2ygd}. Clearly, the maps of multidimensional persistent homology given in Fig. \ref{fig:2ygd} can be employed for deep learning, which is an open field.

Basically each independent parameter that regulates the filtration process  contributes a genuine persistent dimension. When the dimension is higher than two, the result representation is no longer straightforward. Additionally, how to make use of multidimensional persistence in realistic applications is also an interesting problem.

\subsection{Differential geometry theory of surfaces}\label{sec:DGA}

Differential geometry has fruitful applications in physics, particularly, general relativity and has found its success in biomolecular systems as well \cite{Bates:2008,Wei:2009,ZhanChen:2010a}. As stated earlier, in biophysical modeling, surface representation is a crucial subject and commonly used surface definitions lead to geometric singularities. Gaussian surface and general FRI rigidity surface are based on simple geometric ideas. In contrast, differential geometry theory of surfaces gives rise to natural description macromolecular surfaces \cite{Bates:2008,Wei:2009,ZhanChen:2010a}. This approach becomes powerful when it is combined with variation calculus for biomolecular modeling as shown in Section \ref{sec:scalar_field_geometry}. In this section, a brief introduction is given about the differential geometry theory of surfaces using the notations and definitions from Ref.  \cite{Kuhnel:2015}.

\paragraph{Surface elements and immersion}
For an open set $U \subset \mathds{R}^2$, a parametrized surface element is an immersion $f: U \rightarrow \mathds{R}^3$. Here $f$ is also known as a parametrization. One can call the elements of $U$ as parameters and their images under $f$ as points. If the rank of map $f$ is maximal, $f$ is an immersion. The point where the rank is not maximal, it is called a singular point or singularity \cite{Kuhnel:2015}.


One can use the following notations for a parametrized surface element $f:U\rightarrow \mathds{R}^3, u \in U, p=f(u)$. $T_uU$ is the tangent space of $U$ at $u$, $T_uU=\{u\}\times \mathds{R}^2 $; $T_p\mathds{R}^3$ is the tangent space of $\mathds{R}^3$ at $p$, $T_p\mathds{R}^3=\{p\}\times \mathds{R}^3 $; $T_uf$ is the tangent space of $f$ at $p$, $T_uf:=Df|_u(T_uU) \subset T_{f(u)} \mathds{R}^3 $; and $\perp_uf$ is the normal space of $f$ at $p$ $T_uf \oplus \perp_uf =T_{f(u)} \mathds{R}^3 $. The element of $T_uf$ is called tangent vector and the element of $\perp_uf$ is the normal vector \cite{Kuhnel:2015}.

\paragraph{First fundamental form}
The first fundamental form $I$  is the inner product between two tangent vectors $X,Y$ in tangent planes $T_uf$, i.e., $I(X,Y):=<X,Y>$.
For coordinate systems $f(u,v)=\left( x(u,v),y(u,v),z(u,v) \right)$, the first fundamental form can be described by the following tensor matrix \cite{Kuhnel:2015}
$$(g_{ij})=\left( \begin{array}{cc}
        E(u,v) & F(u,v) \\
        F(u,v) & G(u,v)
      \end{array}
    \right)= \left(
                  \begin{array}{cc}
                    I(\frac{\partial f}{\partial u},\frac{\partial f}{\partial u}) & I(\frac{\partial f}{\partial u},\frac{\partial f}{\partial v}) \\
                    I(\frac{\partial f}{\partial u},\frac{\partial f}{\partial v}) & I(\frac{\partial f}{\partial v},\frac{\partial f}{\partial v})  \end{array}
                \right)= \left(
      \begin{array}{cc}
         <\frac{\partial f}{\partial u},\frac{\partial f}{\partial u}> & <\frac{\partial f}{\partial u},\frac{\partial f}{\partial v}> \\
         <\frac{\partial f}{\partial u},\frac{\partial f}{\partial v}> & <\frac{\partial f}{\partial v},\frac{\partial f}{\partial v}>
      \end{array}
    \right)$$
Also one can have the line element
$$ds^2=E(u,v)du^2 +2F(u,v)dudv+G(u,v)dv^2$$
and the surface area
$$dA=\sqrt{g}dudv,$$
where $g= {\rm Det}(g_{ij})$ is the determinant.

\paragraph{Gauss map}
Since each plane is essentially determined by its normal vector, the curvature of the surface can be studied by the variation of the normal vector, i.e., Gauss map. For a surface element $f: U \rightarrow \mathds{R}^3$, the Gauss map is $v:U \rightarrow S^2$ and is defined by the formula \cite{Kuhnel:2015}
$$ v(u_1,u_2):=\frac{\frac{\partial f}{\partial u_1} \times \frac{\partial f}{\partial u_2}}{|\frac{\partial f}{\partial u_1} \times \frac{\partial f}{\partial u_2}|},$$
where $S^2$ denotes the unite sphere $S^2=\{(x,y,z)\in \mathds{R}^3 | x^2+y^2+z^2=1\}$.

\paragraph{Weingarten map}
Let $f: U \rightarrow \mathds{R}^3$ be a surface element with Gauss map $v:U \rightarrow S^2 \in \mathds{R}^3$, and for every $u \in U$ the image plane of the linear map $Dv|_u:T_uU \rightarrow T_{v(u)}\mathds{R}^3$  is parallel to the tangent plane $T_uf$. By canonically identifying $T_{v(u)}\mathds{R}^3 \cong \mathds{R}^3 \cong T_{f(u)}\mathds{R}^3 $, one can have $Dv$ at every point as the map $Dv|_u:T_uU \rightarrow T_uf$. Moreover, by restricting to the image, one may view the map $Df|_u$ as a linear isomorphism $Df|_u:T_uU \rightarrow T_uf$. In this sense the inverse mapping $(Df|_u)^{-1}$ is well-defined and is also an isomorphism. The map $L:=-Dv\circ (Df)^{-1}$ defined point-wisely by
$$ L_u:=-(Dv|_u)\circ (Df|_u)^{-1}: T_uf \rightarrow T_uf $$
is called a Weingarten map or the shape operator of $f$.  This map is independent of the parametrization of $f$, and it is self-adjoint with respect to the first fundamental form I.

\paragraph{Second and third fundamental form}
Let $f: U \rightarrow \mathds{R}^3$ be a surface element with Gauss map $v:U \rightarrow S^2 \in \mathds{R}^3$.  With the shape operator $L$ and tangent vectors $X$ and $Y$, the second fundamental $II$ is given by
$$II(X,Y):=I(LX,Y)$$
and the third fundamental form is
$$III(X,Y):=I(L^2X,Y)=I(LX,LY).$$
$II$ and $III$ are symmetric bilinear forms on $T_uf$ for every $u \in U$.
The three fundamental forms $I$,$II$ and $III$ have a relation as,
$$III-{\rm Tr}(L)II+{\rm Det}(L)I=0.$$
In coordinates $f(u,v)=\left( x(u,v),y(u,v),z(u,v) \right)$, three fundamental forms can be expressed as
\begin{eqnarray}
&&I: g_{ij}=<\frac{\partial f}{\partial u_i},\frac{\partial f}{\partial u_j}>; \\
&&II: h_{ij}=<v,\frac{\partial^2 f}{\partial u_i \partial u_j}>=-<\frac{\partial v}{\partial u_i}, \frac{\partial f}{\partial u_j}>; \\
&&III: e_{ij}=<\frac{\partial v}{\partial u_i},\frac{\partial v}{\partial u_j}>.
\end{eqnarray}

\begin{figure}
\begin{center}
\begin{tabular}{c}
\includegraphics[width=0.5\textwidth]{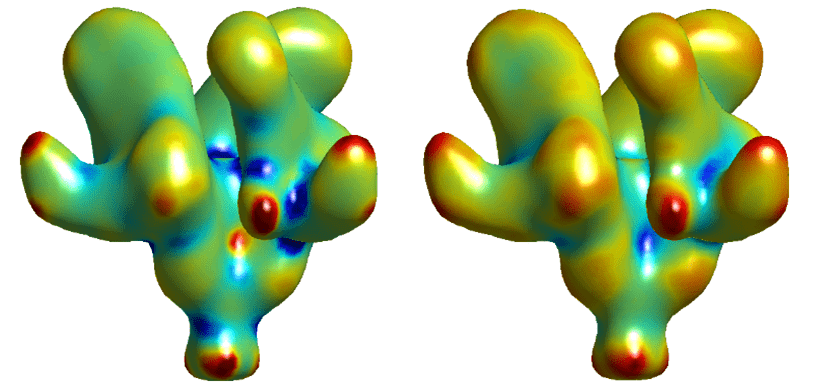}
\end{tabular}
\end{center}
\caption{Illustration of Gaussian curvature (left side) and mean curvature (right side) for an HIV-1 gp 120 trimer structure (EMD-5020).}
\label{fig:curvature_g_k}
\end{figure}

\paragraph{Principal curvature}
For a unite tangent vector $X \in T_uf$ and $I(X,X)=1$, it is a principal curvature direction for $f$ if $II(X,X)$ has a stationary value for $X$, or $X$ is an eigenvalue of the Weingarten map $L$. Further, the corresponding eigenvalue $\kappa$ is the principal curvature.

\paragraph{Gaussian and mean curvature}
Gaussian curvature is the determinate $K={\rm Det}(L)=\kappa_1\cdot \kappa_2$ and mean curvature is the average value $H=\frac{1}{2}{\rm Tr}(L)=\frac{1}{2}(\kappa_1+\kappa_2)$. Gaussian and mean curvatures can be expressed in local coordinates as
\begin{eqnarray}
&&K=\frac{ {\rm Det}(h_{ij})}{{\rm Det}(g_{ij})}=\frac{h_{11}h_{22}-h_{12}^2}{g_{11}g_{22}-g_{12}^2},\\
&&H=\frac{1}{2}\Sigma_i h^i_i=\frac{1}{2}\Sigma_{i,j} h_{ij} g^{ij}= \frac{1}{2 {\rm Det}(g_{ij})}(h_{11}g_{22}-2h_{12}g_{12}+h_{22}g_{11}).
\end{eqnarray}
Gaussian and mean curvatures of an HIV virus fragment are illustrated in Fig. \ref{fig:curvature_g_k}. In biophysics, the region with negative Gaussian is often associated membrane-protein interaction sites, while the region with negative mean curvatures on a  protein surface is commonly regarded as a potential-ligand or protein-drug bonding site \cite{KLXia:2014a}.

\subsection{Differential geometry modeling and computation}\label{sec:scalar_field_geometry}

With the advance of experimental technology, more than a hundred thousand of 3D macromolecular structural data has been accumulated. Differential geometry based surface modeling and curvature measurement are of essential importance to the geometric description and feature recognition of these 3D structural data \cite{Wei:2005, Xu:DSM:2006}.   In this section,   a brief review is given to the differential geometry based modeling of macromolecular surfaces. Two algorithms for curvature calculations are also discussed.

\subsubsection{Minimal molecular surface }\label{Sec:MMSgeneration}

\begin{figure}
\begin{center}
\begin{tabular}{c}
\includegraphics[width=0.5\textwidth]{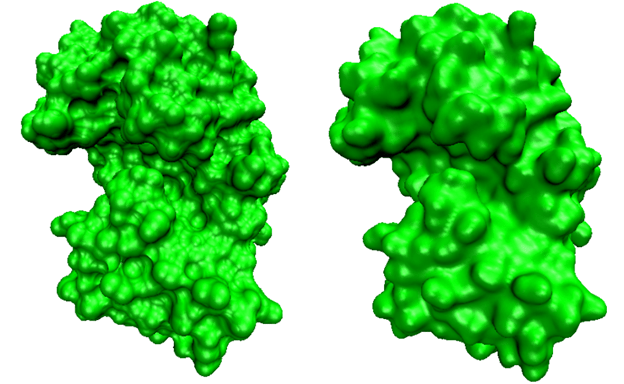}
\end{tabular}
\end{center}
\caption{Comparison of the solvent excluded surface (Left chart) and the minimal molecular surface (Right chart) of protein 1PPL. The minimal molecular surface is free from geometric singularities.}
\label{fig:MMS}
\end{figure}

Minimal molecular surface (MMS) was introduced  to construct surfaces free of  geometric singularities via the variational principle \cite{Bates:2006,Bates:2008}. In this approach, a hypersurface $S$ is defined to represent the biomolecular surface. Basically, one assigns each point with coordinate $(x,y,z)$ a value $S(x,y,z)$, which represents the domain information. It can be viewed as a characteristic function of the macromolecular domain.  By using geometric measure theory, the surface energy functional can be expressed as \cite{Wei:2009}
\begin{eqnarray} \label{eq2surface}
G_{\rm{surface}}= \gamma {\rm{Area}}= \int_{{\mathbb R}^3} \gamma |\nabla S | d{\bf{r}},
\end{eqnarray}
where $\gamma$ is the surface tension. As it is convenient for us to set up the total free functional as a 3D integral in ${\mathbb R}^3$,  one can make use of  the concept of a mean surface area  \cite{Wei:2009,ZhanChen:2011a} and the coarea formula \cite{coarea} on a smooth surface
 \begin{eqnarray} \label{eqarea}
{\rm{Area}}=\int_0^1 \int_{S^{-1}(c)\bigcap\Omega} d\sigma dc = \int_{\Omega} |   \nabla S({\bf{r}})   | d{\bf{r}}, \quad \Omega \subset {\mathbb R}^3.
\end{eqnarray}
{Here the value of  hypersurface function $S$ is distributed between 0 and 1, $S^{-1}(c)$ represents the inverse function of $S$ and $\Omega$ is defined as the whole domain.}
The variation of Eq.  (\ref{eq2surface}) with respect to $S$ leads to the vanishing of surface-tension weighted mean curvature, $\nabla\cdot\left(\gamma \frac{\nabla S}{|\nabla S|}\right)=0$.  The energy minimization of  Eq.  (\ref{eq2surface}) can be realized by the introduction of an artificial time to obtain a generalized Laplace-Beltrami equation
\begin{eqnarray}\label{MeanCF}
   \frac{\partial S}{\partial t}&=&|\nabla S|\nabla\cdot\left(\frac{\gamma\nabla S}{|\nabla S|}\right),
\end{eqnarray}
The final MMS, which is free of geometric singularity, is obtained by extracting an iso-surface from the steady state hypersurface function. During each iteration, one can keep the value of $S$ in the van der Waals surface enclosed domain unchanged. Figure \ref{fig:MMS} illustrates the difference between the solvent excluded surface  and the minimal molecular surface   of protein 1PPL.

In the earlier works, sophisticated  computational algorithms have been developed to accelerate the construction of MMSs for large biomolecules \cite{Bates:2009,ZhanChen:2010a,ZhanChen:2010b,XFeng:2012a}. Differential geometry based molecular surface modeling was extended to solvation analysis \cite{Wei:2009,ZhanChen:2010a}, including level set approaches \cite{Cheng:2007e,Cheng2:2009}, and ion channel transport \cite{Wei:2009,Wei:2012,Wei:2013,DuanChen:2013}. Numerical aspects were examined in the literature \cite{SZhao:2011a,SZhao:2014a}.
Since these approaches work very well for biomolecular structure, function and dynamics, they will attract much attention in the future.
However, a more detailed discussion of  these issues is beyond the scope of the present review.

%
%
%
%
%

\subsubsection{Scalar field curvature evaluation}\label{sec:algorithm}

For a given set of  volumetric biomolecular data, the efficient and accurate computation of curvatures is needed. The evaluation of curvature properties from iso-surface embedded volumetric data has been   studied in geometric modeling, although the related techniques have not received much attention in computational biophysics. In this section,   two popular algorithms are reviewed. Many other elegant methods, including isophote surface based curvature evaluation \cite{Verbeek:1993}, Sander-Zucker approach \cite{Sander:1990}, Direct surface mapping based approach \cite{Stokely:1992}, piece-wise linear manifold techniques \cite{Stokely:1992}, etc, are often used in the computer science community.

\paragraph{Algorithm I}
Essentially, the first and second fundamental forms in the differential geometry are involved in the definition and evaluation of the curvatures.   a brief discussion of the mathematical background is given \cite{Soldea:2006, Bates:2008}.

The surface of interest can be extracted from a level set with iso-value $S_0$, i.e., $S(x,y,z)=S_0$. One can assume $S$ to be non-degenerate, i.e., the norm of the gradient is non-zero at $S(x,y,z)=S_0$. Without loss of generality, one can assume that the projection onto $z$ is non-zero as well. Then the implicit function theorem states that locally, there exists a function $z=f(x,y)$, which parametrizes the surface as ${\bf S}(x,y)=(x,y,f(x,y))$. One can express the iso-value relation as  $S(x,y,f(x,y))=S_0$. The differentiation with respect to $x$ and $y$ variables leads to two more equations
\begin{eqnarray}\nonumber
 S_x(x,y,f(x,y))+S_z(x,y,f(x,y))f_x(x,y)=0,\\\nonumber
 S_y(x,y,f(x,y))+S_z(x,y,f(x,y))f_y(x,y)=0,
\end{eqnarray}
where $f_x(x,y)$ and $f_y(x,y)$ can be given by
\begin{eqnarray} \nonumber
f_x(x,y)=-\frac{S_x(x,y,z)}{S_z(x,y,z)}; \quad {\rm and} \quad f_y(x,y)=-\frac{S_y(x,y,z)}{S_z(x,y,z)}.
\end{eqnarray}
One can define $E(x,y,z), F(x,y,z), G(x,y,z), L(x,y,z), M(x,y,z)$ and $ N(x,y,z)$ below to be the coefficients in the first and second fundamental forms. For simplicity, one can hide  parameter labels. Their values for surface function ${\bf S}=(x,y,f)$ can be given as
\begin{eqnarray}\nonumber
E &=& \langle {\bf S}_x , {\bf S}_x\rangle=1+f_x^2=1+\frac{S_x^2}{S_z^2};\\\nonumber
F &=& \langle {\bf S}_x , {\bf S}_y\rangle=f_xf_y=\frac{S_x S_y}{S_z^2};\\\nonumber
G &=& \langle {\bf S}_y , {\bf S}_y\rangle=1+f_y^2=1+\frac{S_y^2}{S_z^2};\\\nonumber
L &=& \langle {\bf S}_{xx} , {\bf n}\rangle=\frac{2S_xS_zS_{xz}-S_x^2S_{zz}-S_z^2S_{xx}}{g^{\frac{1}{2}}S_z^2};\\\nonumber
M &=& \langle {\bf S}_{xy} , {\bf n}\rangle=\frac{S_xS_zS_{yz}+S_yS_zS_{xz}-S_xS_yS_{zz}-S_z^2S_{xy}}{g^{\frac{1}{2}}S_z^2};\\\nonumber
N &=& \langle {\bf S}_{yy} , {\bf n}\rangle=\frac{2S_yS_zS_{yz}-S_y^2S_{zz}-S_z^2S_{yy}}{g^{\frac{1}{2}}S_z^2}.
\end{eqnarray}
The Gaussian curvature can be expressed as the ratio of the determinants of the second and first fundamental forms,
\begin{eqnarray}\nonumber
&&K=\frac{2S_xS_yS_{xz}S_{yz}+2S_xS_zS_{xy}S_{yz}+2S_yS_zS_{xy}S_{xz}}{g^2}-\frac{2S_xS_zS_{xz}S_{yy}+2S_yS_zS_{xx}S_{yz}+2S_xS_yS_{xy}S_{zz}}{g^2} \\\nonumber
&& \qquad \left. +\frac{S_z^2S_{xx}S_{yy}+S_x^2S_{yy}S_{zz}+S_yS_{xx}S_{zz}}{g^2} -\frac{S_x^2S_{yz}^2+S_y^2S_{xz}^2+S_z^2S_{xy}^2}{g^2} \right..
\end{eqnarray}

Similarly, the mean curvature is given as the average second derivative with respect to the normal direction,
\begin{eqnarray}\label{eq:meanC}
H=\frac{2S_xS_yS_{xz}+2S_xS_zS_{xz}+2S_yS_zS_{yz}-(S_y^2+S_z^2)S_{xx}-(S_x^2+S_z^2)S_{yy}-(S_x^2+S_y^2)S_{zz}}{2g^{\frac{3}{2}}}.
\end{eqnarray}
Note that curvature expressions become analytical when the implicit surface is given by the FRI rigidity density,  Eq.  (\ref{eq:rigidity3}).

\paragraph{Algorithm II}
An alternative  algorithm for the curvature extraction from volumetric data is the Hessian matrix method \cite{Kindlmann:2003}. For  volumetric data $S(x,y,z)$, one defines the surface gradient ${\bf g}$ and surface norm ${\bf n}$.
\begin{eqnarray}
&& {\bf g} = \nabla S= (S_x,S_y,S_z)^T;\\
&& {\bf n  = -\frac{g}{|g|}}.
\end{eqnarray}
Here  ${T}$ denoting the transpose. One further calculates the matrix $\nabla {\bf n}^T$, which can be expressed as
\begin{eqnarray}
&& \nabla {\bf n}^T=- \nabla ( { \frac{{\bf g}}{|{\bf g}|}})=-( { \frac{\nabla {\bf g}^T}{|{\bf g}|}-\frac{{\bf g} \nabla^T |{\bf g}|}{|{\bf g}|^2}}) \\\nonumber
&&  \qquad =-\frac{1}{|{\bf g}|}({H}-\frac{{\bf g} \nabla^T ({\bf g}^T{\bf g})^{\frac{1}{2}}}{|{\bf g}|})=-\frac{1}{|{\bf g}|}({H}-\frac{{\bf g} \nabla^T ({\bf g}^T{\bf g})}{2|{\bf g}|({\bf g}^T{\bf g})^{\frac{1}{2}}}) \\\nonumber
&& \qquad =-\frac{1}{|{\bf g}|}({H}-\frac{{\bf g}(2{\bf g}^TH)}{2|{\bf g}|^2})=-\frac{1}{|{\bf g}|}({ I}-\frac{{\bf g}{\bf g}^T}{|{\bf g}|^2}){H}  \\\nonumber
&& \qquad =-\frac{1}{|{\bf g}|}({ I}-{\bf n}{\bf n}^T){H}=-\frac{1}{|{\bf g}|} {PH},
\end{eqnarray}
where ${I}$ is the identity matrix, matrix $P=I-{\bf nn}^T$ and Hessian matrix ${H}$ is given by
\begin{eqnarray}
{H}=\left[
\begin{array}{ccc}
                    \frac{\partial^2 S}{\partial^2 x}  &\frac{\partial^2  S}{\partial x\partial y} &\frac{\partial^2  S}{\partial x\partial y} \\
                    \frac{\partial^2 S}{\partial x\partial y} &\frac{\partial^2  S}{\partial^2 y}  &\frac{\partial^2  S}{\partial y\partial z}\\
                    \frac{\partial^2 S}{\partial x\partial z} &\frac{\partial^2  S}{\partial y\partial z}  &\frac{\partial^2  S}{\partial^2 z}
\end{array}
          \right].
\end{eqnarray}
Geometrically, ${\bf nn}^T$ project ${\bf n}$ onto a one-dimensional span of ${\bf n}$. Here $I-{\bf nn}^T$ further projects onto the orthogonal space complement to the span of ${\bf n}$, which is the tangent plane.

In general, both ${ P}$ and ${H}$ are symmetric but ${\bf \nabla n}^T$ is not. If ${\bf q_1}$ lies in the tangent plane, $P{\bf q_1}={\bf q_1}$ and ${\bf q_1}^T P={\bf q_1}^T$. Therefore, for   ${\bf q_2}$ and ${\bf q_1}$  in the tangent plane, one has
\begin{eqnarray}
{\bf q_1}^T PH {\bf q_2=q_1}^T H {\bf q_2=q_2}^T H{\bf q_1=q_2}^T PH{\bf q_1}
\end{eqnarray}

The restriction of $\nabla {\bf n}^T=-\frac{1}{|{\bf g}|} { PH}$ to the tangent plane is symmetric and thus there exists an orthonormal basis ${\bf p_1,p_2}$ for the tangent plane in which ${\bf n}^T$ is a 2*2 diagonal matrix. This basis can be easily extended to an orthonormal basis for all $\{{\bf p_1}, {\bf p_2}, {\bf n} \}$. In this basis, the derivative of the surface normal is given by
\begin{eqnarray}
{\bf \nabla n}^T=\left[
\begin{array}{ccc}
                    \kappa_1  &0 &\sigma_1 \\
                    0 &\kappa_2  &\sigma_2\\
                    0 &0  &0
\end{array}
          \right].
\end{eqnarray}
The diagonal term in the  bottom row is zero because no change in normal ${\bf n}$ can lead to a change in length. Motion along ${\bf p_1}$ and ${\bf p_2}$ results in the change of ${\bf n}$ along the same directions, with a ratio of ${\kappa_1}$ and ${\kappa_2}$ respectively. Here ${\bf p_1}$ and ${\bf p_2}$ are the principal curvature directions, while ${\kappa_1}$ and ${\kappa_2}$ are the principal curvatures. When there is a change in normal ${\bf n}$, it  tilts according to ${\sigma_1}$ and ${\sigma_2}$. 

One can further multiply ${\bf \nabla n}^T$ by ${ P}$ to diagonalize the matrix
\begin{eqnarray}
G={\bf \nabla n}^T{ P}= {\bf \nabla n}^T \left[
\begin{array}{ccc}
                    1 &0 &0 \\
                    0 &2  &0\\
                    0 &0  &0
\end{array}
          \right]=\left[
\begin{array}{ccc}
                    \kappa_1  &0 &0 \\
                    0 &\kappa_2  &0\\
                    0 &0  &0
\end{array}
          \right].
\end{eqnarray}

The surface curvature measurements are based on geometry tensor ${ G}$. In a volumetric data set or a scalar field, ${G}$ is known in terms of the Cartesian basis $(x, y, z)$. Matrix invariants provide the leverage to extract the desired curvature values $\kappa_1$ and $\kappa_2$ from ${ G}$, regardless of the coordinate frame of the principal curvature direction. The trace of ${ G}$ is $\kappa_1+ \kappa_2$. The Frobenius norm of ${ G}$, notated $|{ G}|_F$ and defined as $\sqrt{{\rm Tr}({ GG}^T)}$, is $\sqrt{\kappa_1^2+\kappa_2^2}$. Then  $\kappa_1$ and $\kappa_2$  are found with the quadratic formula.

The two principal curvatures can be evaluated by the following procedure.
\begin{enumerate}
\item
Calculate matrix $P=I-{\bf nn}^T$
\item
Evaluate matrix $G=I-\frac{PHP}{|{\bf g}|}$,
\begin{eqnarray}
{ G}=(g_{ij})_{(i,j=1,3)}
\end{eqnarray}
\item
Compute the trace $t$ and Frobenius norm $f$ of matrix ${ G}$;
\begin{eqnarray}
&& t=g_{11}+g_{22}+g_{33};\\
&& f=\|{ G} \|=\sqrt{\sum_i \sum_j g_{ij}^2};\\
&& \kappa_1=\frac{t+\sqrt{2f^2-t^2}}{2};\\
&& \kappa_2=\frac{t-\sqrt{2f^2-t^2}}{2}.
\end{eqnarray}
\end{enumerate}
When the two principal curvatures are available, the Gaussian curvature $K$ and mean curvature $H$ can be obtained as
\begin{eqnarray}
&&K=\kappa_1 \kappa_2;\\
&&H=\frac{\kappa_1+\kappa_2}{2}.
\end{eqnarray}
Essentially, the Hessian matrix method generates the same results as Algorithm I derived from the first and second fundamental form. However, for FRI rigidity density given in  Eq.  (\ref{eq:rigidity3}), Algorithm I is preferred as it is analytical and without matrix diagonalization.

\subsubsection{Analytical minimal molecular surface}\label{sec:density2}

\begin{figure}
\begin{center}
\begin{tabular}{c}
\includegraphics[width=0.5\textwidth]{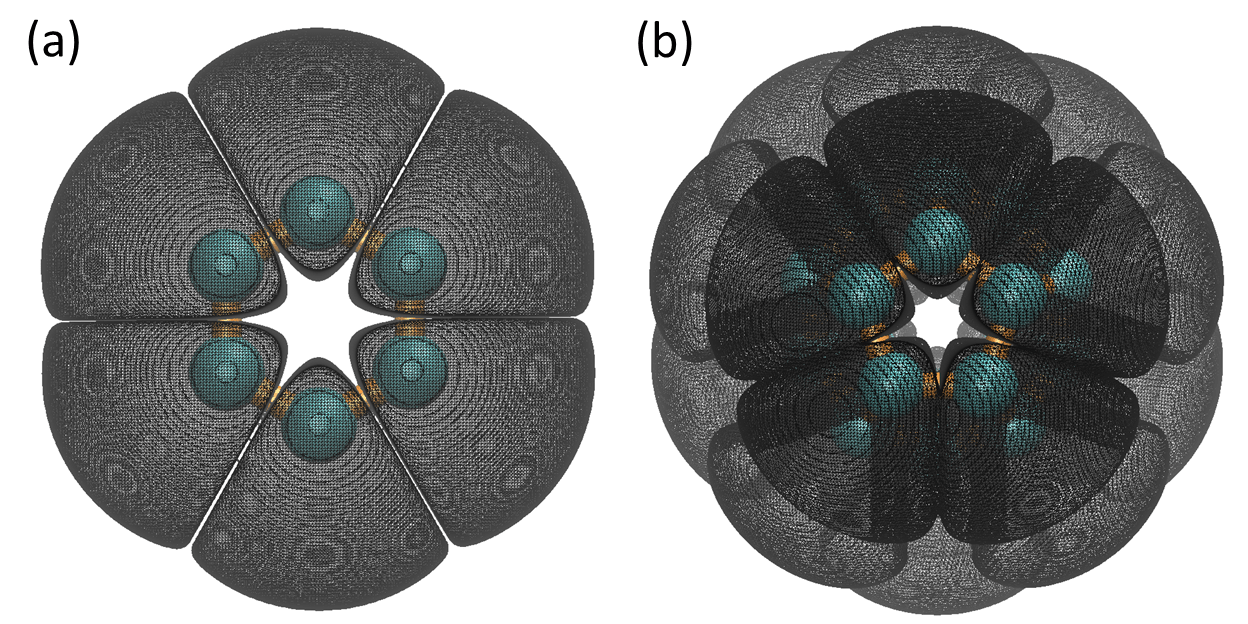}
\end{tabular}
\end{center}
\caption{An illustration of analytical minimal molecular surfaces generated by mean curvature isosurfaces. { (a)} and { (b)} are hexagonal ring  and fullerene $C_{20}$ molecule, respectively. Here the isovalue is chosen as 0.001.
}
\label{fig:minimal_surface1}
%
%
\begin{center}
\begin{tabular}{c}
\includegraphics[width=0.6\textwidth]{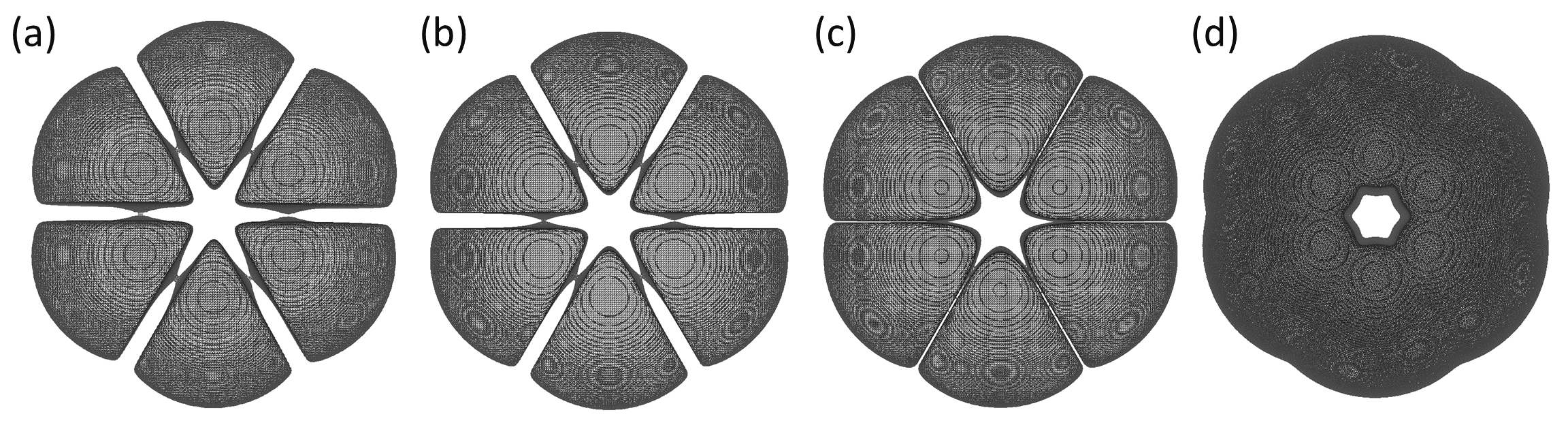}
\end{tabular}
\end{center}
\caption{An illustration of analytical minimal molecular surfaces generated  at different $\sigma$ values. From { (a)} to { (d)}, the $\sigma$ is chosen as 0.5, 0.6, 0.7 and 0.8, respectively. For all   figures,   isosurfaces are extracted at mean curvature isovalue of 0.001.
}
\label{fig:minimal_surface}
\end{figure}

Minimal surfaces are ubiquitous in nature and are a fascinating topic for centuries. The Euler-Lagrange minimization of the hypersurface was proposed to generate minimal molecular surface (MMS) for biomolecules \cite{Bates:2008}. In general, MMS is a result of vanishing mean curvature and is free of geometric singularity. It is a powerful concept in biophysical modeling \cite{Bates:2008,Wei:2009}. Nevertheless, it is still much more computationally expensive to construct MMS than to generate SES. In this work, a new minimal molecular surface, called analytical minimal molecular surface (AMMS) is proposed. In this approach, the rigidity density representation of biomolecular data by Eq. (\ref{eq:rigidity3}) is employed. Then the mean curvature can be {\it analytically} computed by using Eq. (\ref{eq:meanC}). Finally,  AMMS can be constructed by setting the isosurface of the mean curvature to zero (or practically, a number very close zero).

One can consider the hexagonal ring and a fullerene C$_{20}$ molecule in the study of AMMS.
The coordinates used for particles in the hexagonal ring are (0.000,   1.403,   0.000; -1.215,   0.701,   0.000; -1.215,  -0.701,  0.000; 0.000,  -1.403,   0.000; 1.215,  -0.701,   0.000; and 1.215,   0.701,   0.000).
The fullerene C$_{20}$  molecule has a highly symmetric cage structure made of 12 pentagons. Its atomic coordinates  are (1.569,  -0.657,  -0.936; 1.767,   0.643,  -0.472; 0.470,  -0.665,  -1.793; 0.012,   0.648,  -1.826; 0.793,   1.467,  -1.028; -0.487,  -1.482,  -1.216; -1.564,  -0.657,  -0.895; -1.269,   0.649,  -1.277; -0.002,  -1.962,  -0.007; -0.770,  -1.453,   1.036; -1.758,  -0.638,   0.474; 1.288,  -1.450,   0.163; 1.290,  -0.660,   1.305; 0.012,  -0.646,   1.853; 1.583,   0.645,   0.898; 0.485,   1.438,   1.194; -0.503,   0.647,   1.775; -1.606,   0.672,   0.923; -1.296,   1.489,  -0.166; -0.010,   1.973,  -0.006). The fullerene  rigidity density is given by Eq. (\ref{eq:rigidity3}). In both cases, the parameters $w_i$ and $\sigma_i$  are set to $1$ and $0.7$, respectively.

When the mean curvature isovalue equals to zero, the resulting surfaces are usually composed by several non-intersecting surfaces perpendicular to atomic bonds near the BCPs.  When loosing the condition a little bit by setting the isovalue  to 0.001 (or some other very small positive value), a better AMMS can be generated.   Figure \ref{fig:minimal_surface} depicts the AMMS for the hexagonal ring and C$_{20}$ molecule. It is found that for the hexagonal ring,  each atom is enclosed by a surface segment.  These segments are tightly close to each other and only connect  near BCPs.  For the C$_{20}$ molecules,  the surface appears to be  better connected. However, a careful examination reveals the separation as well.  Obviously, these surfaces are not minimal surfaces.

To generate better understanding of the  minimal molecular surface generation, one can explore several $\sigma$ values for the hexagonal ring. Figure \ref{fig:minimal_surface} illustrates the results. From { (a)} to { (d)},  $\sigma$ is chosen as 0.5, 0.6, 0.7 and 0.8, respectively. All the isosurfaces are extracted at mean curvature equals to 0.001. It is seen that a relatively large $\sigma$ value produces a better  minimal surface. The gaps between atomic segments are gradually narrow as the increase of $\sigma$ value and finally disappear, which gives rise to a good quality AMMS.

\subsection{Scalar and vector field topology} \label{sec:scalar_field_topology}

Topological approaches have become  an integral part in data analysis, visualization, and mathematical modeling for volumetric as well as for vector valued data sets. In fact, the results of scalar and vector field topology coincide each other for  gradient vector fields, although the respective mathematical approaches are originated from different fields. Vector field topology is usually developed for analyzing the streamlines of fluid flow generated by the velocity vector during the flow evolution. These techniques can be applied to macromolecular analysis for dealing with cryo-EM data and improving the structural construction.

Three dimensional scalar field data, particular the molecular structural data are widely available from many data sources. To decipher  useful information from these data requires highly efficient methods and algorithms. Mathematical approaches, including topological tools, such as fundamental groups, homology theory, contour tress, Reeb graphs \cite{Dey:2013}, Morse theory \cite{harker:mischaikow:mrozek:nanda, Mischaikow:2013}, etc, have a great potential for revealing the intrinsic  connectivity or structure-function relationship. Topological data analysis (TDA) has gained much attention in the past decade. The traditional atoms in molecules (AIM)  analysis  \cite{Bader:1985,Bader:1990} can be viewed as an application of TDA to electron density analysis. Historically, the topological study of scalar fields, particularly electron densities, has advanced the understanding of molecules and their chemical and biological functions tremendously. The theory of AIM developed by Bader and his coworkers has provided an elegant and feasible approach to define atoms in molecule. AIM can be generalized into a more general theory called quantum chemical topology (QCT), which applies the topological analysis in AIM to the study of other physically meaningful scalar fields  \cite{Popelier:2005}. Mathematically, this topological analysis used in AIM is known as Morse theory. In general, all scalar fields used in QCT are some kind of Morse functions and more can be proposed as long as they satisfy the Mores function constraints. On the other hand, geometric modelings and analysis contribute  a lot to the molecular visualization and structure-function relationship. Various surface definitions from geometric modelings  facilitate the visualization and characterization of molecules. Geometric analysis has contributed enormously to the understanding of structure-function relationship. Among the geometric analysis, surface curvature analysis has provide great insights into solvation analysis, protein-protein interaction, drug design, etc. \cite{KLXia:2014a,DDNguyen:2016c}

In the QCT theory, researchers explore atomic or molecular properties through the analysis of various properties at critical points. In geometric analysis, curvature estimation is normally done on a special surface, such as SES, SAS, Gaussian surface, FRI rigidity surface,  etc. Even though these two approaches are very efficient and powerful in capturing and characterizing atomic and molecular information, enormous information embedded in the electron density scalar field has not been fully utilized. Question is how to improve geometric and topological analysis for biomolecules. Note that in persistent homology analysis,  physical properties are analyzed by a systematic  filtration process. Therefore,  for a given scalar field, a better understanding can be achieved if the topological analysis includes not only some critical points or a special iso-surface, but also a series of iso-surfaces derived from a systematic evolution of iso-values. Further, more information can be extracted if one considers not only simply scalar fields, such as electron density, electron density Laplacian, etc, but also geometric properties like individual eigenvalues and various curvatures. These approaches are developed in  the present work.

In the following, a brief review of some very basic concepts in AIM is given.  Their connection with mathematical theories is discussed.  Then,  some new approaches are presented. Examples of applications are provided.

\subsubsection{Critical points and their classification} \label{sec:CP}

\begin{figure}
\begin{center}
\begin{tabular}{c}
\includegraphics[width=0.6\textwidth]{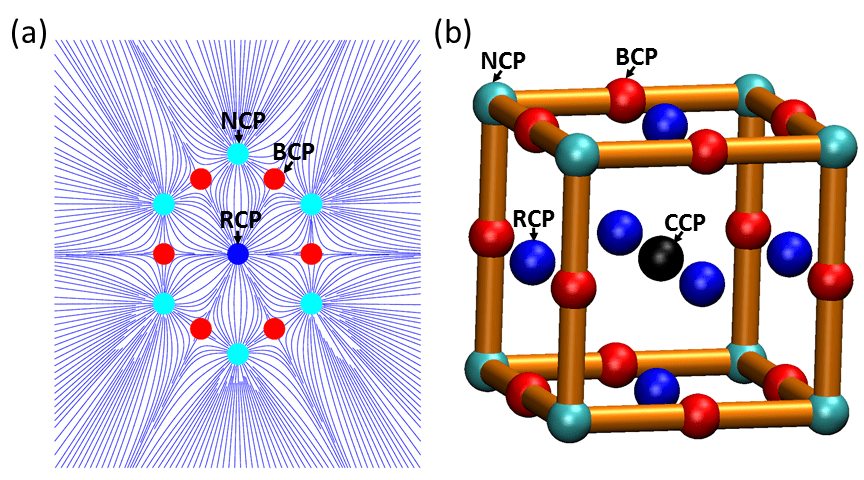}
\end{tabular}
\end{center}
\caption{An illustration of four types of critical points, i.e., nucleic critical point (NCP), bond critical point (BCP), ring critical point (RCP) and cage critical point (CCP).
{\bf (a)} The demonstration of flowlines and CPs for a benzene molecule (Hydrogen atoms are not considered for simplicity).
{\bf (a)} The CPs for a Cubane (Hydrogen atoms are not considered for simplicity).
}
\label{fig:cp}
\end{figure}

Three dimensional molecular electron density data can be collected through experimental tools, like electron microscopy, cryo-electron microscope, etc and also from theoretical models in quantum mechanics. One can denote electron density function $\rho$ and its domain  $\Omega$. A quantitative method to analyze the topology of $\rho$ is to consider its first derivative, i.e., gradient $\nabla \rho$. At certain points, called critical points (CPs), this gradient vanishes. The characteristics of these points is determined by the second derivatives, which form the Hessian of electron density $\rho$.

By diagonalization of  the Hessian   matrix, one can obtain three eigenvalues $\gamma_1 \leq \gamma_2 \leq \gamma_3$.
Their sum equals to the Laplacian of electron density $\rho$. That is, locally  \cite{Bader:1990}
\begin{eqnarray}
\nabla^2 \rho= \gamma_1+\gamma_2 + \gamma_3=\frac{\partial^2 \rho}{\partial^2 x} +\frac{\partial^2 \rho}{\partial^2 y} +\frac{\partial^2 \rho}{\partial^2 z}.
\end{eqnarray}
It can be noticed that the Hessian matrix is symmetric, therefore all the eigenvalues are real. Based on the positive and negative sign of these three eigenvalues, one defines rank and signature to characterize critical points. The rank of a CP is the number of non-zero eigenvalues, and a signature is the algebraic sum of the signs (+1 or -1) of the eigenvalues. In general, a CP is non-degenerate, meaning its three eigenvalues are non-zero (Rank$=3$). A degenerate critical point is unstable in the sense that even a small change in the function will cause it either to vanish or bifurcate into a non-degenerate CP.

Based on the rank and signature value, non-degenerate CPs in three dimensional scalar field can be classified into four basic types, namely,  nucleic critical point (NCP), bond critical point (BCP), ring critical point (RCP) and cage critical point (CCP). Table \ref{tb:cp} lists these values. An illustration of these four types of CPs can be found in Figure \ref{fig:cp}. An NCP is a nucleic center of an atom. It is represented by a cyan-color point. A BCP is a bond center between two atoms and is marked with a red-color point. A RCP is usually found at the center of a ring structure and is colored in blue. A CCP is known as a cage critical point, and can only be found in the center of a cage structure. In general, an NCP is a local maximal point. BCP and RCP both are saddle points. A CCP is a local minimal point. In Sections \ref{sec:eigenvalue} and  \ref{sec:T_C},  it will be demonstrated  that these properties can be used to systematically characterize and analyze the eigenvalue and curvature isosuface information.

Topologically, the general structure connectivity can be characterized by the number of CPs. This is stated in the famous Poincar\'e-Hopf theorem as following:
\begin{eqnarray} \label{Eq:ph}
N_n-N_b+N_r-N_c=\chi (\rho),
\end{eqnarray}
where $N_n$, $N_b$, $N_r$ and $N_c$ are numbers of NCPs, BCPs, RCPs and NCPs, respectively. Here $\chi(\rho)$ is the Euler characteristic. The Poincar\'e index is defined on a vector field. In the AIM theory, one studies the gradient vector field of electron density. Poincar\'e indices for NCP, BCP, RCP and NCP are $ 1, -1, 1$ and $-1$, respectively. More properties of this gradient vector field are  discussed in the following section.

Equation (\ref{Eq:ph}) can also be derived from simplicial complex analysis \cite{KLXia:2014c}. Essentially, NCP can be viewed as a point (0-simplex); BCP corresponds to a straight line (1-simplex); RCP is for a polygon (2-simplex); CCP is then for a polyhedron (3-simplex). In this manner, Euler characteristic can be directly employed and the above equation can be obtained as well.
\begin{table}
\begin{center}
\caption{The critical point can be classified into four basic types including: nucleic critical point (NCP), bond critical point (BCP), ring critical point (RCP) and cage critical point (CCP), as demonstrated in Fig.  \ref{fig:cp}. }
\begin{tabular}{|c|c|c|c|c|c|}
\hline
  &   Rank  &  Signature  & Poincar\'e ~ index  & Simplex  & Property \\
\hline
 NCP   &3  &$-3$  &1  & 0-simplex   & local   maxima   \\
\hline
 BCP  &3  &$-1$  &$-1$ & 1-simplex   &  saddle    \\
\hline
 RCP  &3  &1  &1  & 2-simplex     &  saddle  \\
\hline
 CCP   &3  &3 &$-1$ & 3-simplex     &  local   minima   \\
\hline
\end{tabular}
\label{tb:cp}
\end{center}
\end{table}

\subsubsection{Vector field topology }

The gradient $\nabla \rho$ on the entire domain $\Omega$ forms a vector field. The topological property of this vector field can be explored by tools borrowed from dynamic systems or the mathematical analysis of fluid flows. These tools include integral line, separatrix, and Poincar\'e index, etc.

An integral line is a curve $l(t)$ of a function $f({\bf r})$ that satisfies $\frac{\partial l}{\partial t}=\nabla f(l(t))$. Its origin and destination can be defined as
$${\rm org}(l)=\underset{t \rightarrow - \infty}{\lim} l(t)$$ and $${\rm dest}(l)=\underset{t \rightarrow \infty}{\lim} l(t).$$
Integral lines satisfy the following conditions:
 \begin{enumerate}
\item
Two integral lines are either disjoint or the same.
 \item
Integral lines cover all the manifold.
 \item
The limits, org$(l)$ and dest$(l)$, are critical points.
\end{enumerate}
In general, integral lines represent the gradient flow between critical points. All the integral lines that share the same $org$ form a region called atomic basin. The whole electron density domain is subdivided into many atomic basins. The interface between these attraction basins is called inter-atomic surface (IAS). Mathematical, IAS is the separatrix of a gradient vector field. Each attraction basin includes one and only one atom. This is known as the quantum topological definition of an atom in a molecule.

IAS also satisfies the zero-flux condition $\nabla f({\bf r}) \cdot {\bf n( r)}=0$. Here $\nabla f({\bf r}) $ is a gradient vector and ${\bf n(r)}$ is the normal vector to the IAS.

\paragraph{Morse theory}
It should be noticed that basic concepts like critical point, degeneracy, critical point classification, basin, etc. are the essential part of a mathematical tool called Morse theory \cite{harker:mischaikow:mrozek:nanda, Mischaikow:2013}. In general, the atom in molecular method can be viewed as an application of Morse theory in molecular density field analysis. More specifically, various critical points, including NCP, BCP, RCP, and CCP, are the counterparts of peak point, saddle-1 point, saddle-2 point and valley point, respectively. The separatrix of a gradient vector field just provides a Morse decomposition of the underlying molecular scale field manifold \cite{harker:mischaikow:mrozek:nanda, Mischaikow:2013}.

\subsubsection{Topological characterization of chemical bonds} \label{sec:chemical_bond}

Chemical properties of molecular systems are profoundly determined or influenced by their atomic covalent bonds and noncovalent interactions. Usually, covalent bonds are much stronger and determine the structural integrity of a molecule. Noncovalent interactions are comparably weak but play important roles in macromolecular assembly, protein folding, macromolecular function, etc. Traditionally,  the Laplacian of electron density can be used to interpret noncovalent interactions of a molecular system. Recently,  signed electron density    and  reduced gradient, two scalar fields derived from  electron density, have drawn much attention in quantum chemistry since they enable a qualitative visualization of noncovalent interactions even in complex molecular systems \cite{Johnson:2010,contreras2011analysis,contreras2011nciplot, Gillet:2012,Gunther:2014}. These approaches are reviewed below.

\paragraph{The Laplacian of electron density}
The Laplacian of  electron density $\rho ({\bf r})$ can be used to indicate the electron density concentration and depletion.  Essentially,  the density is locally concentrated where $\nabla^2 \rho({\bf r})<0$, and locally depleted where $\nabla^2 \rho({\bf r})>0$. In this manner, if one defines the function $L({\bf r})=-\nabla^2 \rho({\bf r})$, the maximum in $L({\bf r})$ denotes the maximum in the concentration of the density. The minimum  in $L({\bf r})$ implies a depletion of density. The Laplacian of the electronic charge distribution, $L({\bf r})$, demonstrates the presence of local concentrations of charge in the valence shell of an atom in a molecule. These local maxima faithfully duplicate in number, location, and size of the spatially localized electron pairs of the valence shell electron pair repulsion (VSEPR) model. Thus the Laplacian of the charge density provides a physical basis for the Lewis and VSEPR models \cite{Bader:1988}.


\paragraph{Identifying noncovalent interactions}
The study of the Hessian matrix and its three eigenvalues has yielded many intriguing results \cite{Bader:1990}. It has been found that all eigenvalues ($\gamma_1({\bf r})$, $\gamma_2({\bf r})$ and $\gamma_3({\bf r})$) are negative in the vicinity of the nuclei centers. Away from these centers, the largest eigenvalue $\gamma_3({\bf r})$ becomes positive, and varies along the internuclear axis representing covalent bonds. It is also found that $\gamma_1({\bf r})$ and $\gamma_2({\bf r})$ describe the density variation orthogonal to this internuclear axis. More specially, $\gamma_1({\bf r})$ is always negative even it is away from the nuclei. While $\gamma_2({\bf r})$ can be either positive, meaning attractive interactions concentrating electron charge perpendicular to the bond, or negative, meaning repulsive interactions causing density depletion. Using this localized information, the signed electron density $\widetilde{\rho}({\bf r})$ is defined as
\begin{eqnarray} \label{Eq:sed}
 \widetilde{\rho} ({\bf r})= {\rm sign}(\gamma_2({\bf r}))\rho({\bf r}).
\end{eqnarray}
The signed electron density additionally enables the differentiation of attracting and repulsive interactions.

To further reveal weak noncovalent interactions, the reduced gradient is introduced as following \cite{Gillet:2012,Gunther:2014}
\begin{eqnarray} \label{Eq:reduced_gradient}
s({\bf r})=\frac{1}{2(3\pi^2)^{\frac{1}{3}}}\frac{|\nabla \rho({\bf r})|}{\rho({\bf r})^\frac{4}{3}}.
\end{eqnarray}
The reduced density gradient describes the deviation in atomic densities due to interactions and has found interesting applications in in analyzing biomolecular structure and function \cite{Johnson:2010,Gillet:2012,Gunther:2014}.

\subsection{Geometric-topological (Geo-Topo) fingerprints of scalar fields}\label{sec:GTF}

\subsubsection{Geo-Topo fingerprints of Hessian matrix eigenvalue maps} \label{sec:eigenvalue} %

The QCT and AIM analyses of molecules are limited to special locations and given iso-surfaces. In this work, the Hessian eigenvalue analysis of a molecular density is considered, not only for a few critical points, but also for the whole domain. Additionally, the topology over a series of isosurfaces derived from a systematic evolution of isovalues are analyzed. More specifically, in the topological persistence of Hessian matrix eigenvalues, the topological properties of a series of isosurfaces, generated by varying the isovalue from the smallest to the largest, are systematically studied. Therefore, the isosurface value behaves like a filtration parameter.  The  topological transitions in this series of isosurfaces are emphasized.   Another very important aspect of this analysis is to analyze the behavior of isosurfaces through its relation with the four types of CPs. It should also be noticed that these methods are greatly different from other interatomic surface methods \cite{Pendas:2003,Popelier:1996}, as these models  focus on a special surface.



\begin{figure}
\begin{center}
\begin{tabular}{c}
\includegraphics[width=0.8\textwidth]{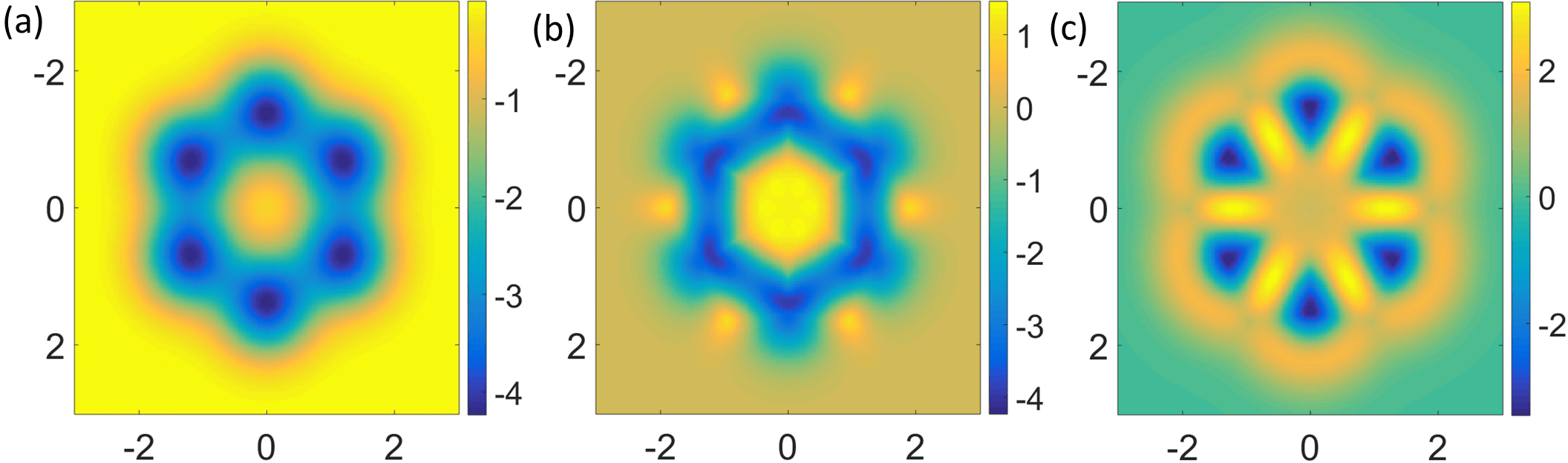}
\end{tabular}
\end{center}
\caption{ Hessian matrix eigenvalue maps of the  hexagonal ring model  at   cross section  $Z=0$.
The behaviors of three eigenvalues, i.e., $\gamma_1$, $\gamma_2$ and $\gamma_3$, are illustrated in { (a)}, { (b)}  and { (c)}, respectively.
}
\label{fig:C6_eigen_2d}
\end{figure}

\begin{figure}
\begin{center}
\begin{tabular}{c}
\includegraphics[width=0.6\textwidth]{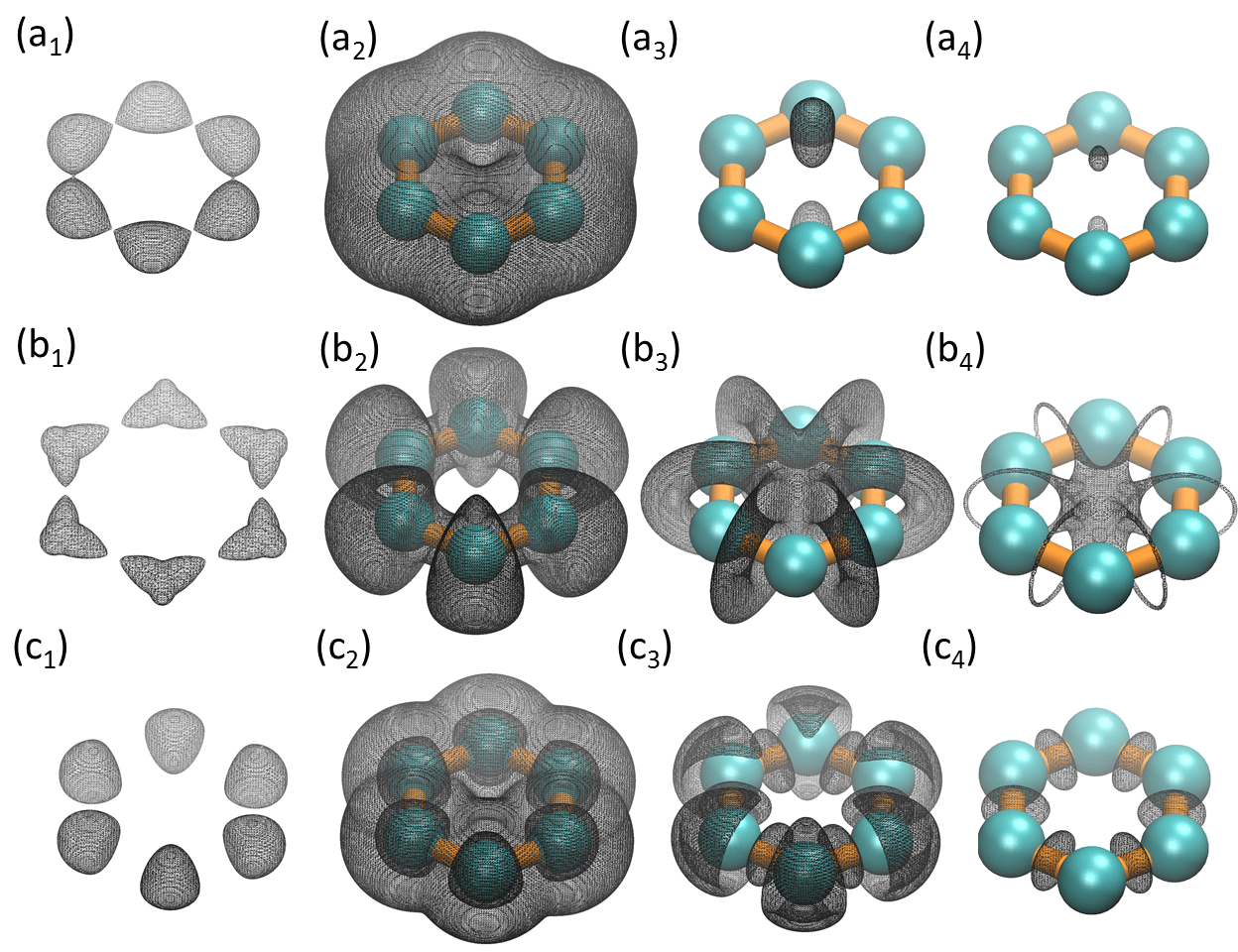}
\end{tabular}
\end{center}
\caption{ Eigenvalue maps obtained from different isovalues (or level-set values)  for a hexagonal ring.
{ (a)} The isosurfaces for the first eigenvalue. The isovalues from { ($a_1$)} to { ($a_4$)} are -3.0, -0.1, 0.0 and 0.1.
{ (b)} The isosurfaces for the second eigenvalue. The isovalues from { ($b_1$)} to { ($b_4$)} are -3.0, -0.1, 0.1 and 1.0.
{ (c)} The isosurfaces for the third eigenvalue. The isovalues from { ($c_1$)} to { ($c_4$)} are -0.1, 1.1, 1.8 and 2.0.
}
\label{fig:C6_eigen_3d}
%
\begin{center}
\begin{tabular}{c}
\includegraphics[width=0.6\textwidth]{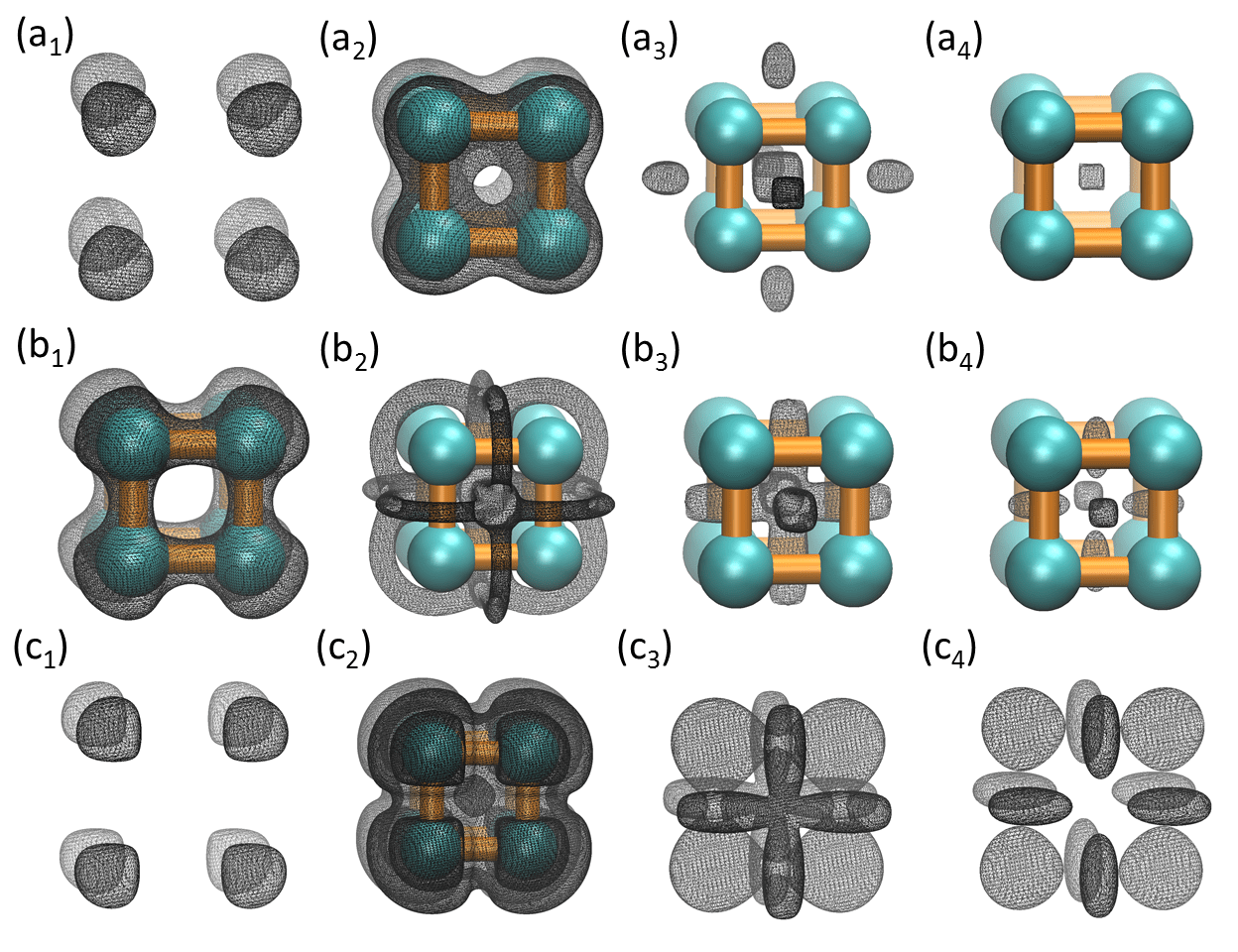}
\end{tabular}
\end{center}
\caption{ Eigenvalue maps obtained from different isovalues (or level-set values) for a cubic structure.
{ (a)} The isosurfaces for the first eigenvalue. The isovalues from { ($a_1$)} to { ($a_4$)} are -3.0, -1.5, 0.1 and 0.9.
{ (b)} The isosurfaces for the second eigenvalue. The isovalues from { ($b_1$)} to { ($b_4$)} are -1.0, 0.5, 1.0 and 1.5.
{ (c)} The isosurfaces for the third eigenvalue. The isovalues from { ($c_1$)} to { ($c_4$)} are -1.0, 1.5, 2.0 and 2.5.
}
\label{fig:Cubane_eigen_3d}
\end{figure}

To illustrate the idea,  two    toy models, i.e., the hexagonal ring discussed earlier and a cubic structure are considered.   The coordinates for the cubic structure are set to (1.245,   0.537,  -0.073; 0.924,  -0.995,   0.024; -0.123,  -0.704,   1.155; 0.199,   0.828,   1.058; 0.123,   0.704,  -1.155; -0.924,   0.995,  -0.024; -1.245,  -0.537,   0.073; and  -0.199,  -0.828,  -1.058). The discrete to continuum mapping,  Eq.  (\ref{eq:rigidity3}), is used to generate  rigidity density.
  The parameters $\sigma$  and $w_i$  are chosen as $0.7$ \AA~ and  $1$ for all particles.  In this approach, Hessian matrix is evaluated at each point of the computational domain and its eigenvalue is obtained everywhere as well, which forms an eigenvalue scalar field.

To have a general idea of the basic distribution of eigenvalues, one can study the Hessian matrix eigenvalue behavior of the hexagonal ring  in a two-dimensional plane $Z=0$, as all its particles are located within this special plane. Results are illustrated in Fig. \ref{fig:C6_eigen_2d}. Three eigenvalues, i.e., $\gamma_1$, $\gamma_2$ and $\gamma_3$, are demonstrated in subfigure {\bf (a)}, {\bf (b)}  and {\bf (c)}, respectively. It can be seen that for regions near NCPs, all three eigenvalues are negative. For regions near BCPs, $\gamma_1$ is always negative. While $\gamma_2$ is negative in the very closed neighborhood and gradual increases to be positive further away. Finally, $\gamma_3$ is always positive. For regions near RCPs, all three eigenvalues are positive. These results are consistent with findings in AIM \cite{Bader:1990}.

To obtain more geometric insights, particularly eigenvalue behaviors on all four types of CPs,  the cubic structure is considered.
In this case, the CCP can be identified as having the positive    $\gamma_1$ and $\gamma_2$.

As stated above, this approach is to extract a series of isosufaces of Hessian matrix through a filtration process. One can carefully observe these eigenvalue isosurface patterns to detect topological transitions. The geometric information is further analyzed and its relation with the four types of CPs is summarized into several unique characteristics, called Geo-Topo fingerprints. Using  hexagonal ring and  cubic structure models, the Geo-Topo fingerprints for eigenvalues are revealed.

Figures \ref{fig:C6_eigen_3d} and \ref{fig:Cubane_eigen_3d} demonstrate four unique patters for each eigenvalue. The subscripts $1$ to $4$ indicate four representative eigenvalue isovalues, from the small to the large. The notations $({\bf a})$, $({\bf b})$ and  $({\bf c})$ represent eigenvalue $\gamma_1$,  $\gamma_2$  and $\gamma_3$, respectively. As stated above, the filtration process   delivers a series of isosurfaces. To avoid confusion, the filtration process always goes from the smallest value to the largest one. Only four representative isosurfaces that capture some unique topological features within the eigenvalue isosurface  series, are selected. Through the comparison of these patterns, some common features can be extracted.
\begin{enumerate}
\item
For all the three eigenvalues, regions around NCPs are always concentrated with negative values.
\item
For $\gamma_1$, as the filtration goes, negative isosurfaces first appears near NCPs. Then they enlarge to incorporate regions near BCPs, being still negative. The regions near RCPs can be analyzed from two different perspectives: i.e., along the ring plane and  perpendicular to the ring plane. Along the ring plane, $\gamma_1$ values gradually increase to about 0 as the eigenvalue isosurface propagate to RCPs. There is a sudden topological transition for the isosuface when its value passes through 0. When becoming positive, eigenvalue isosurfaces form ellipsoids surfaces perpendicular to the ring plane near all RCPs. These ellipsoids usually appear in pairs and symmetric to each other along the ring plane as indicated in Fig. \ref{fig:C6_eigen_3d} $({\bf a})$. Finally, positive isosurface appears near the regions of CCPs.
\item
For $\gamma_2$, as   the filtration process goes, negative isosurfaces first appear near NCPs and then enlarge to incorporate regions near BCPs just like $\gamma_1$. However, positive isosurfaces appear much earlier. They occupy the regions around RCPs and CCPs. More interestingly, they  form a loop around each bond near its BCP. To be more precise, these loops are perpendicular to atom bonds at BCPs and atom bonds pass through them at their centers. The isosurfaces gradually shrink and reduce to regions around RCPs as their values grows.
\item
For $\gamma_3$, its negative isosurfaces concentrate only in regions around NCPs. Positive isosurfaces form a sphere slice around each atom. Isosurface with relative small positive value also appears around the RCPs and CCPs. As the filtration goes further, positive isosurfaces concentrate around regions near BCPs.
\end{enumerate}

\subsubsection{Geo-Topo fingerprints  of curvature maps }\label{sec:T_C}

In this work, the topological analysis of curvature maps are developed. One can still consider the  hexagonal ring and the cubic structure discussed in the last section. The discrete to continuum mapping is carried out to generate the FRI rigidity density. Then curvatures are evaluated at all of the molecular FRI rigidity isosurfaces (or every point in the computational domain) to form a curvature map. At each curvature isovalue, topological analysis is applied.  Here, topological analysis is twofold. One type of topological analysis is to identify topological critical points (i.e., 0-simplex, 1-simplex, 2-simplex, etc.). The other type is to carry out persistent homology analysis to track   topological invariants during the density filtration of the curvature map.

\paragraph{Gaussian and mean curvature  maps}\label{sec:K_H}

\begin{figure}
\begin{center}
\begin{tabular}{c}
\includegraphics[width=0.6\textwidth]{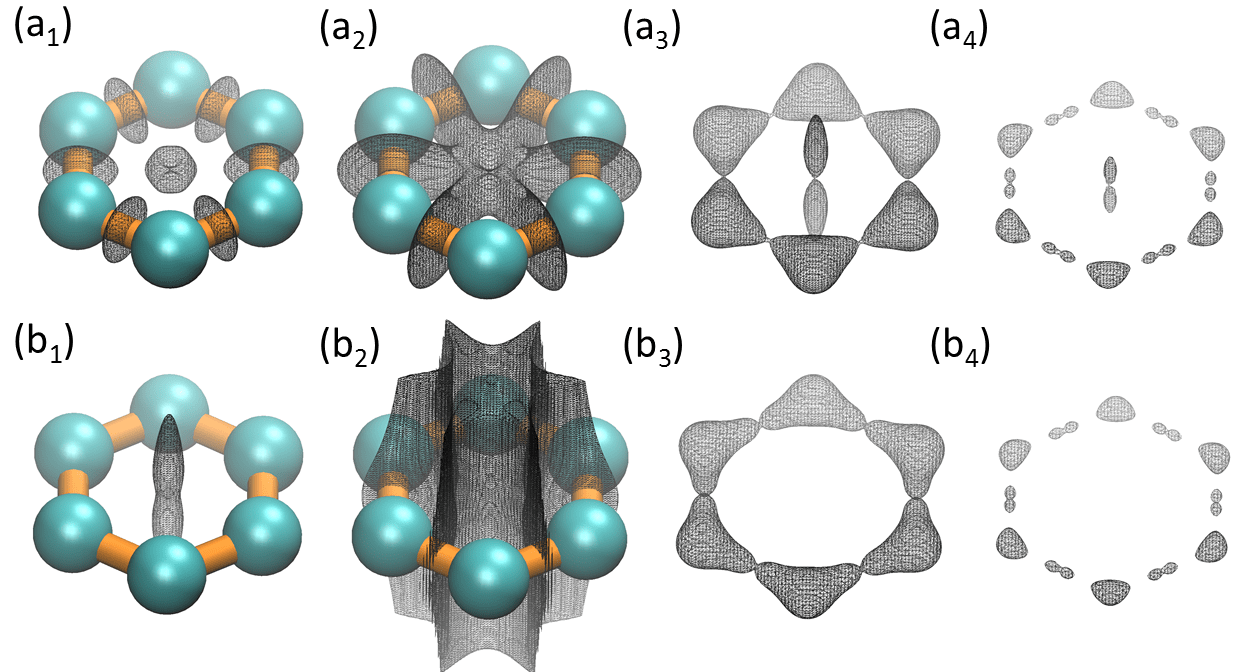}
\end{tabular}
\end{center}
\caption{ Isosurfaces for Gaussian and Mean curvature maps for a hexagonal ring.
{ (a)} The isosurfaces built from same Gaussian curvature. The isovalues from { ($a_{1}$)} to { ($a_{4}$)} are -5.0, -2.0, 5.0 and 30.0.
{ (b)} The isosurfaces built from same Mean curvature. The isovalues from { ($b_{1}$)} to { ($b_{4}$)} are -0.2, 0.001, 3.0 and 6.0.
}
\label{fig:C6_gaussian_mean_3d}
%
\begin{center}
\begin{tabular}{c}
\includegraphics[width=0.6\textwidth]{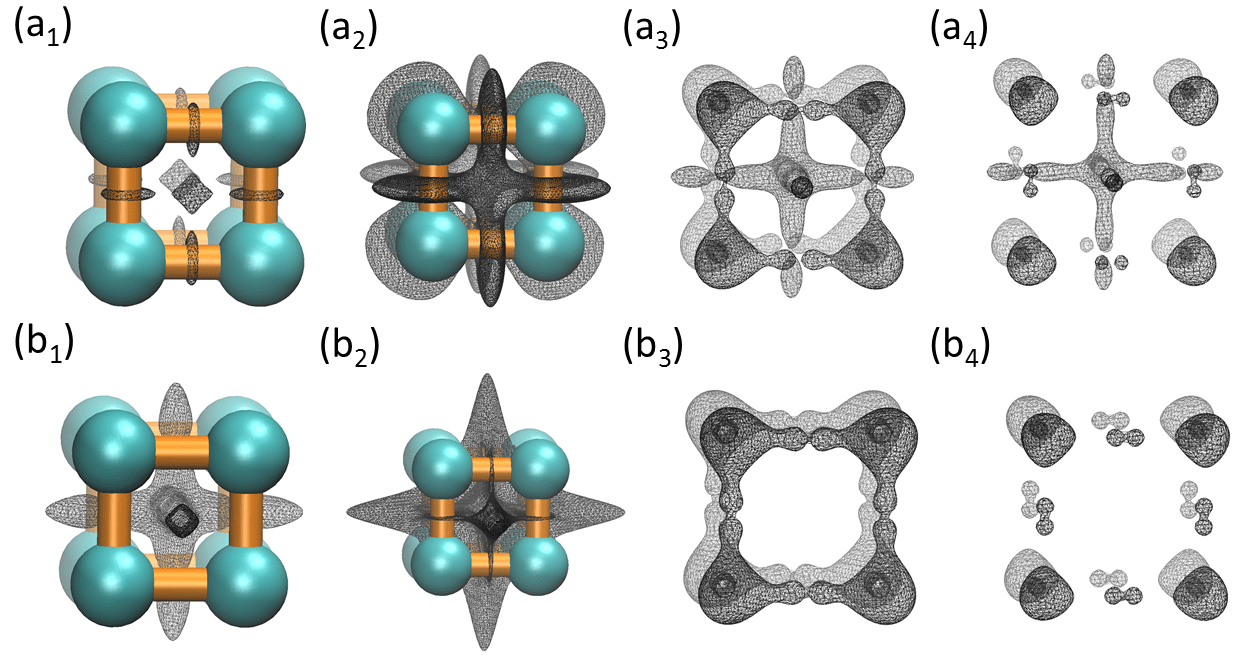}
\end{tabular}
\end{center}
\caption{Isosurfaces of Gaussian and Mean curvature maps for a cubic structure.
{ (a)} The isosurfaces built from same Gaussian curvature. The isovalues from { ($a_{1}$)} to { ($a_{4}$)} are -20.0, -2.0, 10.0 and 20.0.
{ (b)} The isosurfaces built from same Mean curvature. The isovalues from { ($b_{1}$)} to { ($b_{4}$)} are -2.0, -1.0, 3.0 and 4.0.
}
\label{fig:Cubane_gaussian_mean_3d}
\end{figure}

It is  seen from the above analysis that with four types of CPs,  Geo-Topo fingerprints are extracted and the behaviors of the Hessian matrix eigenvalue isosurfaces can been characterized very well. However, before the employment of this technique in Gaussian and mean curvature isosurface analysis, it is  helpful to review a little bit more  the four types of CPs. As stated in Section \ref{sec:CP}, NCPs and CCPs are locally maximal and locally minimal, respectively. Actually, for NCPs, all three eigenvalues have negative signs, while for CCPs, all three eigenvalues have positive signs.  BCPs and RCPs are saddle points, which means that their eigenvalues have both positive and negative values. Geometrically, the curvature is isotropic near NCPs and CCPs, but anisotropic near BCPs and RCPs. So in this analysis, the isosurfaces near  BCPs and RCPs can be divided into two types, namely, A-type or V-type. A-type isosurface means it is along atomic bonds or ring planes. V-type isosurface means it is vertical or perpendicular to the atomic bonds or ring planes. For instance, the isosurface near the RCP in Fig. \ref{fig:C6_gaussian_mean_3d} {\bf ($a_1$)} is an A-type, but it becomes a V-type in Fig. \ref{fig:C6_gaussian_mean_3d} {\bf ($a_3$)}. Another important property is that A-type and V-type isosurfaces are usually with different signs. That is, if A-type of isosurface is obtained from a positive isovlaue,  the associated V-type isosurface can only be  obtained from a negative isosurface. With this notation, one can extract some Geo-Topo fingerprints for Gaussian and mean curvatures. The basic results are summarized as below.

\begin{enumerate}
\item
For a Gaussian curvature field, it has negative A-type isosurfaces near RCPs and negative V-type isosurfaces near BCPs. In contrast, positive isosurfaces enclose regions near BCPs and CCPs. Positive V-type isosurfaces can be found in RCPs and positive A-type isosurfaces are found near BCPs. Finally, positive isosurfaces are found in NCPs.

\item
For a mean curvature field, it has negative isosurfaces near RCPs and CCPs. Negative V-type isosurfaces can be found near BCPs. Positive isosurfaces are found in NCPs. Finally, A-type positive isosurfaces can be found near BCPs.
\end{enumerate}

\paragraph{Maximal and minimal curvature maps} \label{sec:k1k2}

\begin{figure}
\begin{center}
\begin{tabular}{c}
\includegraphics[width=0.6\textwidth]{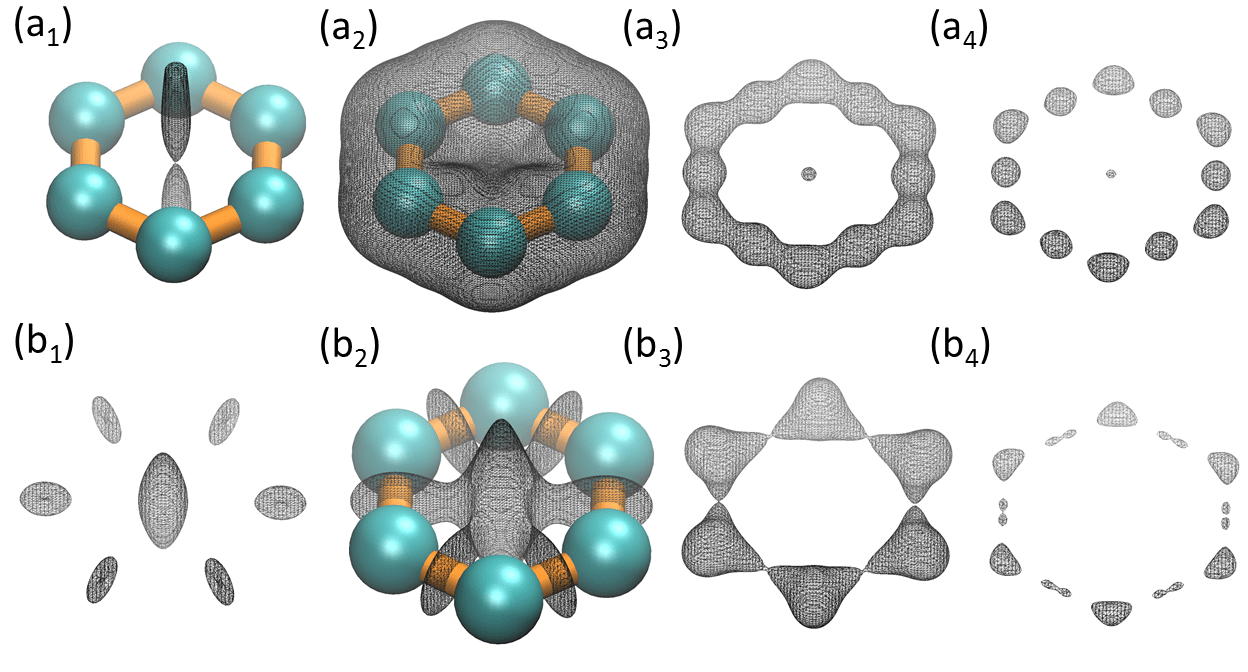}
\end{tabular}
\end{center}
\caption{Isosurfaces for maxmal and minimal curvature maps for a hexagonal ring.
{ (a)} The isosurfaces built from same maximal curvature. The isovalues from { ($b_{1}$)} to { ($b_{4}$)} are -1.0, 1.0, 4.0 and 6.0.
{ (b)} The isosurfaces built from same minimal curvature. The isovalues from { ($a_{1}$)} to { ($a_{4}$)} are -4.0, -2.0, 2.0 and 6.0.
}
\label{fig:C6_k1k2_3d}
%
\begin{center}
\begin{tabular}{c}
\includegraphics[width=0.6\textwidth]{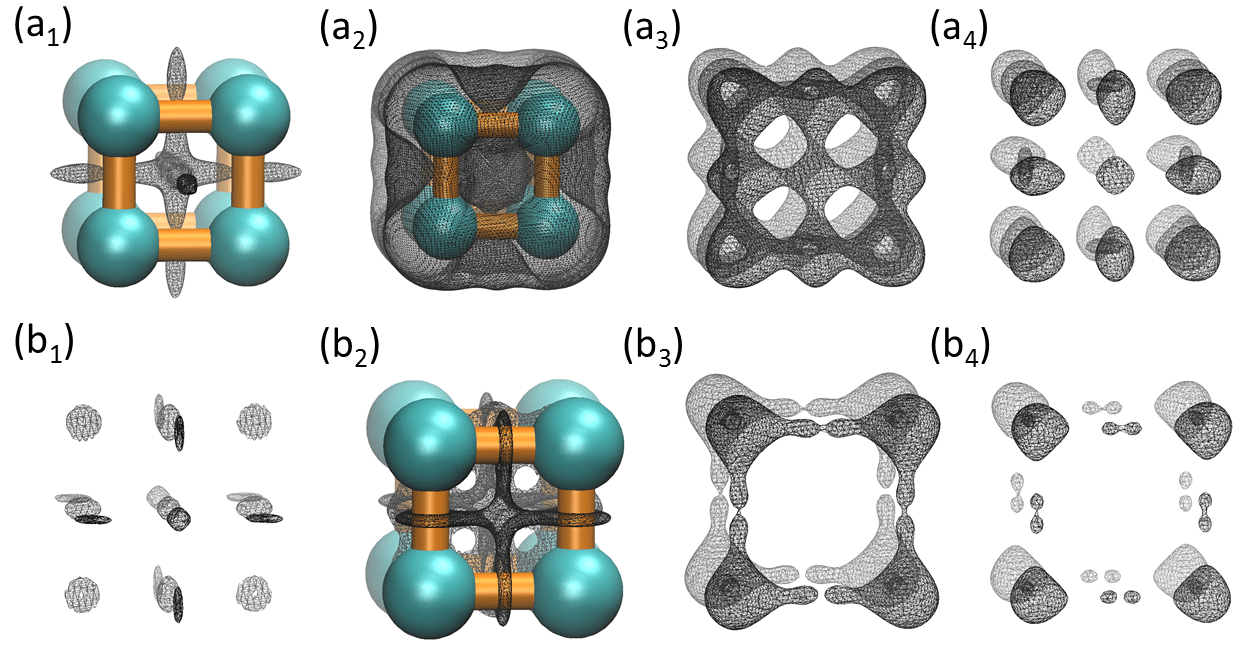}
\end{tabular}
\end{center}
\caption{Isosurfaces for maximal and minimal curvature maps  for a cubic structure.
{\bf (a)} The isosurfaces built from same maximal curvature. The isovalues from {\bf ($b_{1}$)} to {\bf ($b_{4}$)} are -2.0, 1.0, 3.0 and 4.0.
{\bf (b)} The isosurfaces built from same minimal curvature. The isovalues from {\bf ($a_{1}$)} to {\bf ($a_{4}$)} are -10.0, -4.0, 3.0 and 4.0.
}
\label{fig:Cubane_k1k2_3d}
\end{figure}

One can also carry out a throughout investigation of maximal and minimal curvature maps. Isosurfaces near NCPs, BCPs, RCPs and CCPs are analyzed in the same manner as they were done in the above two sections. Figures \ref{fig:C6_k1k2_3d} and \ref{fig:Cubane_k1k2_3d} illustrate results for the hexagonal ring and the cubic structure, respectively. Main results about their Geo-Topo fingerprints are summarized below:

\begin{enumerate}
\item
For maximal curvature map, it has negative isosurfaces near CCPs and negative V-type isosurfaces near RCPs. Positive isosurfaces enclose regions near BCPs and NCPs. Positive A-type isosurfaces can be found in RCPs.

\item
For minimal curvature  map, it has negative isosurfaces near RCPs and negative V-type isosurfaces near BCPs. Negative isosurfaces also enclose region near CCPs. Positive isosurfaces are found in NCPs. Finally, A-type positive isosurfaces can be found near BCPs.
\end{enumerate}

It can be noticed that   all curvature representations, i.e., Gaussian curvature, mean curvature and two principal curvatures, are quite different from Hessian matrix eigenvalue distributions. In general, negative curvatures  usually do  not occur near NCPs, this is particularly true for maximal and mean curvatures. They are more likely to appear near  BCPs, RCPs and CCPs. In contrast, positive isosurfaces are concentrated in the atomic basin (i.e., NCPs).

\begin{table}
\begin{center}
\caption{Geo-Topo fingerprints for eigenvalue and curvature maps. For simplicity, notations ``P'' and ``N'' mean positive and negative, respectively. Loop means the ring structure around the bond (see Section \ref{sec:eigenvalue} for detail). A-type and V-type means the isosurface is {\it along with} or {\it vertical to} atomic bonds or ring planes. Four types of critical points (CPs) are nucleic critical point (NCP), bond critical point (BCP), ring critical point (RCP) and cage critical point (CCP). Here $\gamma_1$, $\gamma_2$ and $\gamma_3$ are three Hessian matrix eigenvalues. Gaussian and mean curvatures are represented by $K$ and $H$, respectively. The two principal curvatures are maximal curvature $\kappa_1$ and minimal curvature $\kappa_2$.}
\begin{tabular}{|c||c|c|c|c|}
\hline
 &   NCP  &  BCP  &  RCP  &  CCP \\
\hline \hline
$\gamma_1$  &N  & N & P  & P\\
\hline
$\gamma_2$ & N  &N; P-Loop  &P & P\\
\hline
$\gamma_3$ & N  &P  & P  &P \\
\hline
$K$        &P  &A-type (P); V-type (N)  & V-type (P); A-type (N)  & P  \\
\hline
$H$        &P  &A-type (P); V-type(N)  &N    &N \\
\hline
$\kappa_1$  &P  &P  &A-type (P); V-type (N)    &N\\
\hline
$\kappa_2$ &P  &A-type (P); V-type (N) &N    &N \\
\hline
\end{tabular}
\label{tb:geometric_fingerprint}
\end{center}
\end{table}

To have a more general understanding of the properties of eigenvalue and curvature maps,  all of the above-mentioned results are summarized in Table \ref{tb:geometric_fingerprint}. To show the consistency of results, one can take some simply tests. For instance, one can compare the results of $\kappa_1$ and $\kappa_2$ with $K$. Since $K=\kappa_1 \kappa_2$, their signs of isosurface at four types of CPs should match with each other. For NCP, $\kappa_1$ and $\kappa_2$ are both positive, and multiplying together yields a positive $K$, exactly as it is found in the table. For all other three types of CPs, they all match very well. 
%
%
%

\subsection{Eigenvector field  analysis} \label{sec:Tensor field}

\begin{figure}
\begin{center}
\begin{tabular}{c}
\includegraphics[width=0.8\textwidth]{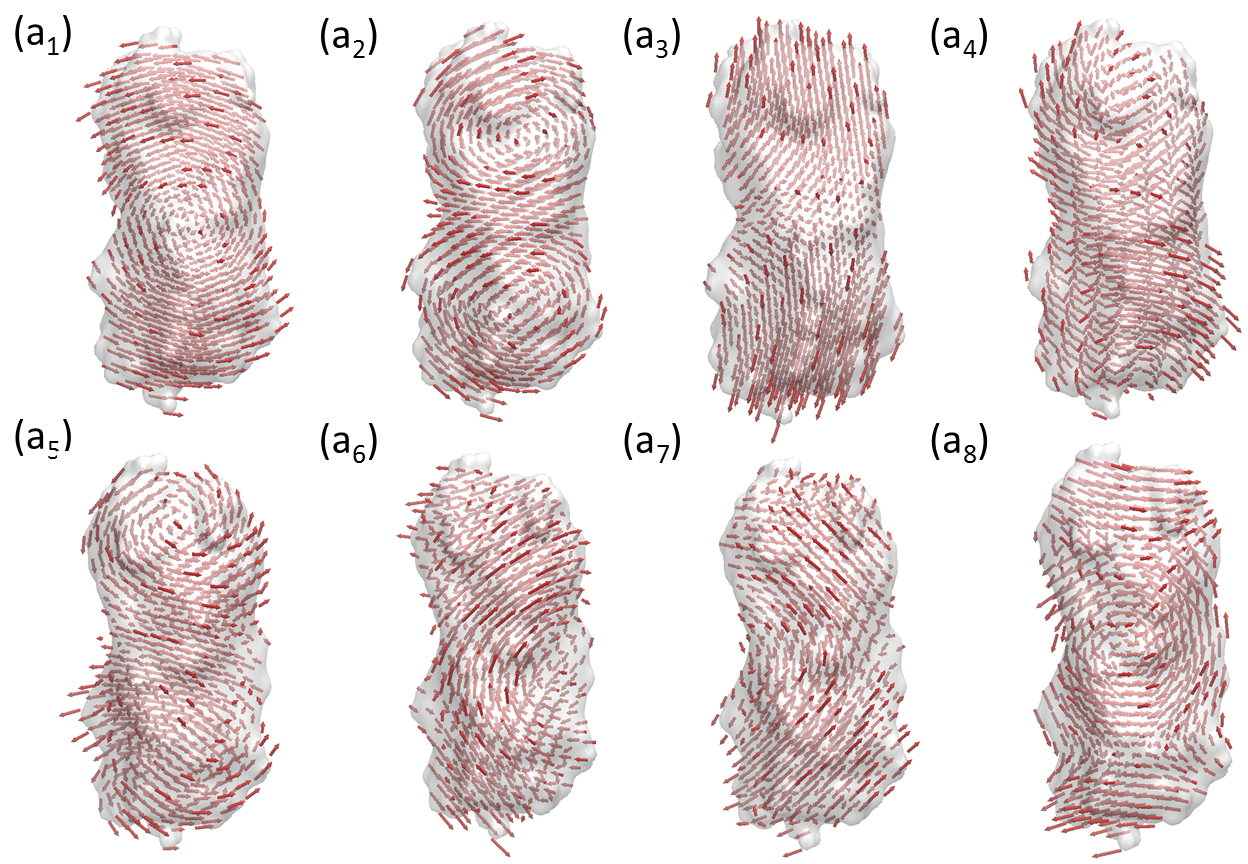}
\end{tabular}
\end{center}
\caption{The 4th to 11th eigenmodes of protein 2XHF. The eigenmodes are evaluated based on the rigidity density of 2XHF. A threshold value as 60\% of the Maximum density is chosen.
}
\label{fig:Density_2XHF}
\end{figure}
Tensor fields widely exist in natural world. For biomolecules, researchers are particular interested in their motions. As stated in the previous sections, various methods, including molecular dynamics, anisotropic network model (ANM), anisotropic FRI (aFRI), etc, have been developed to explore the dynamics of the biomolecular systems. However, all the above mentioned approaches rely on the discrete representation and can not be used in the study of motions of continuous biomolecular density function or data, particularly electron density data, such as cryo-electron microscopy (cryo-EM) maps. To analyze the motion of density profiles, a virtual particle model (VPM) is proposed to systematically explore the anisotropic motions of continuous density.


\subsubsection{Virtual particle model}

Both ANM and aFRI depend on their graph or discrete representation of biomolecules. It is not obvious how to construct a continuum model for a continuous density function to characterize its anisotropic motions.
Previously, vector quantization (VQ) algorithm \cite{Gray:1984vector} is employed to decompose the electron density map of a biological molecule into a set of finite Voronoi cells. It is then combined with ANM to explore the dynamic of the cryo-EM data \cite{Tama:2002exploring,Ming:2002describe}.

In this section,  virtual particles are introduced and defined for each small volume or element of a density data. To be more specific, the domain of the density function can be discretized into many elements. In general, the discretization can be non-uniformed and the elements may have different sizes. One can associate each element with a virtual particle, which is centered at the element center but having a continuous density profile. One can assume that all virtual particles are correlated with each other, but the correlations between them decrease with the distance or follow  prescribed relations.  The anisotropic motions of  virtual particles are obtained. Similar to ANM and aFRI approaches, such anisotropic motions are evaluated from the eigenmodes of the anisotropic Kirchhoff matrix.

One can assume that the density function of interest is given by  $\mu({\bf r})$, which can be either a cryo-EM density map or a rigidity density  computed from atomic coordinates by using the discrete to continuum mapping as shown in Section \ref{Sec:GeometricRep}. One can consider  two virtual particles centered at ${\bf r}_I$ and ${\bf r}_J$ and enclosed by the volume elements of $\Omega_I$ and $\Omega_J$, respectively. A special scaling parameter $\gamma({\bf r}_I,{\bf r}_J,\Omega_1,\Omega_2,\eta_{IJ})$ is proposed as following:
\begin{eqnarray}\label{rigidity_potential8}
 \gamma({\bf r}_I,{\bf r}_J,\Omega_I,\Omega_J,\eta_{IJ})=\left(1+a\int_{\Omega_I}\mu({\bf r}) d{\bf r}\right)\left(1+a\int_{\Omega_J}\mu({\bf r}) d{\bf r}\right)
\Phi(|{\bf r}_I-{\bf r}_J|, \eta_{IJ}),
\end{eqnarray}
where the coefficient $a$ is a normalization factor, $\Phi(|{\bf r}_I-{\bf r}_J|,\eta_{IJ})$ is a FRI correlation function and $ \eta_{IJ}$ is the characteristic length of elements. In contrast, the $\eta_j$ in the discrete to continuum mapping is the characteristic length of atomic distances.


Three realizations of VPM by constructing three anisotropic Kirchhoff matrices are proposed. First, one can modify ANM to construct a VPM anisotropic Kirchhoff matrix. For each matrix element $H_{IJ}$,  a local $3\times3$ Hessian matrix for ANM in Eq. (\ref{eq:multi-kirchoff1}) is formed as
\begin{eqnarray}\label{eq:multi-kirchoff128}
 H_{IJ} = -\frac{\gamma({\bf r}_I,{\bf r}_J,\Omega_I,\Omega_J,\eta_{IJ})}{r^2_{IJ}}\left[ \begin{array}{ccc}
	        (x_J-x_I)(x_J-x_I) &(x_J-x_I)(y_J-y_I) &(x_J-x_I)(z_J-z_I)\\
            (y_J-y_I)(x_J-x_I) &(y_J-y_I)(y_J-y_I) &(y_J-y_I)(z_J-z_I)\\
            (z_J-z_I)(x_J-x_I) &(z_J-z_I)(y_J-y_I) &(z_J-z_I)(z_J-z_I)
	      \end{array}\right]  ~ \forall ~ I \neq J.
 \end{eqnarray}
The diagonal elements are constructed following Eq. (\ref{eq:multi-kirchoff1_diagonal}). The test indicates this ANM based VPM works very well.
However,   the demonstration of this test is skipped.

Additionally, one can modify the aFRI to generate two other realizations of VPM.   It is very natural for one to derive the continuous aFRI by   making use of the local anisotropic matrix $\Phi^{IJ}$ defined in Eqs. (\ref{eq:Anisorigidity1}) and (\ref{eq:afri_local_Hessian}).
However, in general applications, the correlation $\Phi(\|{\bf r}_I-{\bf r}_J \|; \eta_{IJ})$ can be more specifically defined to describe the interaction  between each pair of virtual particles. To construct the final matrix, one can multiply the scaling parameter to the local flexibility Hessian matrix ${\bf F}^{1}$  and  ${\bf F}^{2}$, the corresponding two generalized Hessian matrices can be expressed as following:
\begin{eqnarray}\label{eq:Anisoflexibility8}
	{\bf F}^{1}_{IJ}     =&  - \frac{1}{w_{J}} {\rm adj}(\Phi^{IJ}) \gamma({\bf r}_I,{\bf r}_J,\Omega_I,\Omega_J,\eta_{IJ}),         &\quad  I\neq J;   \\ \label{eq:Anisoflexibilityy4}
	{\bf F}^{1}_{II}=&   \sum_{J=1}^N \frac{1}{w_{J}} {\rm adj}(\Phi^{IJ}) \gamma({\bf r}_I,{\bf r}_J,\Omega_I,\Omega_J,\eta_{IJ}),  &\quad   I=J
\end{eqnarray}
and
\begin{eqnarray}\label{eq:Anisoflexibility58}
	{\bf F}^{2}_{IJ}     =&  - \frac{1}{w_{J}} |\Phi^{IJ}|(J_{3} - \Phi^{IJ}) \gamma({\bf r}_I,{\bf r}_J,\Omega_I,\Omega_J,\eta_{IJ}),     &\quad     I\neq J;    \\ 
	{\bf F}^{2}_{II}=&   \sum_{J=1}^N \frac{1}{w_{J}} |\Phi^{IJ}|(J_3 - \Phi^{IJ}) \gamma({\bf r}_I,{\bf r}_J,\Omega_I,\Omega_J,\eta_{IJ}),  &\quad     I=J,
\end{eqnarray}
where ${\rm adj}(\Phi^{IJ})$ denotes the adjoint of matrix. Here  $J_3$ is a $3\times3$ matrix with every element being  one. 
\begin{figure}
\begin{center}
\begin{tabular}{c}
\includegraphics[width=0.8\textwidth]{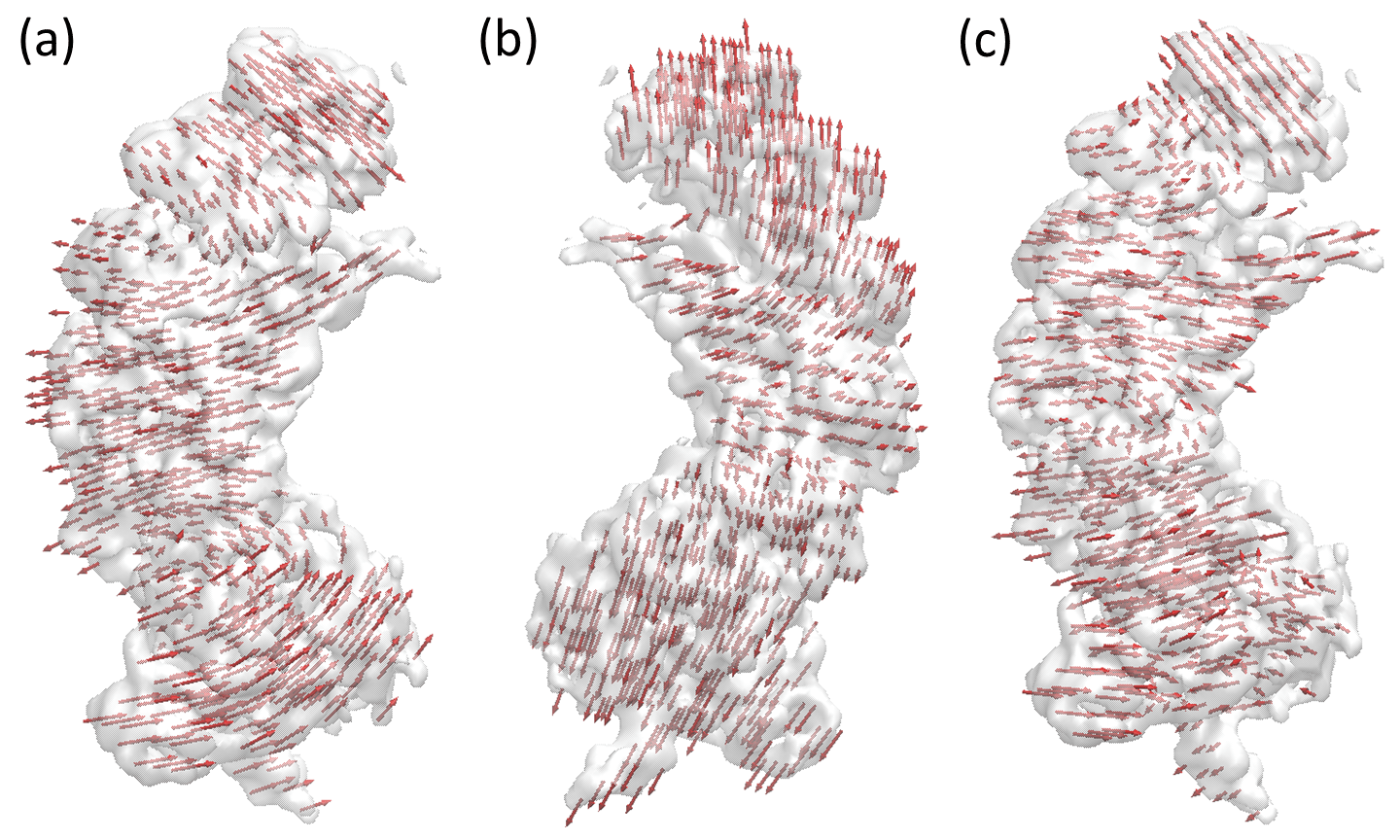}
\end{tabular}
\end{center}
\caption{The first three nontrivial eigenmodes of Cyo-EM data EMD 8295. A threshold value of 0.08 is used in the model to map out the biomolecule. Modes  4, 5 and 6 are demonstrated in (a), (b) and (c), respectively.
}
\label{fig:EM8295}
\end{figure}

\begin{figure}
\begin{center}
\begin{tabular}{c}
\includegraphics[width=0.8\textwidth]{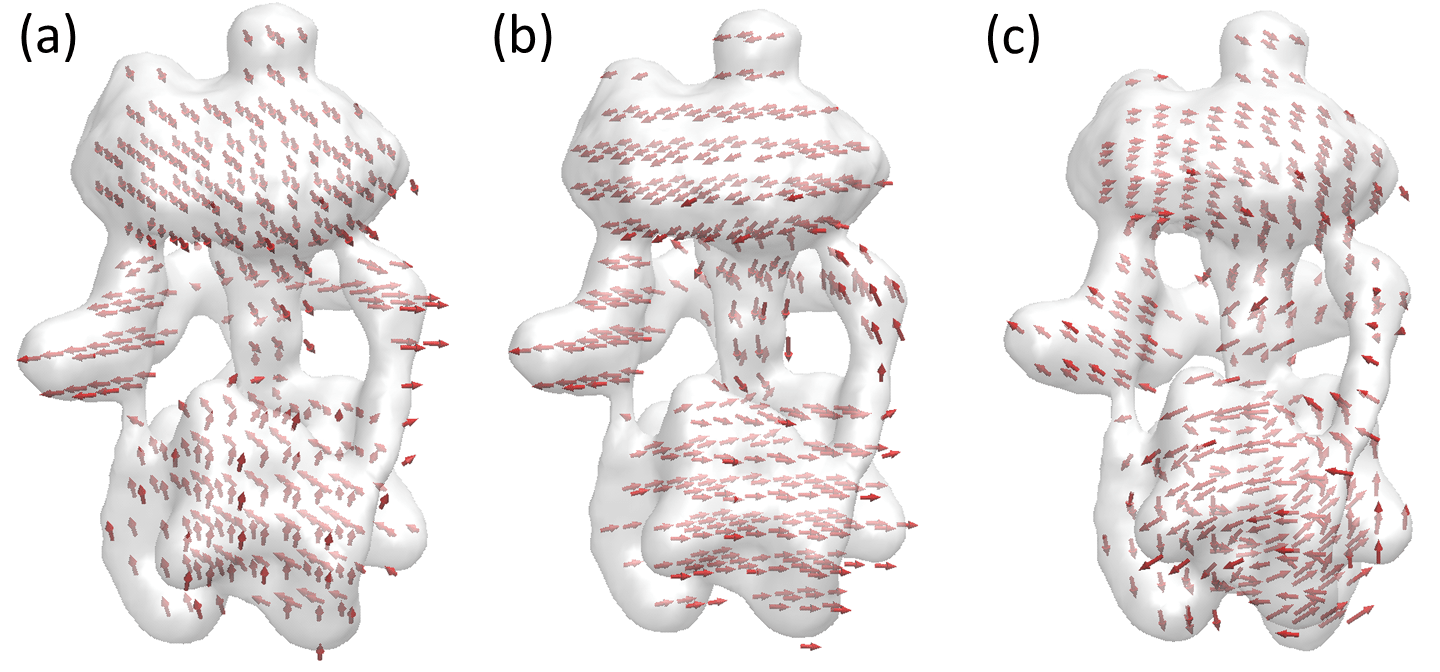}
\end{tabular}
\end{center}
\caption{The first three nontrivial eigenmodes of Cyo-EM data EMD 1590. A threshold value of 0.05 is used in the model. Modes 4,  5 and 6 are demonstrated in (a), (b) and (c), respectively.
}
\label{fig:EM1590}
\end{figure}

It should be noticed that clustering idea in the discrete aFRI matches with the idea of discretization of the continuous density function very well. However, only the simple global form is highlighted in Eqs. (\ref{eq:Anisoflexibility8}) and  (\ref{eq:Anisoflexibility58}).

The proposed model is tested to evaluate the anisotropic motion of density volumetric data. One only needs to consider those density elements in the anisotropic motion analysis that their density values are larger  than a threshold, which is the suggested value for biomolecular visualization. This truncation dramatically reduces the number of entries in final Kirchhoff matrix.

It has been verified that the VPM is able to recover  ANM and aFRI when the elements are very small and each element contains at most one real particle (say a C$_\alpha$).   The demonstration of this result is omitted.

The proposed  method is  illustrated with three examples, i.e., protein 2XHF, cryo-EM maps EMD8295 and EMD3266. For protein 2XHF, the original data contain discrete atomic coordinates. The discrete to continuum mapping is used to generate rigidity density $\mu({\bf r})$. In this transformation,  the Lorentz kernel as in Eq. (\ref{eq:couple_matrix24}) is chosen with $\epsilon=2$ and $\eta_j=1.0$ \AA. To construct VPM Hessian matrix, a threshold is chosen to be $40\%$ of the maximal density value. One can  discretize  the computational domain by using the element size of 3.0 \AA. In the mode analysis, the second aFRI form ${\bf F}^{2}$ is used. The Lorentz kernel in Eq. (\ref{eq:couple_matrix24}) is chosen with $\epsilon=2$ and $\eta_{IJ}=12$\AA. Mode 4 to mode 11 are illustrated  in Fig. \ref{fig:Density_2XHF}. The results are similar to those   obtained with discrete aFRI \cite{DDNguyen:2016b}.

For cryo-EM density map EMD8295,  the data have a dimension of  $326.4*326.4*326.4$\AA$^3$. A mesh of 40*40*40 is used to discretize domain. The threshold of 0.08 is used to result in a total number of 639  nonzero elements in the matrix. Again the second aFRI form ${\bf F}^{2}$ with the Lorentz kernel is used. The parameter used are $\epsilon=2$  and $\eta_{IJ}=40$ \AA. Modes 4, 5 and 6 are illustrated in Fig. \ref{fig:EM8295}.

For EMD1590, the region is of the size $436*436*436$\AA$^3$. A mesh of 25*25*25 is employed for the discretization. The biomolecular domain is chosen by using the threshold of  0.05 and there are 394 nonzero  elements in the final matrix. The second aFRI form ${\bf F}^{2}$ with the Lorentz kernel is used with $\epsilon=2$  and $\eta_{IJ}=60$ \AA~ in matrix construction. Modes 4,  5 and  6 are illustrated in Fig. \ref{fig:EM1590}.

It can be noticed that VPM can be applied to other systems, such as stability analysis of cells, tissues, and some elastic systems with appropriate definitions of correlations functions. However, this aspect is beyond the scope of the present review.

\subsubsection{Eigenvector   analysis}

 Mathematically, vector field can be analyze by Poincar\'e  index \cite{Poincare:1890}, winding number, Morse index \cite{harker:mischaikow:mrozek:nanda, Mischaikow:2013} and more interestingly, the Conley index \cite{Conley:1978,Mischaikow:2002,Chen:2012morse,Manolescu:2013conley}. The essential idea of all these methods is to explore the behavior of the vectors around critical points.

To give a brief introduction of Conley index, one can consider a manifold $M$ and its associated vector field. If one expresses the vector field in terms of a differential equation $\dot{x}=V(x)$. The solution can be expressed as a function $\phi: {\bf R}\times M \rightarrow M$. This solution is a flow that satisfies $\phi(0,x)=x$ for all $x \in M$. One can also define the trajectory  as $\phi ({\bf R},x):=\bigcup_{t \in {\bf R}} \phi(t,x)$. With this setting, an invariant set $S \subset M$ is $\phi (R,S)=S$. Two basic types of invariant sets are fixed points and periodic points, see Table \ref{tb:conley}.
One can define an isolating neighborhood $N \subset M$ as for every $x \in \partial N$, there exists $\epsilon > 0$, such that one has either $\phi((-\epsilon,0),x)\bigcap N =\emptyset$ or $\phi((0,\epsilon),x)\bigcap N =\emptyset$. The exit set of an isolating block $N$ is $L=\{{x \in  \partial N |\phi((0, \epsilon),x)\bigcap N =\emptyset}\}$. The pair $(N,L)$ is called an index pair. The Conley index of an invariant set $S$ is the relative homology of the index pair $(N,L)$, i.e., $CH_*(S):=H_*(N,L)$.

\begin{table}
\begin{center}
\caption{Conley index for various types of invariant sets }
\begin{tabular}{|c|c|}
\hline
 Cases &   Conely index  \\
\hline
 Attracting fixed point   & (1 0 0)   \\
\hline
Saddle fixed point      & (0 1  0)   \\
\hline
Repelling fixed point   &(0 0 1)  \\
\hline
attracting periodic orbit   &(1 1 0)  \\
\hline
Repelling periodic orbit   & (0 1 1) \\
\hline
\end{tabular}
\label{tb:conley}
\end{center}
\end{table}

It should be noticed that the VPM eigenvector field is different from the traditional vector field. For an individual eigenvector, one has local vectors associated all virtual particles. All these local vectors join together to form a unique global vector field, which is able to capture the collective motions of the biomolecule of interest. However, locally, various fixed points can also be identified as demonstrated in Figs. \ref{fig:Density_2XHF}, \ref{fig:EM8295}  and \ref{fig:EM1590}. Rigorous analysis of these eigenvectors is still an open problem.

\subsection{Demonstrations}\label{sec:theory}

In this section, the utility of Geo-Topo algorithms, namely, topological analysis of Hessian matrix eigenvalue and curvature  maps is illustrated by a few case studies.  Additionally, the   persistent homology analysis of molecular Hessian matrix eigenvalue maps is also demonstrated.

\subsubsection{Case studies}\label{sec:density}

Three molecules, a fullerene, an alpha helix and beta sheet, to illustrate Geo-Topo methods are considered.

\paragraph{Fullerene C$_{20}$}
\begin{figure}
\begin{center}
\begin{tabular}{c}
\includegraphics[width=0.6\textwidth]{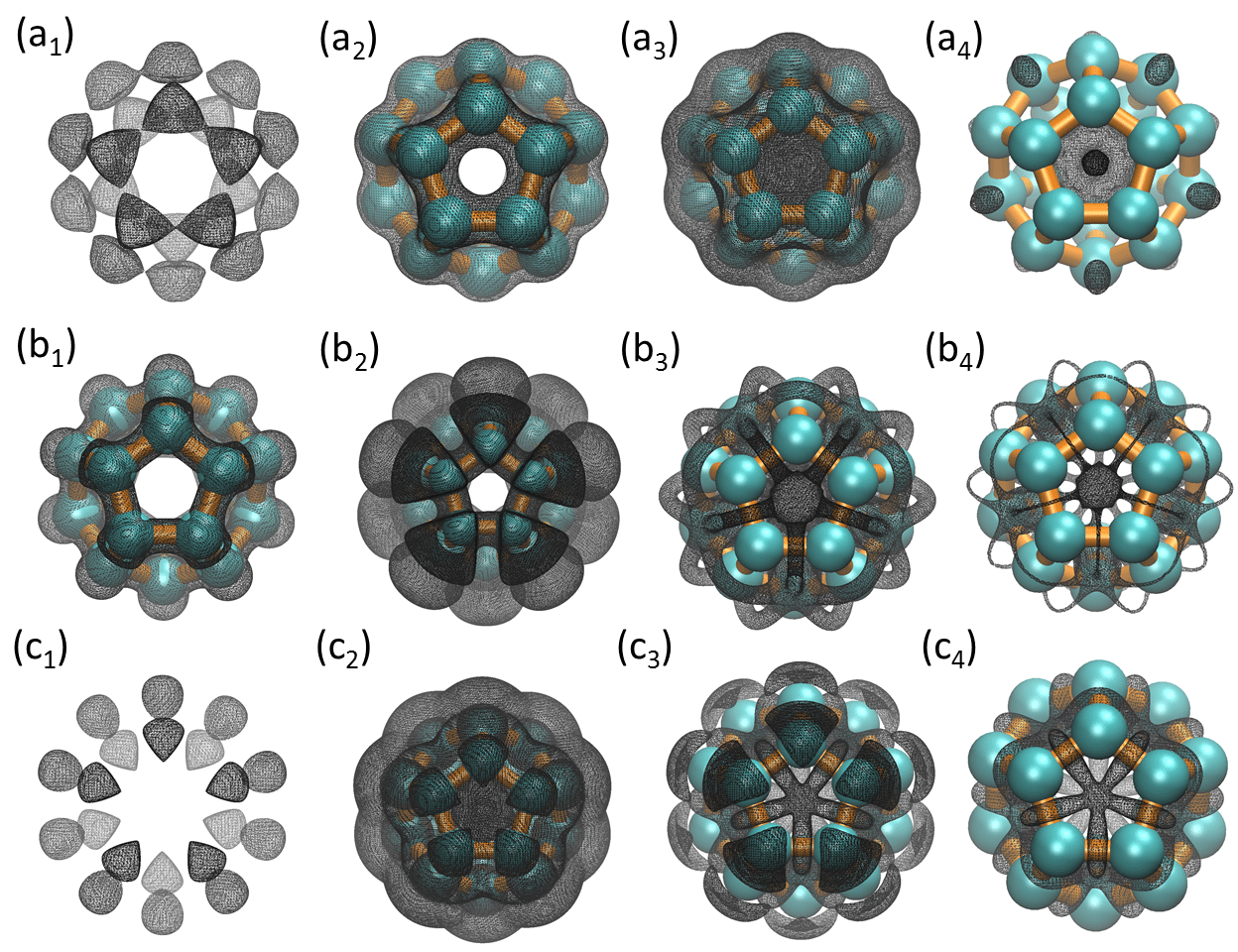}
\end{tabular}
\end{center}
\caption{ Hessian matrix eigenvalue surfaces obtained from different isovalues (or level-set values).
{ (a)} The isosurfaces for the first eigenvalue. The isovalues from { ($a_{1}$)} to { ($a_{4}$)} are -3.0, -2.0, -1.0 and 0.1.
{ (b)} The isosurfaces for the second eigenvalue. The isovalues from { ($b_{1}$)} to { ($b_{4}$)} are -1.0, -0.01, 0.5 and 1.0.
{ (c)} The isosurfaces for the third eigenvalue. The isovalues from { ($c_{1}$)} to { ($c_{4}$)} are  -1.0, 1.0, 1.8 and 2.0.
}
\label{fig:C20_eigen_3d}
\end{figure}

Using the proposed Geo-Topo fingerprint, one can study  more complicated structures. The first one   is the fullerene C$_{20}$ considered in last section.   Again, it rigidity density is described by Eq. (\ref{eq:rigidity3}) and parameter $w_j$ and $\eta_j$ are  set to $1$ and $0.7$, respectively.

First the Hessian matrix eigenvalue isosurfaces of C$_{20}$ is studied. Figure \ref{fig:C20_eigen_3d} has illustrated four representative isosurfaces for each eigenvalue. Subscripts $1$ to $4$ indicate four isovalues from small to large. The notations $({\bf a})$ to $({\bf c})$ represent eigenvalues $\gamma_1$ to $\gamma_3$. One can see that their isosurface behaviors are consistent with descriptions of the Geo-Topo fingerprints summarized in Table \ref{tb:geometric_fingerprint}. State differently, the Geo-Topo fingerprints are able to capture  the essential Geo-Topo properties of  C$_{20}$ isosurfaces.

More specifically, for $\gamma_1$, negative isosurfaces enclose all NCPs and BCPs.   Anisotropic isosurfaces are found near RCPs with negative A-type and positive V-type behaviors. CCPs are enclosed with positive isosurfaces. For $\gamma_2$,  negative isosurfaces enclose all NCPs and bond regions of BCPs. Positive loops  can be found around BCP bonds. RCPs and CCPs are enclosed by positive isosurfaces. For $\gamma_3$,  negative isosurfaces {\it only} enclose atomic basin of NCPs, whereas BCPs, RCPs and CCPs are all enclosed by positive isosurfaces.

\begin{figure}
\begin{center}
\begin{tabular}{c}
\includegraphics[width=0.6\textwidth]{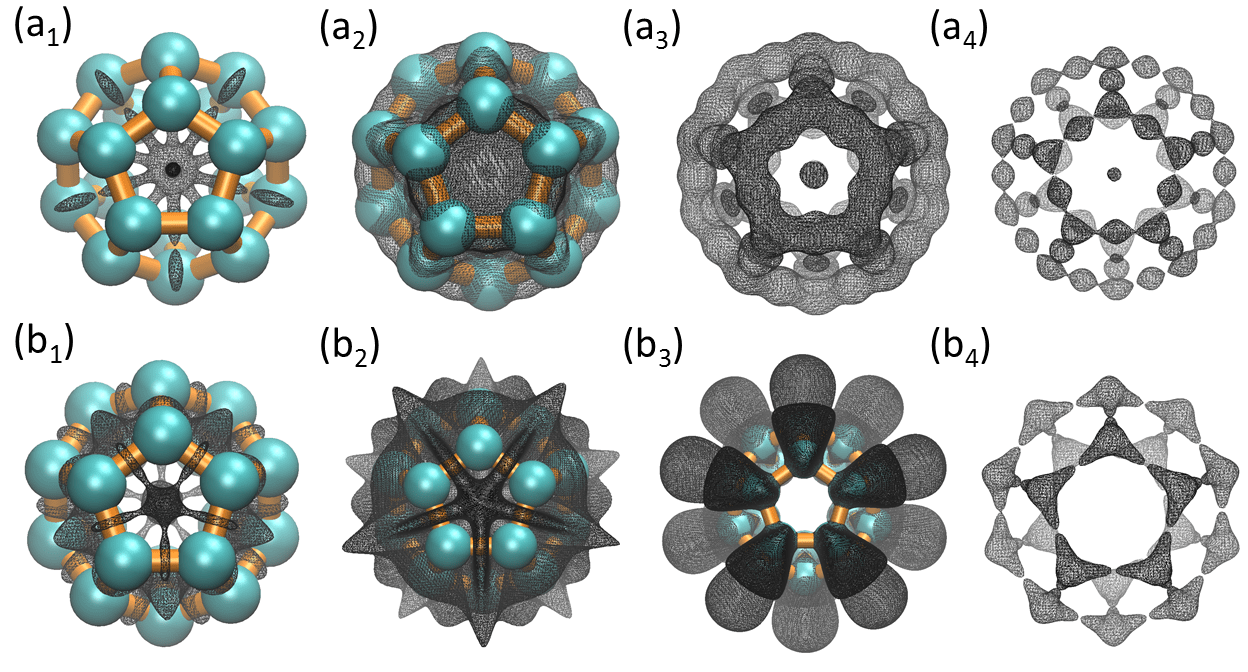}
\end{tabular}
\end{center}
\caption{ Isosurfaces for maximal and minimal curvature maps  of C$_{20}$.
 (a) The isosurfaces built from the  maximal curvature   of C$_{20}$. The isovalues from { ($b_{1}$)} to { ($b_{4}$)} are -3.0, -2.0, 0.5 and 3.0.
{ (b)} The isosurfaces built from  the minimal curvature   of C$_{20}$. The isovalues from { ($a_{1}$)} to { ($a_{4}$)} are -10.0, -5.0, 1.0 and 10.0.
}
\label{fig:C20_k1k2_3d}
%
\begin{center}
\begin{tabular}{c}
\includegraphics[width=0.6\textwidth]{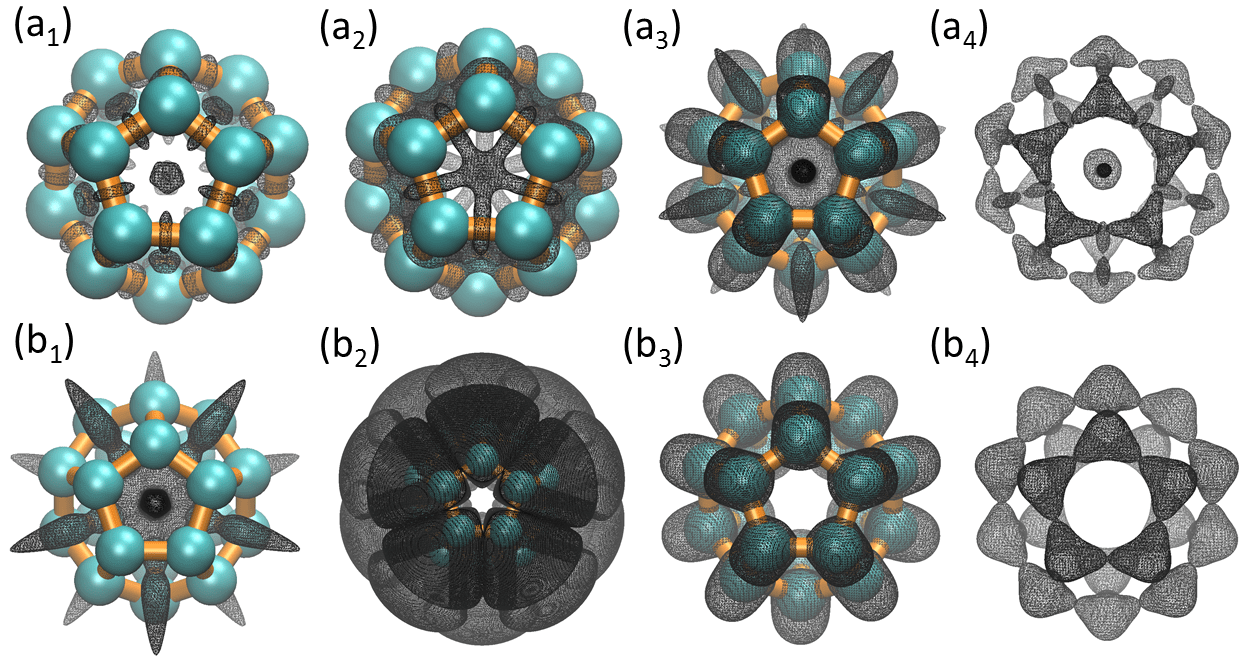}
\end{tabular}
\end{center}
\caption{Isosurfaces for Gaussian and mean curvature maps   of C$_{20}$.
{ (a)} The isosurfaces built from  the Gaussian curvature   of C$_{20}$. The isovalues from { ($a_{1}$)} to { ($a_{4}$)} are -2.0, 2.0, 3.0 and 5.0.
{ (b)} The isosurfaces built from  the Mean curvature   of C$_{20}$. The isovalues from { ($b_{1}$)} to { ($b_{4}$)} are -1.0, 0.001, 1.0 and 2.0.
}
\label{fig:C20_gaussian_mean_3d}
\end{figure}

Figures \ref{fig:C20_k1k2_3d} and \ref{fig:C20_gaussian_mean_3d} illustrates four representative isosurfaces for Gaussian, mean and two principal curvatures. Here the detailed analysis is omitted. However, just as eigenvalue isosurfaces, the curvature isosurfaces can be well-described by the Geo-Topo fingerprints summarized in Table  \ref{tb:geometric_fingerprint}.

\paragraph{An $\alpha$-helix structure}

\begin{figure}
\begin{center}
\begin{tabular}{c}
\includegraphics[width=0.6\textwidth]{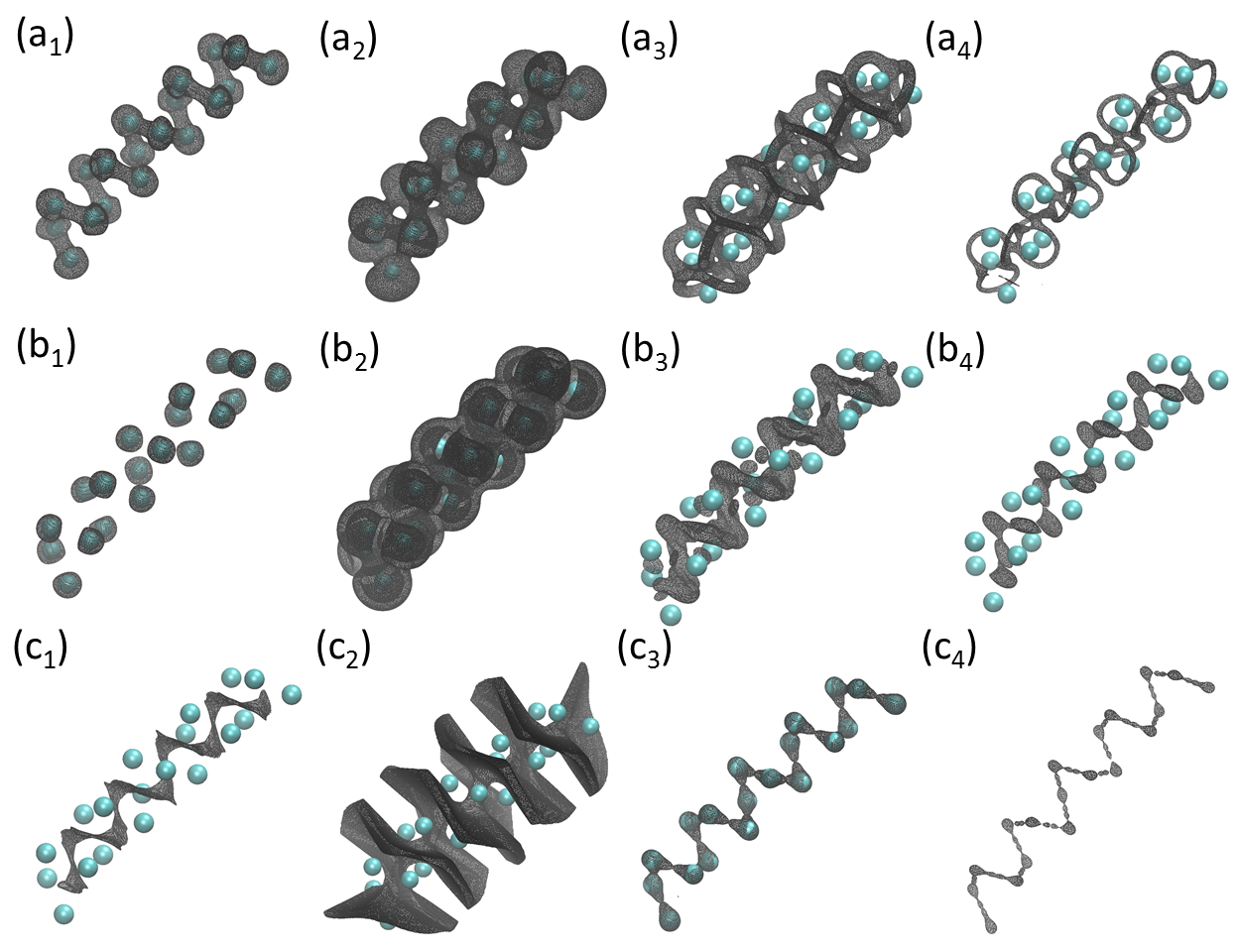}
\end{tabular}
\end{center}
\caption{An illustration of second eigenvalue, third eigenvalue and mean curvature of a coarse-grained representation (C$_{\alpha}$) of an alpha helix.
{ (a)} The isosurfaces for the second eigenvalue. The isovalues from { ($a_{1}$)} to { ($a_{4}$)} are -2.0, -0.05, 0.05 and 0.1.
{ (b)} The isosurfaces for the third eigenvalue. The isovalues from { ($b_{1}$)} to { ($b_{4}$)} are -0.05, 0.2, 0.3 and 0.35.
{ (c)} The isosurfaces for the mean curvature. The isovalues from { ($c_{1}$)} to { ($c_{4}$)} are -1.0, -0.1, 0.1 and 3.0.
}
\label{fig:1c26_helix}
\end{figure}

So far, all cases examined are about highly symmetric molecular structures. In this part, the Geo-Topo analysis is applied to irregular biomolecular structures, particularly protein structure. It is known that there are four distinct levels of protein structure, namely primary structure, which is a sequence of amino acids in the polypeptide chain; secondary structure, which is an $\alpha$-helix or a $\beta$-strand; tertiary structure, which refers to the three-dimensional structure of a monomeric and multimeric protein molecule; and  quaternary structure, which is the three-dimensional structure of a multi-subunit protein complex. The Geo-Topo analysis of proteins for protein secondary structures is demonstrated.

First, one can consider  one of  protein secondary structures, i.e., an $\alpha$-helix segment. This segment is extracted from protein with ID 1C26. The coarse-grained (CG) representation is employed and 19 C$_{\alpha}$ atoms from the $335$th residue to $353$th residue in chain A are used. The distance between two adjacent C$_{\alpha}$ atoms are about $3.8$ \AA~ and the $\eta$ used in the CG rigidity density model is $2.0$ \AA.

In stead of listing all results of eigenvalues and curvatures,  only three geometric parameters of interest are examined, including eigenvalue $\gamma_2$, eigenvalue $\gamma_3$ and mean curvature. For each of them, four representative isosurfaces are extracted. The results are illustrated in Fig. \ref{fig:1c26_helix}.

It can be found that results are very consistent with the Geo-Topo fingerprints.  For $\gamma_2$, positive loops around atomic bonds are found. For $\gamma_3$, large eigenvalues are still concentrated in regions near BCPs. For mean curvature, one can still find   positive A-type and negative V-type isosurfaces near BCPs. Also, large positive values are concentrates around NCPs and indicate the topological connectivity at Figs. \ref{fig:1c26_helix}(a$_1$), (c$_3$) and (c$_4$).

It also should be noticed that unlike the regular symmetric structures studied in  previous sections,   RCPs and CCPs are more complicated in $\alpha$-helix structure. Moreover, since the characteristic distance $\eta$ is chosen as $2.0$ \AA, bond effect (or topological connectivity)  is observed not only between adjacent two atoms but also between atoms in close distance, as indicated in Fig. \ref{fig:1c26_helix}${ (a_3)}$. At meantime, strongest topological connectivity (largest $\gamma_2$ isovalues) is as usual found between  adjacent two atoms as indicated in Fig. \ref{fig:1c26_helix}${(a_4)}$. Interestingly, in the case of $\gamma_3$, positive isosurface forms a strip that is parallel to the backbone of the $\alpha$-helix as demonstrated in Fig. \ref{fig:1c26_helix}${ (b_3)}$. The largest isovalues are concentrated around BCPs not between adjacent two atoms, but  neighboring two atoms in  adjacent two circles as illustrated in Fig. \ref{fig:1c26_helix}${ (b_4)}$. However, if  characteristic distance $\eta$ was chosen as $1.5$ \AA, the largest isovalues of $\gamma_3$ would move to BCP regions between  adjacent two atoms. This is due to the multiscale nature of the model. Further detailed discussion of this multiscale property is beyond the scope of this paper.

\paragraph{A $\beta$-sheet structure}
\begin{figure}
\begin{center}
\begin{tabular}{c}
\includegraphics[width=0.6\textwidth]{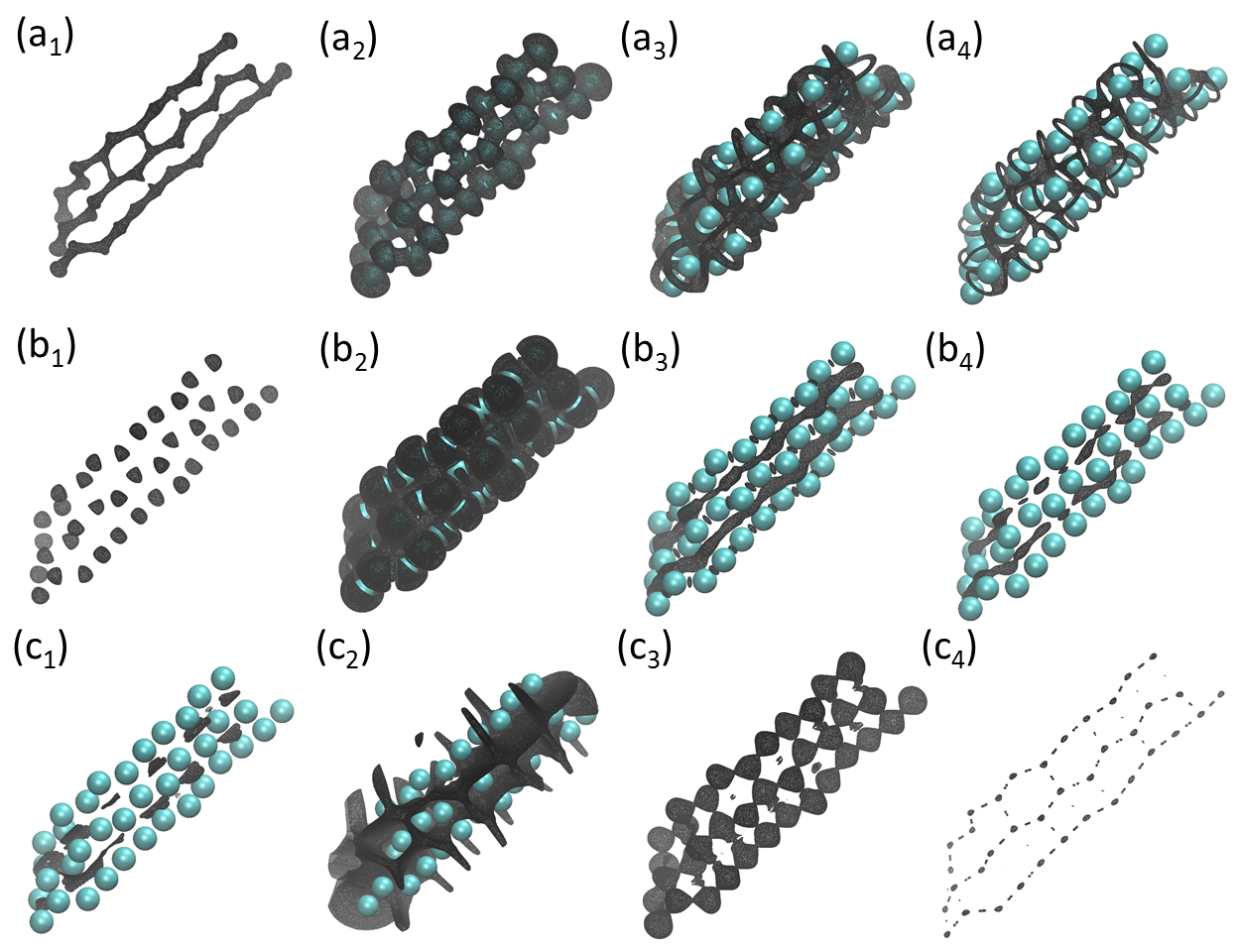}
\end{tabular}
\end{center}
\caption{An illustration of second eigenvalue, third eigenvalue  and mean curvature of a coarse-grained representation (C$_{\alpha}$) of a $\beta$-sheet.
{ (a)} The isosurfaces for the second eigenvalue. The isovalues from { (a$_{1}$)} to { (a$_{4}$)} are -0.3, -0.05, 0.05 and 0.1.
{ (b)} The isosurfaces for the third eigenvalue. The isovalues from { (b$_{1}$)} to { (b$_{4}$)} are -0.1, 0.2, 0.3 and 0.35.
{ (c)} The isosurfaces for the mean curvature. The isovalues from { (c$_{1}$)} to { (c$_{4}$)} are -1.0, -0.1, 0.5 and 3.0.
}
\label{fig:4uw4_sheet}
\end{figure}

Another important protein secondary structure is $\beta$-sheet. In this part, three adjacent $\beta$-strands from protein with ID 4UW4 are considered. Again, the CG representation is used and  three stands include residues from $575$ to $586$, $589$ to $600$ and $603$ to $615$ in chain A. Just as the analysis in the $\alpha$-helix structure, the characteristic distance $\eta$ is chosen as $2.0$ \AA.  One can examine the second eigenvalue $\gamma_2$, the third eigenvalue $\gamma_3$ and mean curvature in the Geo-Topo analysis. For each of them, four representative isosurfaces are extracted. The results are illustrated in Fig. \ref{fig:4uw4_sheet}.

Just as the $\alpha$-helix case, results for $\beta$-sheet are also very consistent with the Geo-Topo fingerprints. Positive loops around  BCPs for $\gamma_2$ can be observed. For  $\gamma_3$,  small negative isosurfaces indicate NCPs. Large $\gamma_3$ eigenvalues are concentrated in regions near BCPs. Positive A-type and negative V-type isosurfaces near BCPs are found in mean curvature. Large positive values are concentrates around NCPs. Further, as the characteristic distance is chosen as $2.0 $ \AA, the bond connection between the stands or sheets are amplified. It can be clearly observed in Fig. \ref{fig:4uw4_sheet} ${ (b_4)}$.


\subsubsection{Persistent homology for scalar field analysis} \label{sec:PHA_scalar_field}

\begin{figure}
\begin{center}
\begin{tabular}{c}
\includegraphics[width=0.9\textwidth]{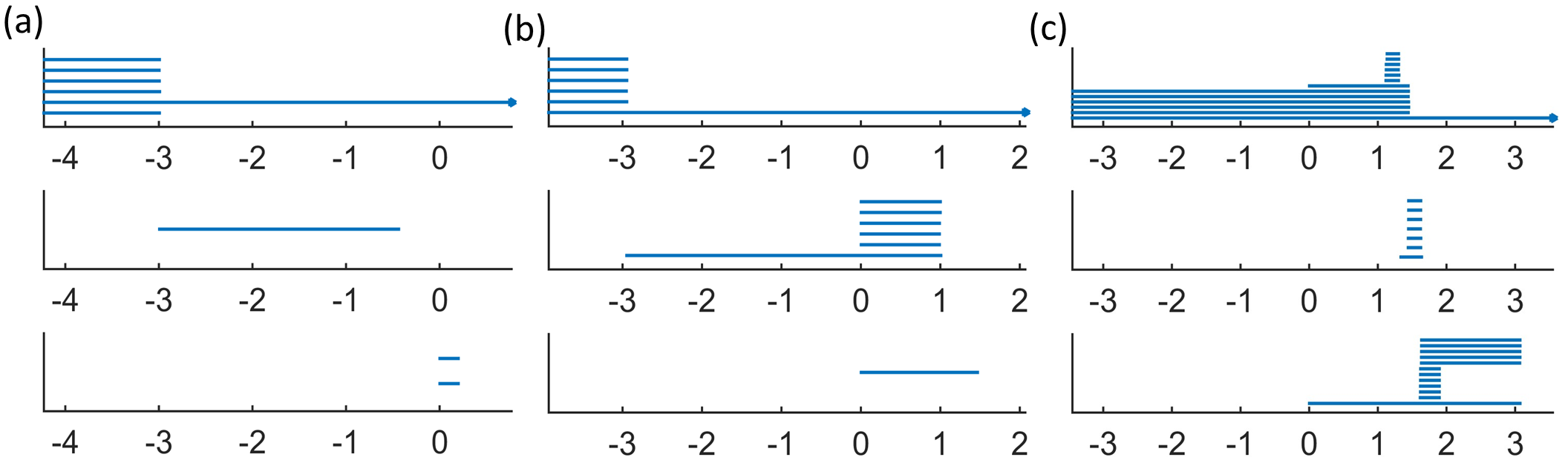}
\end{tabular}
\end{center}
\caption{Barcodes for three eigenvalue maps of benzene molecule. From {  (a)} to {  (c)}, the barcodes are for $\lambda_1$,  $\lambda_2$ and $\lambda_3$, respectively. In each subfigure, from top to bottom, the results are for $\beta_0$, $\beta_1$ and $\beta_2$, respectively.
}
\label{fig:C6_eigen_ph}
\end{figure}

Finally, the topological persistence in the scalar fields generated by Hessian matrix eigenvalues and curvatures is studied.
The hexagonal ring is used for  persistent homology analysis. The barcodes for eigenvalues $\lambda_1$,  $\lambda_2$ and $\lambda_3$ are demonstrated in Figs. \ref{fig:C6_eigen_ph} (a)-(c). In each subfigure, results for $\beta_0$, $\beta_1$ and $\beta_2$ are presented. The filtration goes from the smallest value to the largest in the  persistent homology analysis.

For $\lambda_1$, at very small values, it has $6$ $\beta_0$ bars, i.e., $6$ independent components. Topologically, it represents small $\lambda_1$ negative values concentrating around  $6$ NCPs. As the filtration progresses, a loop is formed, which leads to  a bar  in $\beta_1$. Further down the filtration process, $\beta_0$ isosurface begins to shrink to two balls perpendicular to the RCP. This contributes two bars in $\beta_2$.

For $\lambda_2$, again $6$ $\beta_0$ bars are found in the earliest stage of the filtration. Here $6$ independent components quickly combine to form a loop as filtration progresses. More interesting, individual loops around BCPs  form when filtration value is around the range from 0 to 1. Finally,  isosurfaces shrink into the RCP.

For $\lambda_3$, smallest negative values are concentrated around NCPs and contribute $6$ $\beta_0$ bars at the earliest stage of the filtration. Once the isosurface value becomes positive, a new component emerges due to the generation of a new isosurface at the boundary region. This new isosurface also contributes to a long persisting  $\beta_2$ bar. Further down the filtration, $6$ more $\beta_0$ bars occur, each representing a very narrow ring region around atom bonds. These regions are relatively small and  quickly disappear. After that, $6$ ``hat" regions attaching to the original NCP isosurfaces emerge, contributing to $6$ small loops. They quickly become detached and shrink away. At same time, $6$ isosurfaces form near the BCP and gradually disappear. Together they contribute 12 independent $\beta_2$ bars.

It can be seen that   barcodes in Figs. \ref{fig:C6_eigen_ph} (a)-(c) give a detailed account of the full spectrum of the isosurface evolution process in Figs. \ref{fig:C6_eigen_3d} (a$_1)$-(a$_4)$.

\section{Concluding remarks}

Every field in natural science, engineering, medicine, finance and   social sciences becomes quantitative when it is getting mature. Mathematics is essential for all quantitative fields. Being regarded the last scientific forefront,  biological sciences, particularly molecular biology and structural biology,  have accumulated gigantic among of data in terms of biomolecular structures, activity relations and genetic sequences in the past few decades and are transforming from qualitative and phenomenological to quantitative and predictive. Such a transformation offers unprecedented opportunities for mathematically driven advances in biological sciences \cite{Wei:2016}.

Geometry, topology, and graph theory are some of the core mathematics and have been naturally playing a unique role in molecular biology and molecular biophysics. In this paper, we present a brief review of  geometric, topological, and graph theory apparatuses that are important to the contemporary molecular biology and biophysics. We first discuss the discrete methods and models, including graph theory, Gaussian network models, anisotropic network model, normal mode analysis, flexibility-rigidity index, molecular nonlinear dynamics, spectral graph theory and persistent homology for biomolecular modeling and analysis. Additionally, we  describe  continuous  algorithms  and theories, including, discrete to continuum mapping, multidimensional persistent homology for volumetric data sets,  geometric modeling of biomolecules, differential geometry theory of surfaces,  curvature analysis, atoms in molecule theory, and quantum chemical topology theory. Attention is paid to the connections between  existing biophysical approaches and standard mathematical subjects, such as, Morse theory, Poincar\'e Hopf index,   differential topology, etc.  Open problems and potential new directions are point out in  discussions.

Two new models, namely the analytical minimal molecular surface and virtual particle model, and two new methods, i.e., Hessian eigenvalue map and curvature map, are introduced for biomolecular modeling and analysis. These new approaches were inspirited by the subject under review during our preparation of this review. For simplicity,  only the proof-of-principle applications are demonstrated for all methods, models, theories and algorithms covered in this review.

Selected mathematical topics in geometry, topology, and graph theory   are based on our limited knowledge and understanding of mathematics and molecular biophysics. Many subjects in geometry, topology, and graph theory  that have found much success in molecular biology and biophysics have not been covered in this review.  One of these subjects is knot theory, particularly the DNA knot theory, which is also a very important ingredient of topological modeling of biomolecules \cite{vologodskii:1992,pohl:1980,fuller:1971,ABates:2005,sumners:1992,darcy:2001,arsuaga:2002,buck:2007,IKDarcy:2013,RBrasher:2013,Schlick:1992trefoil}. Knot theory is an area of geometric topology that deals with knots and links. Mathematically, a knot is an embedding of circles or its homeomorphisms in the three-dimensional (3D) Euclidean space, ${\mathbb R}^3$. Physically,  DNA, as a genetic material, exists usually in two very long strands that intertwine  to form chromatins, bind with histones to build nucleosomes,  tie into knots,  and are subjected to successive coiling before package into chromosomes. The loss of knots in chromosomes can cause  Angelman and Prader-Willi syndromes. DNA knot theory has been a very important topic in applied topology. However, in this review, focus is given to   geometric and topological   methods  and models for atomistic biomolecular data. Therefore, DNA knot theory is not covered.

Another relevant subject that have not been considered in this review is  biomolecular interaction network models, including protein interaction networks, metabolic networks and transcriptional regulatory networks. Obviously, part of the mathematical foundation of these models is  graph theory. Protein interaction networks are designed to study protein-protein interactions \cite{rain:2001,giot:2003}. Normally, in these models, proteins are represented as nodes and physical interaction between them are represented by edges. To make the network more reliable, data from different sources are combined together and different rules are applied for the identification of protein interactions. A metabolic network considers all metabolic and physical processes happened within a cell \cite{jeong:2000,overbeek:2000}. This network comprises the chemical reactions of metabolism, the metabolic pathways, as well as the regulatory interactions that guide these reactions. Transcriptional regulatory networks describe the regulatory interactions between genes \cite{lee:2002,salgado:2006}. In this network, each gene is represented by a node and the regulation relations are represented by edges. The exclusion of this subject is also due to our focus on atomistic biomolecular data.

Topological graph theory concerns immersions of graphs as well as the embedding of graphs in surfaces, spatial embeddings of graphs, and graphs as topological spaces. It has had much success in the mathematical modeling of  DNA recombination, DNA-RNA interactions,  protein folding and protein-protein interactions \cite{Angeleska:2009}.  The exclusion of this subject is due to our insufficient knowledge and understanding. For the same reason,  fascinating applications of combinatorics, algebra and tiling theory in the modeling of  viral  capsid self assembly \cite{Jonoska:2009,Twarock:2008,Angeleska:2009, SHarvey:2013,CHeitsch:2014, Sadre-Marandi:2014} have not been covered in our review.

A continuous, differentiable curve can be embedded in a three-dimensional Euclidean space and its kinematic properties, such as the derivatives of  tangent, normal, and binormal unit vectors, can be described by Frenet-Serret formulas in differential geometry \cite{crenshaw1993orientation}. Discrete Frenet-Serret frame offers an efficient description of amino-acid and/or nucleic acid chains  \cite{quine2004mathematical,hu2011discrete}. We believe that  discrete Frenet-Serret frame can be easily applied to RNA chains, microtubules, nucleosomes, chromatins, active  chromosomes and metaphase chromosomes.

An emergent approach is to combine machine learning with geometry, topology and/or graph theory  for analyzing biomolecular data \cite{ZXCang:2015, Kovacev-Nikolic:2016}. Machine learning is a cutting edge  computer science and statistical tool originally developed for pattern recognition and artificial intelligence.  Its combination with mathematical apparatuses leads to extremely powerful approaches to massive biomolecular data challenges, such as the blind predictions of solvation free energies  \cite{BaoWang:2016FFTS,BaoWang:2016HPK} and protein-ligand binding affinity prediction \cite{BaoWang:2016FFTB}.  For example, topological learning algorithm that utilizes exclusively persistent homology and machine learning for protein-ligand binding affinity predictions outperforms all the existing eminent methods in computational biophysics over massive binding data sets \cite{ZXCang:2016b}.    However, this subject is at its early stage and is evolving too fast to have a conclusive review at present.

It is worth mentioning that the geometric, topological and graph theory apparatuses discussed in this review  can be employed in conjugation with   partial differential equation (PDE), which is widely used in computational biophysics,  to model biomolecular systems. Certainly geometric modeling is often a prerequisite in the PDE models of electrostatics, solvation, ion channels, membrane-protein interactions, and biomolecular elasticity \cite{Wei:2009}. As mentioned in Section \ref{sec:spectral},  the graph cut problem can be formulated as a free energy minimization. Such a formulation makes it possible to combine spectral graph theory with PDE approaches for a wide range of biophysical modeling for biomolecular systems, including solvation, ion channel, biomembrane, protein-ligand binding, protein-protein interaction and protein-nucleic acid interaction. Finally,  connection between algebraic topology and differential geometry, including Laplace-Beltrami operator, has been made \cite{BaoWang:2016a}. Essentially, one defines an object function to optimize certain biophysical properties, which leads to a Laplace-Beltrami operator that generates a multiscale representation of the initial data and offers an object-oriented filtration process for persistent homology. The resulting differential geometry based object-oriented persistent homology is able to preserve desirable geometric features in the evolutionary filtration and enhances the corresponding topological persistence. However,  how to design object-oriented persistent homology to  automatically  extract desirable features in the original biomolecular data during the filtration process is still an open problem.

Indeed, the application aspects of geometry, topology and graph theory  have become a driven force for the development of abstract geometry, topology, homology, graph theory in recent years. It is expected that a  versatile variety of pure mathematics concepts, methods and techniques will find their cutting edges in the transcend description of biomolecular structure, function, dynamics and transport. This health interaction between mathematics and molecular bioscience will benefit both fields and attract young researchers for generations to come.

\section*{Acknowledgments}

This work was supported in part by a Start-Up Grant from The Nanyang Technological University (KLX),    NSF IIS- 1302285 (GWW),
NSF DMS-1160352 (GWW),   NIH  R01GM-090208 (GWW) and  MSU Center for Mathematical Molecular Biosciences Initiative (GWW).

\bibliographystyle{plain}

\begin{thebibliography}{100}

\bibitem{AEHW06}
P.~K. Agarwal, H.~Edelsbrunner, J.~Harer, and Y.~Wang.
\newblock Extreme elevation on a 2-manifold.
\newblock {\em Discrete and Computational Geometry (DCG)}, 36(4):553--572,
  2006.

\bibitem{Aldous:2002}
D.~Aldous and J.~Fill.
\newblock Reversible {Markov} chains and random walks on graphs, 2002.

\bibitem{alexandrov:2003}
N.~Alexandrov and I.~Shindyalov.
\newblock {PDP}: protein domain parser.
\newblock {\em Bioinformatics}, 19(3):429--430, 2003.

\bibitem{Allen:1987}
M.~P. Allen and D.~J. Tildesley.
\newblock {\em Computer Simulation of Liquids}.
\newblock Oxford: Clarendon Press, 1987.

\bibitem{Amadei:1993essential}
A.~Amadei, A.~Linssen, and H.~JC Berendsen.
\newblock Essential dynamics of proteins.
\newblock {\em Proteins: Structure, Function, and Bioinformatics},
  17(4):412--425, 1993.

\bibitem{Anfinsen:1973}
C.~B. Anfinsen.
\newblock Einfluss der configuration auf die wirkung den.
\newblock {\em Science}, 181:223 -- 230, 1973.

\bibitem{Angeleska:2009}
A.~Angeleska, N.~Jonoska, and M.~Saito.
\newblock Dna rearrangement through assembly graphs.
\newblock {\em Discrete and Applied Mathematics}, 157:3020--3037, 2009.

\bibitem{arsuaga:2002}
J.~Arsuaga, M.~V{\'a}zquez, S.~Trigueros, D.~W. Sumners, and J.~Roca.
\newblock Knotting probability of {DNA} molecules confined in restricted
  volumes: {DNA} knotting in phage capsids.
\newblock {\em Proceedings of the National Academy of Sciences},
  99(8):5373--5377, 2002.

\bibitem{Atilgan:2001}
A.~R. Atilgan, S.~R. Durrell, R.~L. Jernigan, M.~C. Demirel, O.~Keskin, and
  I.~Bahar.
\newblock Anisotropy of fluctuation dynamics of proteins with an elastic
  network model.
\newblock {\em Biophys. J.}, 80:505 -- 515, 2001.

\bibitem{azran:2006}
A.~Azran and Z.~Ghahramani.
\newblock A new approach to data driven clustering.
\newblock In {\em Proceedings of the 23rd international conference on Machine
  learning}, pages 57--64. ACM, 2006.

\bibitem{Bader:1985}
R.~F. Bader.
\newblock Atoms in molecules.
\newblock {\em Accounts of Chemical Research}, 18(1):9--15, 1985.

\bibitem{Bader:1990}
R.~F. Bader.
\newblock {\em Atoms in molecules}.
\newblock Wiley Online Library, 1990.

\bibitem{Bader:1988}
R.~F. Bader, R.~J. Gillespie, and P.~J. MacDougall.
\newblock A physical basis for the {VSEPR} model of molecular geometry.
\newblock {\em Journal of the American Chemical Society}, 110(22):7329--7336,
  1988.

\bibitem{Bahar:1998}
I.~Bahar, A.~R. Atilgan, M.~C. Demirel, and B.~Erman.
\newblock Vibrational dynamics of proteins: Significance of slow and fast modes
  in relation to function and stability.
\newblock {\em Phys. Rev. Lett}, 80:2733 -- 2736, 1998.

\bibitem{Bahar:1997}
I.~Bahar, A.~R. Atilgan, and B.~Erman.
\newblock Direct evaluation of thermal fluctuations in proteins using a
  single-parameter harmonic potential.
\newblock {\em Folding and Design}, 2:173 -- 181, 1997.

\bibitem{Baker:2005}
N.~A. Baker.
\newblock Improving implicit solvent simulations: a {Poisson}-centric view.
\newblock {\em Current Opinion in Structural Biology}, 15(2):137--43, 2005.

\bibitem{Baker:2001}
N.~A. Baker, D.~Sept, S.~Joseph, M.~J. Holst, and J.~A. McCammon.
\newblock Electrostatics of nanosystems: Application to microtubules and the
  ribosome.
\newblock {\em Proceedings of the National Academy of Sciences of the United
  States of America}, 98(18):10037--10041, 2001.

\bibitem{ABates:2005}
A.~D. Bates and A.~Maxwell.
\newblock {\em {DNA} topology}.
\newblock Oxford University Press, USA, 2005.

\bibitem{Bates:2009}
P.~W. Bates, Z.~Chen, Y.~H. Sun, G.~W. Wei, and S.~Zhao.
\newblock Geometric and potential driving formation and evolution of
  biomolecular surfaces.
\newblock {\em J. Math. Biol.}, 59:193--231, 2009.

\bibitem{Bates:2006}
P.~W. Bates, G.~W. Wei, and S.~Zhao.
\newblock The minimal molecular surface.
\newblock {\em arXiv:q-bio/0610038v1}, [q-bio.BM], 2006.

\bibitem{Bates:2008}
P.~W. Bates, G.~W. Wei, and Shan Zhao.
\newblock Minimal molecular surfaces and their applications.
\newblock {\em Journal of Computational Chemistry}, 29(3):380--91, 2008.

\bibitem{Dipha}
U.~Bauer, M.~Kerber, and J.~Reininghaus.
\newblock Distributed computation of persistent homology.
\newblock {\em Proceedings of the Sixteenth Workshop on Algorithm Engineering
  and Experiments (ALENEX)}, 2014.

\bibitem{Beketayev:2011}
K.~Beketayev, G.~H. Weber, M.~Haranczyk, P.T. Bremer, M.~Hlawitschka, and
  B.~Hamann.
\newblock Topology-based visualization of transformation pathways in complex
  chemical systems.
\newblock In {\em Computer Graphics Forum}, volume~30, pages 663--672. Wiley
  Online Library, 2011.

\bibitem{Belkin:2003}
M.~Belkin.
\newblock {\em Problems of learning on manifolds}.
\newblock PhD thesis, The University of Chicago, 2003.

\bibitem{Belkin:2005}
M.~Belkin and P.~Niyogi.
\newblock Towards a theoretical foundation for {Laplacian}-based manifold
  methods.
\newblock In {\em International Conference on Computational Learning Theory},
  pages 486--500. Springer, 2005.

\bibitem{Bendich:2010}
Paul Bendich, Herbert Edelsbrunner, and Michael Kerber.
\newblock Computing robustness and persistence for images.
\newblock {\em IEEE Transactions on Visualization and Computer Graphics},
  16:1251--1260, 2010.

\bibitem{BH11}
Paul Bendich and John Harer.
\newblock Persistent intersection homology.
\newblock {\em Foundations of Computational Mathematics (FOCM)},
  11(3):305--336, 2011.

\bibitem{BVP15}
Janine Bennett, Fabien Vivodtzev, and Valerio Pascucci, editors.
\newblock {\em Topological and statistical methods for complex data: {T}ackling
  large-scale, high-dimensional and multivariate data spaces}.
\newblock Mathematics and Visualization. Springer-Verlag Berlin Heidelberg,
  2015.

\bibitem{Bergstrom}
C.A.S. Bergstrom, M.~Strafford, L.~Lazorova, A.~Avdeef, K.~Luthman, and
  P.~Artursson.
\newblock Absorption classification of oral drugs based on molecular surface
  properties.
\newblock {\em J. Medicinal Chem.}, 46:558--570, 2003.

\bibitem{Biasotti:2008}
S.~Biasotti, L.~De~Floriani, B.~Falcidieno, P.~Frosini, D.~Giorgi, C.~Landi,
  L.~Papaleo, and M.~Spagnuolo.
\newblock Describing shapes by geometrical-topological properties of real
  functions.
\newblock {\em ACM Computing Surveys}, 40(4):12, 2008.

\bibitem{Biegler:2002}
F.~Biegler-K{\"o}nig and J.~Sch{\"o}nbohm.
\newblock Update of the {AIM}2000-program for atoms in molecules.
\newblock {\em Journal of computational chemistry}, 23(15):1489--1494, 2002.

\bibitem{Blinn:1982}
J.~Blinn.
\newblock A generalization of algebraic surface drawing.
\newblock {\em ACM Transactions on Graphics}, 1(3):235--256, 1982.

\bibitem{RBrasher:2013}
R.~Brasher, R.~G. Scharein, and M.~Vazquez.
\newblock New biologically motivated knot table.
\newblock {\em Biochemical Society Transactions}, 41:606--611, 2013.

\bibitem{BHPP14}
P.~T. Bremer, V.~Pascucci I.~Hotz, and R.~Peikert, editors.
\newblock {\em Topological methods in data analysis and visualization III:
  {T}heory, algorithms and applications}.
\newblock Mathematics and Visualization. Springer International Publishing,
  2014.

\bibitem{Brooks:1983harmonic}
B.~Brooks and M.~Karplus.
\newblock Harmonic dynamics of proteins: normal modes and fluctuations in
  bovine pancreatic trypsin inhibitor.
\newblock {\em Proceedings of the National Academy of Sciences},
  80(21):6571--6575, 1983.

\bibitem{Brooks:1983}
B.~R. Brooks, R.~E. Bruccoleri, B.~D. Olafson, D.J. States, S.~Swaminathan, and
  M.~Karplus.
\newblock Charmm: A program for macromolecular energy, minimization, and
  dynamics calculations.
\newblock {\em J. Comput. Chem.}, 4:187--217, 1983.

\bibitem{Brooks:1995harmonic}
B.~R. Brooks, D.~Jane{\v{z}}i{\v{c}}, and M.~Karplus.
\newblock Harmonic analysis of large systems. {I. Methodology}.
\newblock {\em Journal of computational chemistry}, 16(12):1522--1542, 1995.

\bibitem{Bubenik:2015statistical}
Peter Bubenik.
\newblock Statistical topological data analysis using persistence landscapes.
\newblock {\em Journal of Machine Learning Research}, 16(1):77--102, 2015.

\bibitem{Bubenik:2007}
Peter Bubenik and Peter~T. Kim.
\newblock A statistical approach to persistent homology.
\newblock {\em Homology, Homotopy and Applications}, 19:337--362, 2007.

\bibitem{buck:2007}
D.~Buck and E.~Flapan.
\newblock Predicting knot or catenane type of site-specific recombination
  products.
\newblock {\em Journal of molecular biology}, 374(5):1186--1199, 2007.

\bibitem{ZXCang:2016a}
Z.~X. Cang and G.~W. Wei.
\newblock {Feature functional theory-mutation predictor (FFT-MP) for the blind
  prediction of protein mutation energy changes}.
\newblock {\em Submitted}, 2016.

\bibitem{ZXCang:2016b}
Z.~X. Cang and G.~W. Wei.
\newblock {Persistent topology based scoring function (T-Score) for the blind
  prediction of protein-ligand binding affinities }.
\newblock {\em Submitted}, 2016.

\bibitem{ZXCang:2015}
Zixuan Cang, Lin Mu, Kedi Wu, Kris Opron, Kelin Xia, and Guo-Wei Wei.
\newblock A topological approach to protein classificationy.
\newblock {\em Molecular based Mathematical Biologys}, 3:140--162, 2015.

\bibitem{Carlsson:2009}
G.~Carlsson.
\newblock Topology and data.
\newblock {\em Am. Math. Soc}, 46(2):255--308, 2009.

\bibitem{Carlsson:2008}
G.~Carlsson, T.~Ishkhanov, V.~Silva, and A.~Zomorodian.
\newblock On the local behavior of spaces of natural images.
\newblock {\em International Journal of Computer Vision}, 76(1):1--12, 2008.

\bibitem{Carlsson:2009theory}
G.~Carlsson and A.~Zomorodian.
\newblock The theory of multidimensional persistence.
\newblock {\em Discrete Computational Geometry}, 42(1):71--93, 2009.

\bibitem{CZOG05}
G.~Carlsson, A.~Zomorodian, A.~Collins, and L.~J. Guibas.
\newblock Persistence barcodes for shapes.
\newblock {\em International Journal of Shape Modeling}, 11(2):149--187, 2005.

\bibitem{zigzag}
Gunnar Carlsson and Vin De~Silva.
\newblock Zigzag persistence.
\newblock {\em Foundations of computational mathematics}, 10(4):367--405, 2010.

\bibitem{CSM09}
Gunnar Carlsson, Vin de~Silva, and Dmitriy Morozov.
\newblock Zigzag persistent homology and real-valued functions.
\newblock In {\em Proc. 25th Annu. ACM Sympos. Comput. Geom.}, pages 247--256,
  2009.

\bibitem{SLChan:1998}
S.~L. Chan and E.~O. Purisima.
\newblock Molecular surface generation using marching tetrahedra.
\newblock {\em J. Computat. Chem.}, 11:1268--1277, 1998.

\bibitem{ChangHW:2013}
H.~W. Chang, S.~Bacallado, V.~S. Pande, and G.~E. Carlsson.
\newblock Persistent topology and metastable state in conformational dynamics.
\newblock {\em PLos ONE}, 8(4):e58699, 2013.

\bibitem{CCG09}
Fr{\'e}d{\'e}ric Chazal, David Cohen-Steiner, Marc Glisse, Leonidas~J. Guibas,
  and Steve Oudot.
\newblock Proximity of persistence modules and their diagrams.
\newblock In {\em Proc. 25th ACM Sympos. on Comput. Geom.}, pages 237--246,
  2009.

\bibitem{CGOS11}
Fr{\'e}d{\'e}ric Chazal, Leonidas~J. Guibas, Steve~Y. Oudot, and Primoz Skraba.
\newblock Persistence-based clustering in riemannian manifolds.
\newblock In {\em Proceedings of the 27th annual ACM symposium on Computational
  geometry}, SoCG '11, pages 97--106, 2011.

\bibitem{DuanChen:2011a}
Duan Chen, Zhan Chen, Changjun Chen, W.~H. Geng, and G.~W. Wei.
\newblock {MIBPB}: A software package for electrostatic analysis.
\newblock {\em J. Comput. Chem.}, 32:657 -- 670, 2011.

\bibitem{DuanChen:2012a}
Duan Chen, Zhan Chen, and G.~W. Wei.
\newblock Quantum dynamics in continuum for proton transport {II: Variational}
  solvent-solute interface.
\newblock {\em International Journal for Numerical Methods in Biomedical
  Engineering}, 28:25 -- 51, 2012.

\bibitem{DuanChen:2012b}
Duan Chen and G.~W. Wei.
\newblock Quantum dynamics in continuum for proton transport---{Generalized}
  correlation.
\newblock {\em J Chem. Phys.}, 136:134109, 2012.

\bibitem{DuanChen:2013}
Duan Chen and G.~W. Wei.
\newblock Quantum dynamics in continuum for proton transport {I: Basic}
  formulation.
\newblock {\em Commun. Comput. Phys.}, 13:285--324, 2013.

\bibitem{Chen:2012morse}
Guoning Chen, Qingqing Deng, Andrzej Szymczak, Robert~S Laramee, and Eugene
  Zhang.
\newblock Morse set classification and hierarchical refinement using conley
  index.
\newblock {\em IEEE transactions on visualization and computer graphics},
  18(5):767--782, 2012.

\bibitem{MXChen:2011}
Minxin Chen and Benzhuo Lu.
\newblock Tmsmesh: A robust method for molecular surface mesh generation using
  a trace technique.
\newblock {\em J Chem. Theory and Comput.}, 7:203--212, 2011.

\bibitem{MXChen:2012}
Minxin Chen, Bin Tu, and Benzhuo Lu.
\newblock Triangulated manifold meshing method preserving molecular surface
  topology.
\newblock {\em J. Mole. Graph. Model.}, 38:411--418, 2012.

\bibitem{WYChen:2010}
Wenyu Chen, Jianmin Zheng, and Yiyu Cai.
\newblock Kernel modeling for molecular surfaces using a uniform solution.
\newblock {\em Computer Aided Design}, 42:267--278, 2010.

\bibitem{ZhanChen:2010a}
Z.~Chen, N.~A. Baker, and G.~W. Wei.
\newblock Differential geometry based solvation models {I}: Eulerian
  formulation.
\newblock {\em J. Comput. Phys.}, 229:8231--8258, 2010.

\bibitem{ZhanChen:2010b}
Z.~Chen, N.~A. Baker, and G.~W. Wei.
\newblock Differential geometry based solvation models {II}: Lagrangian
  formulation.
\newblock {\em J. Math. Biol.}, 63:1139-- 1200, 2011.

\bibitem{ZhanChen:2011a}
Z.~Chen and G.~W. Wei.
\newblock Differential geometry based solvation models {III}: Quantum
  formulation.
\newblock {\em J. Chem. Phys.}, 135:194108, 2011.

\bibitem{Cheng:2007e}
L.~T. Cheng, Joachim Dzubiella, Andrew~J. McCammon, and B.~Li.
\newblock Application of the level-set method to the implicit solvation of
  nonpolar molecules.
\newblock {\em Journal of Chemical Physics}, 127(8), 2007.

\bibitem{Cheng2:2009}
Li-Tien Cheng, Yang Xie, Joachim Dzubiella, J.~Andrew McCammon, Jianwei Che,
  and Bo~Li.
\newblock {Coupling the level-set method with molecular mechanics for
  variational implicit solvation of nonpolar molecules}.
\newblock {\em J. Chem. Theory Comput.}, 5:257--266, 2009.

\bibitem{Chiti:2006}
F.~Chiti and C.~M. Dobson.
\newblock Protein misfolding, functional amyloid, and human disease.
\newblock {\em Annu. Rev. Biochem.}, 75:333 -- 366, 2006.

\bibitem{ChungOverview}
F.~Chung.
\newblock {\em Spectral graph theory}.
\newblock American Mathematical Society, 1997.

\bibitem{Cioslowski:1999}
J.~Cioslowski and G.~H. Liu.
\newblock Topology of electron-electron interactions in atoms and molecules.
  {II}. the correlation cage.
\newblock {\em The Journal of chemical physics}, 110(4):1882--1887, 1999.

\bibitem{CEH07}
David Cohen-Steiner, Herbert Edelsbrunner, and John Harer.
\newblock Stability of persistence diagrams.
\newblock {\em Discrete {\&} Computational Geometry}, 37(1):103--120, 2007.

\bibitem{CEH09}
David Cohen-Steiner, Herbert Edelsbrunner, and John Harer.
\newblock Extending persistence using poincar{\'e} and lefschetz duality.
\newblock {\em Foundations of Computational Mathematics}, 9(1):79--103, 2009.

\bibitem{CEHM09}
David Cohen-Steiner, Herbert Edelsbrunner, John Harer, and Dmitriy Morozov.
\newblock Persistent homology for kernels, images, and cokernels.
\newblock In {\em Proceedings of the Twentieth Annual ACM-SIAM Symposium on
  Discrete Algorithms}, SODA 09, pages 1011--1020, 2009.

\bibitem{Conley:1978}
Charles Conley.
\newblock {\em Isolated invariant sets and the Morse index. CBMS Regional
  Conference Series in Mathematics, 38}.
\newblock American Mathematical Society,Providence, R.I, 1978.

\bibitem{Connolly85}
M.~L. Connolly.
\newblock Depth buffer algorithms for molecular modeling.
\newblock {\em J. Mol. Graphics}, 3:19--24, 1985.

\bibitem{contreras2011nciplot}
Julia Contreras-Garc{\'\i}a, Erin~R Johnson, Shahar Keinan, Robin Chaudret,
  Jean-Philip Piquemal, David~N Beratan, and Weitao Yang.
\newblock Nciplot: a program for plotting noncovalent interaction regions.
\newblock {\em Journal of chemical theory and computation}, 7(3):625--632,
  2011.

\bibitem{contreras2011analysis}
Julia Contreras-Garc{\'\i}a, Weitao Yang, and Erin~R Johnson.
\newblock Analysis of hydrogen-bond interaction potentials from the electron
  density: integration of noncovalent interaction regions.
\newblock {\em The Journal of Physical Chemistry A}, 115(45):12983--12990,
  2011.

\bibitem{crenshaw1993orientation}
Hugh~C Crenshaw and Leah Edelstein-Keshet.
\newblock Orientation by helical motion—ii. changing the direction of the
  axis of motion.
\newblock {\em Bulletin of mathematical biology}, 55(1):213--230, 1993.

\bibitem{Crowley}
P.B. Crowley and A.~Golovin.
\newblock Cation{-}pi interactions in protein-protein interfaces.
\newblock {\em Proteins {-} Struct. Func. Bioinf.}, 59:231--239, 2005.

\bibitem{Cui:2002}
Q.~Cui.
\newblock Combining implicit solvation models with hybrid quantum
  mechanical/molecular mechanical methods: A critical test with glycine.
\newblock {\em Journal of Chemical Physics}, 117(10):4720, 2002.

\bibitem{QCui:2010}
Q.~Cui and I.~Bahar.
\newblock {\em Normal mode analysis: theory and applications to biological and
  chemical systems}.
\newblock Chapman and Hall/CRC, 2010.

\bibitem{QCui:2004}
Q.~Cui, G.~J. Li, J.~Ma, and M.~Karplus.
\newblock A normal mode analysis of structural plasticity in the biomolecular
  motor f(1)-atpase.
\newblock {\em J. Mol. Biol.}, 340(2):345 -- 372, 2004.

\bibitem{Dabaghian:2012}
Y.~Dabaghian, F.~Memoli, L.~Frank, and G.~Carlsson.
\newblock A topological paradigm for hippocampal spatial map formation using
  persistent homology.
\newblock {\em PLoS Comput Biol}, 8(8):e1002581, 08 2012.

\bibitem{darcy:2001}
I.~K. Darcy.
\newblock Biological distances on {DNA} knots and links: applications to {XER}
  recombination.
\newblock {\em Journal of Knot Theory and its Ramifications}, 10(02):269--294,
  2001.

\bibitem{IKDarcy:2013}
I.~K. Darcy and M.~Vazquez.
\newblock Determining the topology of stable { protein-DNA} complexes.
\newblock {\em Biochemical Society Transactions}, 41:601--605, 2013.

\bibitem{Darnell:2008}
S.~J. Darnell, L.~LeGault, and J.~C. Mitchell.
\newblock {KFC} server: interactive forecasting of protein interaction hot
  spots.
\newblock {\em Nucleic Acids Research}, 36:W265--W269, 2008.

\bibitem{DasGupta2016}
Bhaskar DasGupta and Jie Liang.
\newblock {\em Models and Algorithms for Biomolecules and Molecular Networks}.
\newblock John Wiley \& Sons, 2016.

\bibitem{SMV11}
Vin de~Silva, Dmitriy Morozov, and Mikael Vejdemo-Johansson.
\newblock Persistent cohomology and circular coordinates.
\newblock {\em Discrete and Comput. Geom.}, 45:737--759, 2011.

\bibitem{Decherchi:2013}
S.~Decherchi and W.~Rocchia.
\newblock {A general and robust ray-casting-based algorithm for triangulating
  surfaces at the nanoscale}.
\newblock {\em PLoS ONE}, 8:e59744, 2013.

\bibitem{Demerdash:2009}
O.~N.~A. Demerdash, M.~D. Daily, and J.~C. Mitchell.
\newblock Structure-based predictive models for allosteric hot spots.
\newblock {\em PLOS Computational Biology}, 5:e1000531, 2009.

\bibitem{Demerdash:2012}
Omar N.~A. Demerdash and Julie~C. Mitchell.
\newblock {Density-cluster NMA: A new protein decomposition technique for
  coarse-grained normal mode analysis}.
\newblock {\em {Proteins:Structure Function and Bioinformatics}},
  {80}({7}):{1766--1779}, {JUL} {2012}.

\bibitem{Dey:2008}
T.~K. Dey, K.~Y. Li, J.~Sun, and C.~S. David.
\newblock Computing geometry aware handle and tunnel loops in 3d models.
\newblock {\em ACM Trans. Graph.}, 27, 2008.

\bibitem{DFW14}
Tamal~K Dey, Fengtao Fan, and Yusu Wang.
\newblock Computing topological persistence for simplicial maps.
\newblock In {\em Proc. 30th Annu. Sympos. Comput. Geom. (SoCG)}, pages
  345--354, 2014.

\bibitem{Dey:2013}
Tamal~K. Dey and Y.~S. Wang.
\newblock Reeb graphs: Approximation and persistence.
\newblock {\em Discrete and Computational Geometry}, 49(1):46--73, 2013.

\bibitem{DiFabio:2011}
Barbara Di~Fabio and Claudia Landi.
\newblock A mayer-vietoris formula for persistent homology with an application
  to shape recognition in the presence of occlusions.
\newblock {\em Foundations of Computational Mathematics}, 11:499--527, 2011.

\bibitem{Ding:2001min}
C.~H.~Q. Ding, X.~F. He, H.~Y. Zha, M.~Gu, and H.~D. Simon.
\newblock A min-max cut algorithm for graph partitioning and data clustering.
\newblock In {\em Data Mining, 2001. ICDM 2001, Proceedings IEEE International
  Conference on}, pages 107--114. IEEE, 2001.

\bibitem{Dragan}
A.I. Dragan, C.M. Read, E.N. Makeyeva, E.I. Milgotina, M.E.A. Churchill,
  C.~Crane{-}Robinson, and P.L. Privalov.
\newblock Dna binding and bending by hmg boxes: Energetic determinants of
  specificity.
\newblock {\em J. Mol. Biol.}, 343:371--393, 2004.

\bibitem{Duncan:1993}
B.~S. Duncan and A.~J. Olson.
\newblock Shape analysis of molecular surfaces.
\newblock {\em Biopolymers}, 33:231--238, 1993.

\bibitem{Edelsbrunner:2008persistent}
H.~Edelsbrunner and J.~Harer.
\newblock Persistent homology-a survey.
\newblock {\em Contemporary mathematics}, 453:257--282, 2008.

\bibitem{Edelsbrunner:2002}
H.~Edelsbrunner, D.~Letscher, and A.~Zomorodian.
\newblock Topological persistence and simplification.
\newblock {\em Discrete Comput. Geom.}, 28:511--533, 2002.

\bibitem{Edelsbrunner:1994}
H.~Edelsbrunner and E.~P. Mucke.
\newblock Three-dimensional alpha shapes.
\newblock {\em Physical Review Letters}, 13:43--72, 1994.

\bibitem{edelsbrunner:2010}
Herbert Edelsbrunner and John Harer.
\newblock {\em Computational topology: an introduction}.
\newblock American Mathematical Soc., 2010.

\bibitem{coarea}
H.~Federer.
\newblock {Curvature Measures}.
\newblock {\em Trans. Amer. Math. Soc.}, 93:418--491, 1959.

\bibitem{XFeng:2013b}
X.~Feng, K.~L. Xia, Y.~Y. Tong, and G.~W. Wei.
\newblock Multiscale geometric modeling of macromolecules {II:} lagrangian
  representation.
\newblock {\em Journal of Computational Chemistry}, 34:2100--2120, 2013.

\bibitem{XFeng:2012a}
Xin Feng, Kelin Xia, Yiying Tong, and Guo-Wei Wei.
\newblock Geometric modeling of subcellular structures, organelles and large
  multiprotein complexes.
\newblock {\em International Journal for Numerical Methods in Biomedical
  Engineering}, 28:1198--1223, 2012.

\bibitem{Fera:2004rag}
D.~Fera, N.~Kim, N.~Shiffeldrim, J.~Zorn, U.~Laserson, H.~H. Gan, and
  T.~Schlick.
\newblock {RAG}: {RNA}-as-graphs web resource.
\newblock {\em BMC bioinformatics}, 5(1):1, 2004.

\bibitem{Flory:1976}
P.~J. Flory.
\newblock Statistical thermodynamics of random networks.
\newblock {\em Proc. Roy. Soc. Lond. A,}, 351:351 -- 378, 1976.

\bibitem{Fortunato:2010}
S.~Fortunato.
\newblock Community detection in graphs.
\newblock {\em Physics reports}, 486(3):75--174, 2010.

\bibitem{Fro90}
Patrizio Frosini.
\newblock A distance for similarity classes of submanifolds of a {Euclidean}
  space.
\newblock {\em BUllentin of Australian Mathematical Society}, 42(3):407--416,
  1990.

\bibitem{Frosini:1999}
Patrizio Frosini and Claudia Landi.
\newblock Size theory as a topological tool for computer vision.
\newblock {\em Pattern Recognition and Image Analysis}, 9(4):596--603, 1999.

\bibitem{Frosini:2013}
Patrizio Frosini and Claudia Landi.
\newblock Persistent betti numbers for a noise tolerant shape-based approach to
  image retrieval.
\newblock {\em Pattern Recognition Letters}, 34:863--872, 2013.

\bibitem{Fujishiro:2000}
Issei Fujishiro, Yuriko Takeshima, Taeko Azuma, and Shigeo Takahashi.
\newblock Volume data mining using 3d field topology analysis.
\newblock {\em IEEE Computer Graphics and Applications}, 20(5):46--51, 2000.

\bibitem{fuller:1971}
F.~B. Fuller.
\newblock The writhing number of a space curve.
\newblock {\em Proceedings of the National Academy of Sciences},
  68(4):815--819, 1971.

\bibitem{Gameiro:2014}
M.~Gameiro, Y.~Hiraoka, S.~Izumi, M.~Kramar, K.~Mischaikow, and V.~Nanda.
\newblock Topological measurement of protein compressibility via persistence
  diagrams.
\newblock {\em Japan Journal of Industrial and Applied Mathematics}, 32:1--17,
  2014.

\bibitem{Gan:2004rag}
H.~H. Gan, D.~Fera, J.~Zorn, N.~Shiffeldrim, M.~Tang, U.~Laserson, N.~Kim, and
  T.~Schlick.
\newblock {RAG}: {RNA}-as-graphs database—concepts, analysis, and features.
\newblock {\em Bioinformatics}, 20(8):1285--1291, 2004.

\bibitem{Garcia:1992large}
A.~E. Garc{\'\i}a.
\newblock Large-amplitude nonlinear motions in proteins.
\newblock {\em Physical review letters}, 68(17):2696, 1992.

\bibitem{Geng:2011}
W.~Geng and G.~W. Wei.
\newblock Multiscale molecular dynamics using the matched interface and
  boundary method.
\newblock {\em J Comput. Phys.}, 230(2):435--457, 2011.

\bibitem{Geng:2007a}
Weihua Geng, Sining Yu, and G.~W. Wei.
\newblock Treatment of charge singularities in implicit solvent models.
\newblock {\em Journal of Chemical Physics}, 127:114106, 2007.

\bibitem{Gerek:2010}
Z.~Nevin Gerek and S.~Banu Ozkan.
\newblock A flexible docking scheme to explore the binding selectivity of pdz
  domains.
\newblock {\em Protein Science}, 19:914--928, 2010.

\bibitem{Ghrist:2008}
R.~Ghrist.
\newblock Barcodes: {The} persistent topology of data.
\newblock {\em Bull. Amer. Math. Soc.}, 45:61--75, 2008.

\bibitem{Gillet:2012}
N.~Gillet, R.~Chaudret, J.~Contreras-Garc{\i}́a, W.~T. Yang, B.~Silvi, and
  J.~P. Piquemal.
\newblock Coupling quantum interpretative techniques: another look at chemical
  mechanisms in organic reactions.
\newblock {\em Journal of chemical theory and computation}, 8(11):3993--3997,
  2012.

\bibitem{Gine:2006empirical}
E.~Gin{\'e} and V.~Koltchinskii.
\newblock Empirical graph {Laplacian} approximation of {Laplace Beltrami}
  operators: Large sample results.
\newblock In {\em High dimensional probability}, pages 238--259. Institute of
  Mathematical Statistics, 2006.

\bibitem{giot:2003}
L.~Giot, J.~S. Bader, C.~Brouwer, A.~Chaudhuri, B.~Kuang, Y.~Li, Y.L. Hao, C.E.
  Ooi, B.~Godwin, E.~Vitols, et~al.
\newblock A protein interaction map of drosophila melanogaster.
\newblock {\em science}, 302(5651):1727--1736, 2003.

\bibitem{Go:1983}
N.~Go, T.~Noguti, and T.~Nishikawa.
\newblock Dynamics of a small globular protein in terms of low-frequency
  vibrational modes.
\newblock {\em Proc. Natl. Acad. Sci.}, 80:3696 -- 3700, 1983.

\bibitem{Grant:1995}
J.~A. Grant and B.~T. Pickup.
\newblock A gaussian description of molecular shape.
\newblock {\em Journal of Physical Chemistry}, 99:3503--3510, 1995.

\bibitem{Grant:2007}
J.~A. Grant, B.~T. Pickup, M.~T. Sykes, C.~A. Kitchen, and A.~Nicholls.
\newblock The {Gaussian Generalized Born} model: application to small
  molecules.
\newblock {\em Physical Chemistry Chemical Physics}, 9:4913--22, 2007.

\bibitem{Zap}
J.~Andrew Grant, Barry~T. Pickup, and Anthony Nicholls.
\newblock A smooth permittivity function for {{Poisson-Boltzmann}} solvation
  methods.
\newblock {\em Journal of Computational Chemistry}, 22(6):608--640, 2001.

\bibitem{Gray:1984vector}
R.~Gray.
\newblock Vector quantization.
\newblock {\em IEEE Assp Magazine}, 1(2):4--29, 1984.

\bibitem{Gunther:2014}
D.~G{\"u}nther, A.~Jacobson, J.~Reininghaus, H.~P. Seidel, O.~Sorkine-Hornung,
  and T.~Weinkauf.
\newblock Fast and memory-efficient topological denoising of {2D and 3D} scalar
  fields.
\newblock {\em IEEE Transactions on Visualization and Computer Graphics},
  20:12, 2014.

\bibitem{Guo:2003}
J.~T. Guo, D.~Xu, D.~Kim, and Y.~Xu.
\newblock Improving the performance of {DomainParser} for structural domain
  partition using neural network.
\newblock {\em Nucleic Acids Research}, 31(3):944--952, 2003.

\bibitem{Hagen:1992new}
L.~Hagen and A.~B. Kahng.
\newblock New spectral methods for ratio cut partitioning and clustering.
\newblock {\em IEEE transactions on computer-aided design of integrated
  circuits and systems}, 11(9):1074--1085, 1992.

\bibitem{Halle:2002}
B.~Halle.
\newblock Flexibility and packing in proteins.
\newblock {\em PNAS}, 99:1274--1279, 2002.

\bibitem{harker:mischaikow:mrozek:nanda}
Shaun Harker, Konstantin Mischaikow, Marian Mrozek, and Vidit Nanda.
\newblock Discrete morse theoretic algorithms for computing homology of
  complexes and maps.
\newblock {\em Foundations of Computational Mathematics}, pages 1--34, 2013.

\bibitem{SHarvey:2013}
S.~C. Harvey, Y.~Zeng, and C.~E. Heitsch.
\newblock The icosahedral rna virus as a grotto: organizing the genome into
  stalagmites and stalactite.
\newblock {\em J Biol Phys}, Chapter 7:163--172, 2013.

\bibitem{Hayward:2008normal}
S.~Hayward and B.~L. De~Groot.
\newblock Normal modes and essential dynamics.
\newblock {\em Molecular Modeling of Proteins}, pages 89--106, 2008.

\bibitem{Hein:2005graphs}
M.~Hein, J.~Y. Audibert, and U.~Von~Luxburg.
\newblock From graphs to manifolds--weak and strong pointwise consistency of
  graph {Laplacian}.
\newblock In {\em International Conference on Computational Learning Theory},
  pages 470--485. Springer, 2005.

\bibitem{CHeitsch:2014}
C.~Heitsch and S.~Poznanovic.
\newblock Combinatorial insights into rna secondary structure, in { N. Jonoska
  and M. Saito}, editors.
\newblock {\em Discrete and Topological Models in Molecular Biology}, Chapter
  7:145--166, 2014.

\bibitem{Henkelman:2006}
G.~Henkelman, A.~Arnaldsson, and H.~J{\'o}nsson.
\newblock A fast and robust algorithm for {Bader} decomposition of charge
  density.
\newblock {\em Computational Materials Science}, 36(3):354--360, 2006.

\bibitem{Hinsen:1998}
K.~Hinsen.
\newblock Analysis of domain motions by approximate normal mode calculations.
\newblock {\em Proteins}, 33:417 -- 429, 1998.

\bibitem{Hinsen:2008}
K.~Hinsen.
\newblock Structural flexibility in proteins: impact of the crystal
  environment.
\newblock {\em Bioinformatics}, 24:521 -- 528, 2008.

\bibitem{holm:1996}
L.~Holm and C.~Sander.
\newblock Mapping the protein universe.
\newblock {\em Science}, 273(5275):595, 1996.

\bibitem{Holst:1994}
Michael Holst.
\newblock {\em The {Poisson-Boltzmann} equation: Analysis and multilevel
  numerical solution.}
\newblock PhD thesis, California Institute of Technology, 1994.

\bibitem{Horak:2009}
D.~Horak, S~Maletic, and M.~Rajkovic.
\newblock Persistent homology of complex networks.
\newblock {\em Journal of Statistical Mechanics: Theory and Experiment},
  2009(03):P03034, 2009.

\bibitem{GHu:1998}
G.~Hu, J.~H. Yang, and W.~J. Liu.
\newblock Instability and controllability of linearly coupled oscillators:
  Eigenvalue analysis.
\newblock {\em Phys. Rev. E}, 58:4440-- 4453, 1998.

\bibitem{Hu:2013method}
H.~Y. Hu, T.~Laurent, M.~A. Porter, and A.~L. Bertozzi.
\newblock A method based on total variation for network modularity optimization
  using the {MBO} scheme.
\newblock {\em SIAM Journal on Applied Mathematics}, 73(6):2224--2246, 2013.

\bibitem{hu2011discrete}
Shuangwei Hu, Martin Lundgren, and Antti~J Niemi.
\newblock Discrete frenet frame, inflection point solitons, and curve
  visualization with applications to folded proteins.
\newblock {\em Physical Review E}, 83(6):061908, 2011.

\bibitem{Hyon:2010}
YunKyong Hyon, Bob Eisenberg, and Chun Liu.
\newblock {A mathematical model of the hard sphere repulsion in ionic
  solutions}.
\newblock {\em Communications in Mathematical Sciences}, 9:459--475, 2010.

\bibitem{Jackson}
R.M. Jackson and M.J. Sternberg.
\newblock Dna binding and bending by hmg boxes: Energetic determinants of
  specificity.
\newblock {\em J. Mol. Biol.}, 250:258--275, 1995.

\bibitem{Jain:2010}
A.~K. Jain.
\newblock Data clustering: 50 years beyond k-means.
\newblock {\em Pattern recognition letters}, 31(8):651--666, 2010.

\bibitem{jeong:2000}
H.~Jeong, B.~Tombor, R.~Albert, Z.~N. Oltvai, and A.~L. Barab{\'a}si.
\newblock The large-scale organization of metabolic networks.
\newblock {\em Nature}, 407(6804):651--654, 2000.

\bibitem{Johnson:2010}
E.~R. Johnson, S.~Keinan, P.~Mori-Sanchez, J.~Contreras-Garcia, A.~J. Cohen,
  and W.T. Yang.
\newblock Revealing noncovalent interactions.
\newblock {\em Journal of the American Chemical Society}, 132(18):6498--6506,
  2010.

\bibitem{Jonoska:2009}
N.~Jonoska and G.~McColm.
\newblock Complexity classes for self-assembling flexible tiles.
\newblock {\em Theoretical Computer Science}, 410:332--346, 2009.

\bibitem{Kaczynski:2004}
T.~Kaczynski, K.~Mischaikow, and M.~Mrozek.
\newblock {\em Computational homology}.
\newblock Springer-Verlag, 2004.

\bibitem{kaczynski:mischaikow:mrozek:04}
Tomasz Kaczynski, Konstantin Mischaikow, and Marian Mrozek.
\newblock {\em Computational Homology}, volume 157 of {\em Applied Mathematical
  Sciences}.
\newblock Springer-Verlag, New York, 2004.

\bibitem{Kasson:2007}
P.~M. Kasson, A.~Zomorodian, S.~Park, N.~Singhal, L.~J. Guibas, and V.~S.
  Pande.
\newblock Persistent voids a new structural metric for membrane fusion.
\newblock {\em Bioinformatics}, 23:1753--1759, 2007.

\bibitem{Keith:1993}
T.~A. Keith and R.~F. Bader.
\newblock Topological analysis of magnetically induced molecular current
  distributions.
\newblock {\em The Journal of chemical physics}, 99(5):3669--3682, 1993.

\bibitem{Keith:1996}
T.~A. Keith, R.~F. Bader, and Y.~Aray.
\newblock Structural homeomorphism between the electron density and the virial
  field.
\newblock {\em International journal of quantum chemistry}, 57(2):183--198,
  1996.

\bibitem{Keskin:2002}
O.~Keskin, I.~Bahar, D.~Flatow, D.~G. Covell, and R.~L. Jernigan.
\newblock Molecular mechanisms of chaperonin groel-groes function.
\newblock {\em Biochem.}, 41:491 -- 501, 2002.

\bibitem{Kim2004:candidates}
N.~Kim, N.~Shiffeldrim, H.~H. Gan, and T.~Schlick.
\newblock Candidates for novel {RNA} topologies.
\newblock {\em Journal of molecular biology}, 341(5):1129--1144, 2004.

\bibitem{Kindlmann:2003}
G.~Kindlmann, R.~Whitaker, T.~Tasdizen, and T.~M{\"o}ller.
\newblock Curvature-based transfer functions for direct volume rendering:
  methods and applications.
\newblock {\em Proc. IEEE Visualization}, 2003.

\bibitem{Kitao:1991effects}
A.~Kitao, F.~Hirata, and N.~G{\=o}.
\newblock The effects of solvent on the conformation and the collective motions
  of protein: normal mode analysis and molecular dynamics simulations of
  melittin in water and in vacuum.
\newblock {\em Chemical physics}, 158(2-3):447--472, 1991.

\bibitem{Kohout:2004}
M.~Kohout, K.~Pernal, F.~R. Wagner, and Y.~Grin.
\newblock Electron localizability indicator for correlated wavefunctions. {I.
  Parallel-spin} pairs.
\newblock {\em Theoretical Chemistry Accounts}, 112(5-6):453--459, 2004.

\bibitem{Koltun:1965}
W.~L. Koltun.
\newblock Precision space-filling atomic models.
\newblock {\em Biopolymers}, 3:667--679, 1965.

\bibitem{Kondrashov:2007}
D.~A. Kondrashov, A.~W. Van~Wynsberghe, R.~M. Bannen, Q.~Cui, and Jr. G.~N.
  Phillips.
\newblock Protein structural variation in computational models and
  crystallographic data.
\newblock {\em Structure}, 15:169 -- 177, 2007.

\bibitem{Kovacev-Nikolic:2016}
Violeta Kovacev-Nikolic, Peter Bubenik, Dragan Nikoli\'c, and Giseon Heo.
\newblock Using persistent homology and dynamical distances to analyze protein
  binding.
\newblock {\em Stat. Appl. Genet. Mol. Biol.}, 15(1):19--38, 2016.

\bibitem{Krishnamoorthy:2007}
Bala Krishnamoorthy, Scott Provan, and Alexander Tropsha.
\newblock A topological characterization of protein structure.
\newblock In {\em Data Mining in Biomedicine, Springer Optimization and Its
  Applications}, pages 431--455, 2007.

\bibitem{Kuhn}
L.A. Kuhn, M.~A. Siani, M.~E. Pique, C.~L. Fisher, E.~D. Getzoff, and J.~A.
  Tainer.
\newblock The interdependence of protein surface topography and bound water
  molecules revealed by surface accessibility and fractal density measures.
\newblock {\em J. Mol. Biol.}, 228:13--22, 1992.

\bibitem{Kuhnel:2015}
W.~K{\"u}hnel.
\newblock {\em Differential Geometry: Curves-Surfaces-Manifolds, Student Math}.
\newblock American mathematical Society, 2015.

\bibitem{Kundu:2002}
S.~Kundu, J.~S. Melton, D.~C. Sorensen, and Jr. G.~N. Phillips.
\newblock Dynamics of proteins in crystals: comparison of experiment with
  simple models.
\newblock {\em Biophys. J.}, 83:723 -- 732, 2002.

\bibitem{SKundu:2004}
S.~Kundu, D.~C. Sorensen, and G.~N. Phillips.
\newblock Automatic domain decomposition of proteins by a {Gaussian} network
  model.
\newblock {\em Proteins: Structure, Function, and Bioinformatics},
  57(4):725--733, 2004.

\bibitem{LLi:2013}
Z.~Zhang L.~Li, C.~Li and Emil Alexov.
\newblock On the dielectric "constant'' of proteins: Smooth dielectric function
  for macromolecular modeling and its implementation in {DelPhi}.
\newblock {\em J. Chem. Theory Comput.}, 9:2126--2136, 2013.

\bibitem{Lafon:2004}
S.~S. Lafon.
\newblock {\em Diffusion maps and geometric harmonics}.
\newblock PhD thesis, Yale University, 2004.

\bibitem{Leboeuf:1999}
M.~Leboeuf, A.~M. K{\"o}ster, K.~Jug, and D.~R. Salahub.
\newblock Topological analysis of the molecular electrostatic potential.
\newblock {\em The Journal of chemical physics}, 111(11):4893--4905, 1999.

\bibitem{Lee:1971}
B.~Lee and F.~M. Richards.
\newblock The interpretation of protein structures: estimation of static
  accessibility.
\newblock {\em J Mol Biol}, 55(3):379--400, 1971.

\bibitem{LeeH:2012}
H~Lee, H.~Kang, M.~K. Chung, B.~Kim, and D.~S. Lee.
\newblock Persistent brain network homology from the perspective of dendrogram.
\newblock {\em Medical Imaging, IEEE Transactions on}, 31(12):2267--2277, Dec
  2012.

\bibitem{lee:2002}
T.~I. Lee, N.~J. Rinaldi, F.~Robert, D.~T. Odom, Z.~Bar-Joseph, G.~K. Gerber,
  N.~M. Hannett, C.~T. Harbison, C.~M. Thompson, I.~Simon, et~al.
\newblock Transcriptional regulatory networks in saccharomyces cerevisiae.
\newblock {\em science}, 298(5594):799--804, 2002.

\bibitem{Levitt:1983normal}
M.~Levitt, C.~Sander, and P.~S. Stern.
\newblock The normal modes of a protein: Native bovine pancreatic trypsin
  inhibitor.
\newblock {\em International Journal of Quantum Chemistry}, 24(S10):181--199,
  1983.

\bibitem{Levitt:1985}
M.~Levitt, C.~Sander, and P.~S. Stern.
\newblock Protein normal-mode dynamics: Trypsin inhibitor, crambin,
  ribonuclease and lysozyme.
\newblock {\em J. Mol. Biol.}, 181(3):423 -- 447, 1985.

\bibitem{DWLi:2009}
D.~W. Li and R.~Br{\"u}schweiler.
\newblock All-atom contact model for understanding protein dynamics from
  crystallographic b-factors.
\newblock {\em Biophysical journal}, 96(8):3074--3081, 2009.

\bibitem{LiGH:2002}
G.~H. Li and Q.~Cui.
\newblock A coarse-grained normal mode approach for macromolecules: an
  efficient implementation and application to {Ca(2+)-ATPase}.
\newblock {\em Bipohys. J.}, 83:2457 -- 2474, 2002.

\bibitem{JLi:2013}
J.~Li, P.~Mach, and P.~Koehl.
\newblock Measuring the shapes of macromolecules - and why it matters.
\newblock {\em Comput Struct Biotechnol J.}, 8:e201309001, 2013.

\bibitem{LiLin:2014}
Lin Li, Chuan Li, and Emil Alexov.
\newblock On the modeling of polar component of solvation energy using smooth
  gaussian-based dielectric function.
\newblock {\em Journal of Theoretical and Computational Chemistry},
  13:10.1142/S0219633614400021, 2014.

\bibitem{LiCata}
V.J. LiCata and N.M. Allewell.
\newblock Functionally linked hydration changes in escherichia coli aspartate
  transcarbamylase and its catalytic subunit.
\newblock {\em Biochemistry}, 36:10161--10167, 1997.

\bibitem{CPLin:2008}
C.~P. Lin, S.~W. Huang, Y.~L. Lai, S.~C. Yen, C.~H. Shih, C.~H. Lu, C.~C.
  Huang, and J.~K. Hwang.
\newblock Deriving protein dynamical properties from weighted protein contact
  number.
\newblock {\em Proteins: Structure, Function, and Bioinformatics},
  72(3):929--935, 2008.

\bibitem{ESES:2015}
Beibei Liu, Bao Wang, Rundong Zhao, Yiying Tong, and Guo~Wei Wei.
\newblock {ESES: software for Eulerian solvent excluded surface}.
\newblock {\em {Preprint}}, 2015.

\bibitem{XuLiu:2012}
Xu~Liu, Zheng Xie, and Dongyun Yi.
\newblock A fast algorithm for constructing topological structure in large
  data.
\newblock {\em Homology, Homotopy and Applications}, 14:221--238, 2012.

\bibitem{Livesay:2004}
D.~R. Livesay, S.~Dallakyan, G.~G. Wood, and D.~J. Jacobs.
\newblock {A flexible approach for understanding protein stability}.
\newblock {\em {FEBS Letters}}, {576}:{468--476}, {2004}.

\bibitem{Lopez:2014normal}
J.~R. L{\'o}pez-Blanco, O.~Miyashita, F.~Tama, and P.~Chac{\'o}n.
\newblock Normal mode analysis techniques in structural biology.
\newblock {\em eLS}, 2014.

\bibitem{Lorenz:1963}
E.~N. Lorenz.
\newblock Deterministic nonperiodic flow.
\newblock {\em Journal of the Atmospheric Sciences}, 20:130-- 141, 1963.

\bibitem{Lovasz:1993}
L.~Lov{\'a}sz.
\newblock Random walks on graphs.
\newblock {\em Combinatorics, Paul erdos is eighty}, 2:1--46, 1993.

\bibitem{JMa:2005}
J.~P. Ma.
\newblock Usefulness and limitations of normal mode analysis in modeling
  dynamics of biomolecular complexes.
\newblock {\em Structure}, 13:373 -- 180, 2005.

\bibitem{PMach:2011}
P.~Mach and P.~Koehl.
\newblock Geometric measures of large biomolecules: Surface, volume, and
  pockets.
\newblock {\em J. Comp. Chem.}, 32:3023--3038, 2011.

\bibitem{Manolescu:2013conley}
Ciprian Manolescu.
\newblock The conley index, gauge theory, and triangulations.
\newblock {\em Journal of Fixed Point Theory and Applications}, 13(2):431--457,
  2013.

\bibitem{McCammon:1977}
J.~A. McCammon, B.~R. Gelin, and M.~Karplus.
\newblock Dynamics of folded proteins.
\newblock {\em Nature}, 267:585--590, 1977.

\bibitem{meila:2001}
M.~Meila and J.~B. Shi.
\newblock A random walks view of spectral segmentation.
\newblock In {\em In Tenth International Workshop on Artificial Intelligence
  and Statistics {AISTATS}}, 2001.

\bibitem{Merkurjev:2013}
E.~Merkurjev, T.~Kostic, and A.~L. Bertozzi.
\newblock An {MBO} scheme on graphs for classification and image processing.
\newblock {\em SIAM Journal on Imaging Sciences}, 6(4):1903--1930, 2013.

\bibitem{Mezey:1981}
P.~G. Mezey.
\newblock Catchment region partitioning of energy hypersurfaces,{I}.
\newblock {\em Theoretica chimica acta}, 58(4):309--330, 1981.

\bibitem{Ming:2002describe}
D.~M. Ming, Y.~F. Kong, M.~A. Lambert, Z.~Huang, and J.~P. Ma.
\newblock How to describe protein motion without amino acid sequence and atomic
  coordinates.
\newblock {\em Proceedings of the National Academy of Sciences},
  99(13):8620--8625, 2002.

\bibitem{Mischaikow:1999}
K.~Mischaikow, M~Mrozek, J.~Reiss, and A.~Szymczak.
\newblock Construction of symbolic dynamics from experimental time series.
\newblock {\em Physical Review Letters}, 82:1144--1147, 1999.

\bibitem{Mischaikow:2013}
K.~Mischaikow and V.~Nanda.
\newblock Morse theory for filtrations and efficient computation of persistent
  homology.
\newblock {\em Discrete and Computational Geometry}, 50(2):330--353, 2013.

\bibitem{Mischaikow:2002}
Konstantin Mischaikow and Marian Mrozek.
\newblock {\em Conley index. Chapter 9 in Handbook of Dynamical Systems, vol 2,
  pp 393-460}.
\newblock Elsevier, 2002.

\bibitem{Mohar:1997some}
B.~Mohar.
\newblock Some applications of {Laplace} eigenvalues of graphs.
\newblock In {\em Graph symmetry}, pages 225--275. Springer, 1997.

\bibitem{Mohar:1991laplacian}
B.~Mohar, Y.~Alavi, G.~Chartrand, and O.~R. Oellermann.
\newblock The laplacian spectrum of graphs.
\newblock {\em Graph theory, combinatorics, and applications}, 2(871-898):12,
  1991.

\bibitem{murzin:1995}
A.~G. Murzin, S.~E. Brenner, T.~Hubbard, and C.~Chothia.
\newblock {SCOP}: a structural classification of proteins database for the
  investigation of sequences and structures.
\newblock {\em Journal of molecular biology}, 247(4):536--540, 1995.

\bibitem{Perseus}
Vidit Nanda.
\newblock Perseus: the persistent homology software.
\newblock Software available at \url{http://www.sas.upenn.edu/~vnanda/perseus}.

\bibitem{NKWH07}
V.~Natarajan, P.~Koehl, Y.~Wang, and B.~Hamann.
\newblock Visual analysis of biomolecular surfaces.
\newblock In L.~Linsen, H.~Hagen, and B.~Hamann, editors, {\em Mathematical
  Methods for Visualization in Medicine and Life Science}, pages 237--256.
  Springer Verlag, 2008.

\bibitem{Newman:2006}
M.~E.~J. Newman.
\newblock Modularity and community structure in networks.
\newblock {\em Proceedings of the national academy of sciences},
  103(23):8577--8582, 2006.

\bibitem{Newman:2004}
M.~E.~J. Newman and M.~Girvan.
\newblock Finding and evaluating community structure in networks.
\newblock {\em Physical review E}, 69(2):026113, 2004.

\bibitem{ng:2002}
A.~Y. Ng, M.~I. Jordan, and Y.~Weiss.
\newblock On spectral clustering: {Analysis} and an algorithm.
\newblock {\em Advances in neural information processing systems}, 2:849--856,
  2002.

\bibitem{DDNguyen:2016c}
Duc~D Nguyen and G.~W. Wei.
\newblock The impact of surface area, volume, curvature and lennard-jones
  potential to solvation modeling.
\newblock {\em Journal of Computational Chemistry}, submitted 2016.

\bibitem{DDNguyen:2016b}
Duc~D Nguyen, K.~L. Xia, and G.~W. Wei.
\newblock Generalized flexibility-rigidity index.
\newblock {\em Journal of Chemical Physics}, 144:234106, 2016.

\bibitem{Niyogi:2011}
P.~Niyogi, S.~Smale, and S.~Weinberger.
\newblock A topological view of unsupervised learning from noisy data.
\newblock {\em SIAM Journal on Computing}, 40:646--663, 2011.

\bibitem{Onuchic:1997}
J.~N. Onuchic, Z.~Luthey-Schulten, and P.~G. Wolynes.
\newblock Theory of protein folding: The energy landscape perspective.
\newblock {\em Annu. Rev. Phys. Chem}, 48:545--600, 1997.

\bibitem{Opron:2014}
K.~Opron, K.~L. Xia, and G.~W. Wei.
\newblock Fast and anisotropic flexibility-rigidity index for protein
  flexibility and fluctuation analysis.
\newblock {\em Journal of Chemical Physics}, 140:234105, 2014.

\bibitem{Opron:2016a}
Kristopher Opron, K.~L. Xia, Z.~Burton, and G.~W. Wei.
\newblock Flexibility-rigidity index for protein-nucleic acid flexibility and
  fluctuation analysis.
\newblock {\em Journal of Computational Chemistry}, 37:1283--1295, 2016.

\bibitem{Opron:2015a}
Kristopher Opron, K.~L. Xia, and G.~W. Wei.
\newblock Communication: Capturing protein multiscale thermal fluctuations.
\newblock {\em Journal of Chemical Physics}, 142(211101), 2015.

\bibitem{orengo:1997}
C.~A. Orengo, A.~D. Michie, S.~Jones, D.~T. Jones, M.~B. Swindells, and J.~M.
  Thornton.
\newblock {CATH}--a hierarchic classification of protein domain structures.
\newblock {\em Structure}, 5(8):1093--1109, 1997.

\bibitem{OS13}
Steve~Y. Oudot and Donald~R. Sheehy.
\newblock {Z}igzag {Z}oology: {R}ips {Z}igzags for {H}omology {I}nference.
\newblock In {\em Proc. 29th Annual Symposium on Computational Geometry}, pages
  387--396, June 2013.

\bibitem{overbeek:2000}
R.~Overbeek, N.~Larsen, G.~D. Pusch, M.~D’Souza, E.~Selkov~Jr, N.~Kyrpides,
  M.~Fonstein, N.~Maltsev, and E.~Selkov.
\newblock {WIT}: integrated system for high-throughput genome sequence analysis
  and metabolic reconstruction.
\newblock {\em Nucleic Acids Research}, 28(1):123--125, 2000.

\bibitem{Pachauri:2011}
D.~Pachauri, C.~Hinrichs, M.K. Chung, S.C. Johnson, and V.~Singh.
\newblock Topology-based kernels with application to inference problems in
  alzheimer's disease.
\newblock {\em Medical Imaging, IEEE Transactions on}, 30(10):1760--1770, Oct
  2011.

\bibitem{JKPark:2013}
J.~K. Park, Robert Jernigan, and Zhijun Wu.
\newblock Coarse grained normal mode analysis vs. refined gaussian network
  model for protein residue-level structural fluctuations.
\newblock {\em Bulletin of Mathematical Biology}, 75:124 --160, 2013.

\bibitem{Pecora:1997}
L.~M. Pecora, T.~L. Carroll, G.~A. Johnson, and D.~J. Mar.
\newblock Fundamentals of synchronization in chaotic systems, concepts and
  applications.
\newblock {\em Chaos}, 7:520-- 543, 1997.

\bibitem{Pendas:2003}
A.~M. Pend{\'a}s and V.~Lua{\~n}a.
\newblock Curvature of interatomic surfaces. i. fundamentals.
\newblock {\em The Journal of chemical physics}, 119(15):7633--7642, 2003.

\bibitem{Perea:2015b}
J.~A. Perea, A.~Deckard, S.~B. Haase, and J.~Harer.
\newblock Sw1pers: Sliding windows and 1-persistence scoring; discovering
  periodicity in gene expression time series data.
\newblock {\em BMC Bioinformatics}, 16:257, 2015.

\bibitem{Perea:2015a}
J.~A. Perea and J.~Harer.
\newblock Sliding windows and persistence: An application of topological
  methods to signal analysis.
\newblock {\em Foundations of Computational Mathematics}, 15:799--838, 2015.

\bibitem{GRASP2}
D.~Petrey and B.~Honig.
\newblock {GRASP2}: Visualization, surface properties, and electrostatics of
  macromolecular structures and sequences.
\newblock {\em Methods in Enzymology}, 374:492--509, 2003.

\bibitem{pohl:1980}
W.~F. Pohl.
\newblock {DNA} and differential geometry.
\newblock {\em The Mathematical Intelligencer}, 3(1):20--27, 1980.

\bibitem{Poincare:1890}
J.~H. Poincar\'{e}.
\newblock Sur le probleme des trois corps et les \'{e}quations de la dynamique.
  divergence des s�ries de m. lindstedt.
\newblock {\em Acta Mathematica}, 13:A3--A270, 1890.

\bibitem{Popelier:1996}
P.~L. Popelier.
\newblock On the differential geometry of interatomic surfaces.
\newblock {\em Canadian journal of chemistry}, 74(6):829--838, 1996.

\bibitem{Popelier:2005}
P.~L. Popelier.
\newblock Quantum chemical topology: on bonds and potentials.
\newblock In {\em Intermolecular forces and clusters I}, pages 1--56. Springer,
  2005.

\bibitem{Popelier:2000}
P.~L. Popelier, F.~M. Aicken, and S.~E. O'Brien.
\newblock Atoms in molecules.
\newblock {\em Chemical Modelling: Applications and Theory}, 1:143--198, 2000.

\bibitem{Quine:2006intensity}
J.~R. Quine, S.~Achuthan, T.~Asbury, R.~Bertram, M.S. Chapman, J.~Hu, and T.A.
  Cross.
\newblock Intensity and mosaic spread analysis from {PISEMA} tensors in
  solid-state {NMR}.
\newblock {\em Journal of Magnetic Resonance}, 179(2):190--198, 2006.

\bibitem{quine2004mathematical}
JR~Quine, Timothy~A Cross, Michael~S Chapman, and Richard Bertram.
\newblock Mathematical aspects of protein structure determination with nmr
  orientational restraints.
\newblock {\em Bulletin of mathematical biology}, 66(6):1705--1730, 2004.

\bibitem{Rader:2005}
A.~J. Rader, D.~H. Vlad, and I.~Bahar.
\newblock Maturation dynamics of bacteriophage hk97 capsid.
\newblock {\em Structure}, 13:413 -- 421, 2005.

\bibitem{rain:2001}
J.~C. Rain, L.~Selig, H.~De~Reuse, V.~Battaglia, C.~Reverdy, S.~Simon,
  G.~Lenzen, F.~Petel, J.~Wojcik, V.~Sch{\"a}chter, Y.~Chemama, A.~Labigne, and
  P.~Legrain.
\newblock The protein-protein interaction map of {Helicobacter} pylori.
\newblock {\em Nature}, 409(6817):211--215, 2001.

\bibitem{Raschke}
T.M. Raschke, J.~Tsai, and M.~Levitt.
\newblock Quantification of the hydrophobic interaction by simulations of the
  aggregation of small hydrophobic solutes in water.
\newblock {\em Proc. Natl. Acad. Sci. USA}, 98:5965--5969, 2001.

\bibitem{Richards:1977}
F.~M. Richards.
\newblock Areas, volumes, packing, and protein structure.
\newblock {\em Annual Review of Biophysics and Bioengineering}, 6(1):151--176,
  1977.

\bibitem{Rieck:2012}
Bastian Rieck, Hubert Mara, and Heike Leitte.
\newblock Multivariate data analysis using persistence-based filtering and
  topological signatures.
\newblock {\em IEEE Transactions on Visualization and Computer Graphics},
  18:2382--2391, 2012.

\bibitem{Robins:1999}
Vanessa Robins.
\newblock Towards computing homology from finite approximations.
\newblock In {\em Topology Proceedings}, volume~24, pages 503--532, 1999.

\bibitem{Rocchia:2002}
W.~Rocchia, S.~Sridharan, A.~Nicholls, E~Alexov, A~Chiabrera, and B.~Honig.
\newblock {Rapid grid-based construction of the molecular surface and the use
  of induced surface charge to calculate reaction field energies: Applications
  to the molecular systems and geometric objects}.
\newblock {\em Journal of Computational Chemistry}, 23:128 -- 137, 2002.

\bibitem{Sadre-Marandi:2014}
F.~Sadre-Marandi, J.~Liu, S.~Tavener, and C.~Chen.
\newblock Generating vectors for the lattice structures of tubular and conical
  viral capsids.
\newblock {\em Mol. Based Math. Biol.}, 2:128--140, 2014.

\bibitem{salgado:2006}
H.~Salgado, A.~Santos-Zavaleta, S.~Gama-Castro, M.~Peralta-Gil, M.~I.
  Pe{\~n}aloza-Sp{\'\i}nola, A.~Mart{\'\i}nez-Antonio, P.~D. Karp, and
  J.~Collado-Vides.
\newblock The comprehensive updated regulatory network of escherichia coli
  k-12.
\newblock {\em BMC bioinformatics}, 7(1):1, 2006.

\bibitem{Sander:1990}
P.~T. Sander and S.~W. Zucker.
\newblock Inferring surface trace and differential structure from {3D} images.
\newblock {\em IEEE Transactions on Pattern Analysis and Machine Intelligence},
  12(9):833--854, 1990.

\bibitem{Sanner:1996}
M.~F. Sanner, A.~J. Olson, and J.~C. Spehner.
\newblock Reduced surface: An efficient way to compute molecular surfaces.
\newblock {\em Biopolymers}, 38:305--320, 1996.

\bibitem{Schlick:1992trefoil}
T.~Schlick and W.~K. Olson.
\newblock Trefoil knotting revealed by molecular dynamics simulations of
  supercoiled {DNA}.
\newblock {\em Science}, 257(5073):1110--1115, 1992.

\bibitem{Schmider:2000}
H.~L. Schmider and A.~D. Becke.
\newblock Chemical content of the kinetic energy density.
\newblock {\em Journal of Molecular Structure: THEOCHEM}, 527(1):51--61, 2000.

\bibitem{Schroder:2005}
M~Schroder and R.~J. Kaufman.
\newblock The mammalian unfolded protein response.
\newblock {\em Annual Review of Biochemistry}, 74:739 -- 789, 2005.

\bibitem{shi:2000}
J.~B. Shi and J.~Malik.
\newblock Normalized cuts and image segmentation.
\newblock {\em {IEEE} Transactions on pattern analysis and machine
  intelligence}, 22(8):888--905, 2000.

\bibitem{XShi:2011}
X.~Shi and P.~Koehl.
\newblock Geometry and topology for modeling biomolecular surfaces.
\newblock {\em Far East J. Applied Math.}, 50:1--34, 2011.

\bibitem{Silva:2005}
V.~D. Silva and R~Ghrist.
\newblock Blind swarms for coverage in 2-d.
\newblock In {\em In Proceedings of Robotics: Science and Systems}, page~01,
  2005.

\bibitem{Silvi:1994}
B.~Silvi and A.~Savin.
\newblock Classification of chemical bonds based on topological analysis of
  electron localization functions.
\newblock {\em Nature}, 371(6499):683--686, 1994.

\bibitem{singh:2007}
G.~Singh, F.~M{\'e}moli, and G.~E. Carlsson.
\newblock Topological methods for the analysis of high dimensional data sets
  and { 3D} object recognition.
\newblock In {\em SPBG}, pages 91--100, 2007.

\bibitem{Singh:2008}
G.~Singh, F.~Memoli, T.~Ishkhanov, G.~Sapiro, G.~Carlsson, and D.~L. Ringach.
\newblock Topological analysis of population activity in visual cortex.
\newblock {\em Journal of Vision}, 8(8), 2008.

\bibitem{Skjaven:2009}
L.~Skjaerven, S.~M. Hollup, and N.~Reuter.
\newblock Normal mode analysis for proteins.
\newblock {\em Journal of Molecular Structure: Theochem.}, 898:42 -- 48, 2009.

\bibitem{Soldea:2006}
O.~Soldea, G.~Elber, and E.~Rivlin.
\newblock Global segmentation and curvature analysis of volumetric data sets
  using trivariate b-spline functions.
\newblock {\em IEEE Trans. on PAMI}, 28(2):265 -- 278, 2006.

\bibitem{GSong:2007}
G.~Song and R.~L. Jernigan.
\newblock vgnm: a better model for understanding the dynamics of proteins in
  crystals.
\newblock {\em J. Mol. Biol.}, 369(3):880 -- 893, 2007.

\bibitem{Spolar}
R.~S. Spolar and M.~T. Record~Jr.
\newblock Coupling of local folding to site{-}specific binding of proteins to
  dna.
\newblock {\em Science}, 263:777--784, 1994.

\bibitem{Stokely:1992}
E.~M. Stokely and S.~Y. Wu.
\newblock Surface parametrization and curvature measurement of arbitrary {3-D}
  objects: five practical methods.
\newblock {\em IEEE Transactions on pattern analysis and machine Intelligence},
  14(8):833--840, 1992.

\bibitem{Strombom:2007}
D.~Strombom.
\newblock {Persistent Homology in the cubical setting}.
\newblock {\em Master's Thesis, Lulea University of Technology}, 2007.

\bibitem{sumners:1992}
D.~W. Sumners.
\newblock Knot theory and {DNA}.
\newblock In {\em Proceedings of Symposia in Applied Mathematics}, volume~45,
  pages 39--72, 1992.

\bibitem{Tama:2005}
F.~Tama and C.~K. Brooks~III.
\newblock Diversity and identity of mechanical properties of icosahedral viral
  capsids studied with elastic network normal mode analysis.
\newblock {\em J. Mol. Biol.}, 345:299 -- 314, 2005.

\bibitem{Tama:2001}
F.~Tama and Y.~H. Sanejouand.
\newblock Conformational change of proteins arising from normal mode
  calculations.
\newblock {\em Protein Eng.}, 14:1 -- 6, 2001.

\bibitem{Tama:2003}
F.~Tama, M.~Valle, J.~Frank, and C.~K. Brooks~III.
\newblock Dynamic reorganization of the functionally active ribosome explored
  by normal mode analysis and cryo-electron microscopy.
\newblock {\em Proc. Natl Acad. Sci.}, 100(16):9319 -- 9323, 2003.

\bibitem{Tama:2002exploring}
F.~Tama, W.~Wriggers, and C.~L. Brooks.
\newblock Exploring global distortions of biological macromolecules and
  assemblies from low-resolution structural information and elastic network
  theory.
\newblock {\em Journal of molecular biology}, 321(2):297--305, 2002.

\bibitem{Tasumi:1982}
M.~Tasumi, H.~Takenchi, S.~Ataka, A.~M. Dwidedi, and S.~Krimm.
\newblock Normal vibrations of proteins: Glucagon.
\newblock {\em Biopolymers}, 21:711 -- 714, 1982.

\bibitem{javaPlex}
Andrew Tausz, Mikael Vejdemo-Johansson, and Henry Adams.
\newblock Javaplex: A research software package for persistent (co)homology.
\newblock Software available at \url{http://code.google.com/p/javaplex}, 2011.

\bibitem{Tirion:1996}
M.~M. Tirion.
\newblock Large amplitude elastic motions in proteins from a single-parameter,
  atomic analysis.
\newblock {\em Phys. Rev. Lett.}, 77:1905 -- 1908, 1996.

\bibitem{Twarock:2008}
R.~Twarock and N.~Jonoska.
\newblock Blueprints for dodecahedral dna cages.
\newblock {\em Journal of Physics A: Mathematical and Theoretical}, 41:304043
  --304057, 2008.

\bibitem{Uversky:2008}
V.~Uversky and A.~K. Dunker.
\newblock Controlled chaos.
\newblock {\em Sceince}, 322:1340 -- 1341, 2008.

\bibitem{Verbeek:1993}
P.~W. Verbeek and L.~J. Van~Vliet.
\newblock Curvature and bending energy in digitized {2D and 3D} images.
\newblock In {\em 8th Scandinavian Conference on Image Analysis, Tromso,
  Norway}, 1993.

\bibitem{veretnik:2007}
S.~Veretnik and I.~Shindyalov.
\newblock Computational methods for domain partitioning of protein structures.
\newblock In {\em Computational Methods for Protein Structure Prediction and
  Modeling}, pages 125--145. Springer New York, 2007.

\bibitem{Volkmann:2010}
N.~Volkmann.
\newblock Methods for segmentation and interpretation of electron tomographic
  reconstructions.
\newblock In {\em Methods Enzymol}, volume 483, pages 31--46, 2010.

\bibitem{vologodskii:1992}
A.~Vologodskii.
\newblock {\em Topology and Physics of Circular {DNA}}.
\newblock CRC Press, 1992.

\bibitem{von:2007tutorial}
U.~Von~Luxburg.
\newblock A tutorial on spectral clustering.
\newblock {\em Statistics and computing}, 17(4):395--416, 2007.

\bibitem{BaoWang:2015a}
B.~Wang and G.~W. Wei.
\newblock Parameter optimization in differential geometry based solvation
  models.
\newblock {\em Journal Chemical Physics}, 143:134119, 2015.

\bibitem{BaoWang:2016a}
B.~Wang and G.~W. Wei.
\newblock Object-oriented persistent homology.
\newblock {\em Journal of Computational Physics}, 305:276--299, 2016.

\bibitem{BaoWang:2016FFTS}
Bao Wang, Chengzhang Wang, and G.~W. Wei.
\newblock Feature functional theory - solvation predictor (fft-sp) for the
  blind prediction of solvation free energy.
\newblock {\em Journal of Chemical Information and Modeling}, submitted 2016.

\bibitem{BaoWang:2016HPK}
Bao Wang, Zhixiong Zhao, and G.~W. Wei.
\newblock Automatic parametrization of non-polar implicit solvent models for
  the blind prediction of solvation free energies.
\newblock {\em Journal of Chemical Physics}, 145:124110, 2016.

\bibitem{BaoWang:2016FFTB}
Bao Wang, Zhixiong Zhao, and G.~W. Wei.
\newblock Feature functional theory - binding predictor (fft-bp) for the blind
  prediction of binding free energy.
\newblock {\em Journal of Chemical Thoery and Computation}, submitted 2016.

\bibitem{BeiWang:2011}
Bei Wang, Brian Summa, Valerio Pascucci, and M.~Vejdemo-Johansson.
\newblock Branching and circular features in high dimensional data.
\newblock {\em IEEE Transactions on Visualization and Computer Graphics},
  17:1902--1911, 2011.

\bibitem{LinWang:2015}
Lin Wang, Lin Li, and Emil Alexov.
\newblock {pKa} predictions for proteins, {RNAs and DNAs with the Gaussian}
  dielectric function using {DelPhiPKa}.
\newblock {\em Proteins}, 83:2186--2197, 2015.

\bibitem{YWang:2004}
Y.~Wang, A.~J. Rader, I.~Bahar, and R.~L. Jernigan.
\newblock Global ribosome motions revealed with elastic network model.
\newblock {\em J. Struct. Biol.}, 147:302 -- 314, 2004.

\bibitem{Warshel:1976}
A.~Warshel and M.~Levitt.
\newblock Theoretical studies of enzymic reactions: Dielectric, electrostatic
  and steric stabilization of the carbonium ion in the reaction of lysozyme.
\newblock {\em Journal of Molecular Biology}, 103:227--249, 1976.

\bibitem{GWei:2000}
G.~W. Wei.
\newblock Wavelets generated by using discrete singular convolution kernels.
\newblock {\em Journal of Physics A: Mathematical and General}, 33:8577 --
  8596, 2000.

\bibitem{Wei:2009}
G.~W. Wei.
\newblock Differential geometry based multiscale models.
\newblock {\em Bulletin of Mathematical Biology}, 72:1562 -- 1622, 2010.

\bibitem{Wei:2005}
G.~W. Wei, Y.~H. Sun, Y.~C. Zhou, and M.~Feig.
\newblock Molecular multiresolution surfaces.
\newblock {\em arXiv:math-ph/0511001v1}, pages 1 -- 11, 2005.

\bibitem{Wei:2013}
Guo~Wei Wei.
\newblock Multiscale, multiphysics and multidomain models {I: Basic} theory.
\newblock {\em Journal of Theoretical and Computational Chemistry},
  12(8):1341006, 2013.

\bibitem{Wei:2016}
Guo~Wei Wei.
\newblock Mathematical molecular bioscience and biophysics.
\newblock {\em SIAM News}, 49(7), September 2016.

\bibitem{Wei:2012}
Guo-Wei Wei, Qiong Zheng, Zhan Chen, and Kelin Xia.
\newblock Variational multiscale models for charge transport.
\newblock {\em SIAM Review}, 54(4):699 -- 754, 2012.

\bibitem{White:1999}
S.~H. White and W.~C. Wimley.
\newblock Membrane protein folding and stability: Physical principles.
\newblock {\em Annual Review of Biophysics and Biomolecular Structure},
  28:319--365, 1999.

\bibitem{Whitley:2012}
David Whitley.
\newblock {\em Analysing molecular surface properties, in Drug design
  strategies: computational techniques and applications, ed. Lee Banting and
  Tim Clark}.
\newblock Royal Society of Chemistry, 2012.

\bibitem{KLXia:2014a}
K.~L. Xia, X.~Feng, Y.~Y. Tong, and G.~W. Wei.
\newblock Multiscale geometric modeling of macromolecules i: Cartesian
  representation.
\newblock {\em Journal of Computational Physics}, 275:912--936, 2014.

\bibitem{KLXia:2015a}
K.~L. Xia, X.~Feng, Y.~Y. Tong, and G.~W. Wei.
\newblock Persistent homology for the quantitative prediction of fullerene
  stability.
\newblock {\em Journal of Computational Chemsitry}, 36:408--422, 2015.

\bibitem{KLXia:2013d}
K.~L. Xia, K.~Opron, and G.~W. Wei.
\newblock Multiscale multiphysics and multidomain models --- { Flexibility} and
  rigidity.
\newblock {\em Journal of Chemical Physics}, 139:194109, 2013.

\bibitem{KLXia:2015f}
K.~L. Xia, K.~Opron, and G.~W. Wei.
\newblock Multiscale {Gaussian network model (mGNM)} and multiscale anisotropic
  network model {(mANM)}.
\newblock {\em Journal of Chemical Physics}, 143:204106, 2015.

\bibitem{KLXia:2014b}
K.~L. Xia and G.~W. Wei.
\newblock Molecular nonlinear dynamics and protein thermal uncertainty
  quantification.
\newblock {\em Chaos}, 24:013103, 2014.

\bibitem{KLXia:2014c}
K.~L. Xia and G.~W. Wei.
\newblock Persistent homology analysis of protein structure, flexibility and
  folding.
\newblock {\em International Journal for Numerical Methods in Biomedical
  Engineerings}, 30:814--844, 2014.

\bibitem{KLXia:2015c}
K.~L. Xia and G.~W. Wei.
\newblock Multidimensional persistence in biomolecular data.
\newblock {\em Journal Computational Chemistry}, 36:1502--1520, 2015.

\bibitem{KLXia:2015b}
K.~L. Xia and G.~W. Wei.
\newblock Persistent topology for {cryo-EM} data analysis.
\newblock {\em International Journal for Numerical Methods in Biomedical
  Engineering}, 31:e02719, 2015.

\bibitem{KLXia:2015e}
K.~L. Xia, Z.~X. Zhao, and G.~W. Wei.
\newblock Multiresolution persistent homology for excessively large
  biomolecular datasets.
\newblock {\em Journal of Chemical Physics}, 143:134103, 2015.

\bibitem{KLXia:2015d}
K.~L. Xia, Z.~X. Zhao, and G.~W. Wei.
\newblock Multiresolution topological simplification.
\newblock {\em Journal Computational Biology}, 22:1--5, 2015.

\bibitem{CXu:2003}
C.~Xu, D.~Tobi, and I.~Bahar.
\newblock Allosteric changes in protein structure computed by a simple
  mechanical model: hemoglobin t <--> r2 transition.
\newblock {\em J. Mol. Biol.}, 333:153 -- 168, 2003.

\bibitem{Xu:DSM:2006}
Guoliang Xu, Qing Pan, and Chandrajit~L. Bajaj.
\newblock Discrete surface modeling using partial differential equations.
\newblock {\em Computer Aided Geometric Design}, 23(2):125--145, 2006.

\bibitem{LWYang:2008}
L.~W. Yang and C.~P. Chng.
\newblock Coarse-grained models reveal functional dynamics--{I}. elastic
  network models--theories, comparisons and perspectives.
\newblock {\em Bioinformatics and Biology Insights}, 2:25 -- 45, 2008.

\bibitem{Yang:2006}
{Lee-Wei} Yang, A~Rader, Xiong Liu, Cristopher Jursa, Shann Chen, Hassan
  Karimi, and Ivet Bahar.
\newblock {oGNM:} online computation of structural dynamics using the gaussian
  network model.
\newblock {\em Nucleic Acids Research}, 34(Web Server issue):W24--W31, 2006.

\bibitem{YaoY:2009}
Y.~Yao, J.~Sun, X.~H. Huang, G.~R. Bowman, G.~Singh, M.~Lesnick, L.~J. Guibas,
  V.~S. Pande, and G.~Carlsson.
\newblock Topological methods for exploring low-density states in biomolecular
  folding pathways.
\newblock {\em The Journal of Chemical Physics}, 130:144115, 2009.

\bibitem{Yu:2007a}
S.~N. Yu and G.~W. Wei.
\newblock Three-dimensional matched interface and boundary {(MIB)} method for
  treating geometric singularities.
\newblock {\em J. Comput. Phys.}, 227:602--632, 2007.

\bibitem{ZYu:2008b}
Z.~Yu, M.~Holst, T.~Hayashi, C.~L. Bajaj, M.~H. Ellisman, J.~A. McCammon, and
  M.~Hoshijima.
\newblock Three-dimensional geometric modeling of membrane-bound organelles in
  ventricular myocytes: Bridging the gap between microscopic imaging and
  mathematical simulation.
\newblock {\em Journal of Structural Biology}, 164:304--313, 2008.

\bibitem{ZYu:2008}
Z.~Y. Yu, M.~Holst, Y.~Cheng, and J.~A. McCammon.
\newblock Feature-preserving adaptive mesh generation for molecular shape
  modeling and simulation.
\newblock {\em Journal of Molecular Graphics and Modeling}, 26:1370--1380,
  2008.

\bibitem{zelnik:2005}
Lihi Zelnik-manor and Pietro Perona.
\newblock Self-tuning spectral clustering.
\newblock In {\em Advances in Neural Information Processing Systems 17}, pages
  1601--1608. MIT Press, 2004.

\bibitem{YZhang:2009a}
Y.~Zhang, H.~Yu, J.~H. Qin, and B.~C. Lin.
\newblock A microfluidic dna computing processor for gene expression analysis
  and gene drug synthesisn.
\newblock {\em Biomicrofluidics}, 3(044105), 2009.

\bibitem{SZhao:2011a}
Shan Zhao.
\newblock Pseudo-time-coupled nonlinear models for biomolecular surface
  representation and solvation analysis.
\newblock {\em International Journal for Numerical Methods in Biomedical
  Engineering}, 27:1964--1981, 2011.

\bibitem{SZhao:2014a}
Shan Zhao.
\newblock Operator splitting {ADI} schemes for pseudo-time coupled nonlinear
  solvation simulations.
\newblock {\em Journal of Computational Physics}, 257:1000 -- 1021, 2014.

\bibitem{QZheng:2012}
Q.~Zheng, S.~Y. Yang, and G.~W. Wei.
\newblock { Molecular surface generation using PDE transform}.
\newblock {\em International Journal for Numerical Methods in Biomedical
  Engineering}, 28:291--316, 2012.

\bibitem{QZheng:2011a}
Qiong Zheng, Duan Chen, and G.~W. Wei.
\newblock Second-order {Poisson-Nernst-Planck} solver for ion transport.
\newblock {\em Journal of Comput. Phys.}, 230:5239 -- 5262, 2011.

\bibitem{QZheng:2011b}
Qiong Zheng and G.~W. Wei.
\newblock {Poisson-Boltzmann-Nernst-Planck model}.
\newblock {\em Journal of Chemical Physics}, 134:194101, 2011.

\bibitem{WZheng:2007}
W.~Zheng, B.~R. Brooks, and D.~Thirumalai.
\newblock Allosteric transitions in the chaperonin groel are captured by a
  dominant normal mode that is most robust to sequence variations.
\newblock {\em Biophys. J.}, 93:2289 -- 2299, 2007.

\bibitem{WZheng:2003}
W.~J. Zheng and S.~Doniach.
\newblock A comparative study of motor-protein motions by using a simple
  elastic-network model.
\newblock {\em Proc. Natl. Acad. Sci. USA.}, 100(23):13253 -- 13258, 2003.

\bibitem{Zhou:2008d}
Y.~C. Zhou, M.~J. Holst, and J.~A. McCammon.
\newblock A nonlinear elasticity model of macromolecular conformational change
  induced by electrostatic forces.
\newblock {\em Journal of Mathematical Analysis and Applications},
  340:135--164, 2008.

\bibitem{Zomorodian:2005}
A.~Zomorodian and G.~Carlsson.
\newblock Computing persistent homology.
\newblock {\em Discrete Comput. Geom.}, 33:249--274, 2005.

\end{thebibliography}

\end{document}